\documentclass[fleqn,usenatbib]{mnras}

\usepackage{graphicx}
\usepackage{subfigure}
\usepackage{epsfig}
\usepackage{multirow}
\usepackage{color}

\usepackage{url}
\usepackage{breakurl}
\usepackage{gensymb}
\usepackage{amsmath}
\usepackage{float}
\usepackage{multicol}
\usepackage{textcomp}
\usepackage{adjustbox,lipsum}
\usepackage{placeins}

\DeclareMathOperator{\sech}{sech}

\makeatletter
\long\def\@makecaption#1#2{%
  \vskip\abovecaptionskip
  \sbox\@tempboxa{{\bf #1.} #2}%
  \ifdim \wd\@tempboxa >\hsize
    {\bf #1.} #2\par
  \else
    \global \@minipagefalse
    \hb@xt@\hsize{\hfil\box\@tempboxa\hfil}%
  \fi
  \vskip\belowcaptionskip}
\makeatother

\title[\vspace{-0.5cm} Bulge and Disk Evolution in HST
Galaxies]{\vspace{-0.5cm} Measuring the Evolution of Bulge, Disk and
  Colour Gradients in HST Observations of Galaxies with 3D Modelling}
\author[N. Welikala et al.]
{\parbox{\textwidth}{\vspace{-0.5cm}N.~Welikala$^{\Edinburgh}$, 
L.~Miller$^{\Oxford}$, A.~N.~Taylor$^{\Edinburgh}$, G.~Congedo$^{\Edinburgh}$\thanks{E-mail:giuseppe.congedo@ed.ac.uk}
}\vspace{1cm}\\
\parbox{\textwidth}{
$^{\Edinburgh}$ Institute for Astronomy, University of Edinburgh, Royal
Observatory, Blackford Hill, Edinburgh, EH9 3HJ, UK\\
$^{\Oxford}$ Department of Physics, University of Oxford, Denys Wilkinson Building, Keble Road, OX1 3RH, UK\\
}}
\def\Edinburgh{1}
\def\Oxford{2}

\begin{document}

\maketitle

\label{firstpage}

\begin{abstract}
We measure galaxy structural properties and colour gradients using HST
images to trace the evolution of galaxy components. We
  develop and apply a physically motivated 3D bulge+disk modelling
  framework, jointly fitting the components in 2505
galaxies in GOODS-South across seven bands (bvizYJH) to $I_{\rm
  AB}=25.5$, accounting for different component ellipticities and
inclination-dependent dust extinction. Extinction strongly affects
structural parameters and colour gradients in $\sim$26\% of the
sample---primarily edge-on galaxies with central obscuration (B-band
face-on optical depth $\tau \simeq 4$) that reveal clear bulge
components in the near-infrared. Despite irregular morphologies, the
model captures observed colour gradients well. Bulges at $z\sim1$
differ markedly from $z\sim0$, with typical S\'ersic index $n \sim
1.0$ and bulge-to-disc size ratio $R_{\rm e}/h_{\rm d} \sim 0.15$,
suggesting most galaxies host pseudo-bulges formed via secular
evolution. Galaxy ellipticity correlates strongly with disk
scale-length. We trace bulge and disk evolution from $z\sim0$ to $z
\sim 2.5$: bulges are redder than disks (observed-frame) at $z < 1.4$,
but colours converge at higher redshifts and fainter
magnitudes. Redder galaxies show redder cores relative to their
outskirts, and brighter galaxies have redder
cores. These results provide new measurements of bulge
  and disk structural properties at $z\sim1$ and observational
  constraints on their evolution, while also providing empirically
  motivated input distributions for assessing colour gradient
  systematics in weak lensing surveys.
\end{abstract}


\section{INTRODUCTION}
\label{sec:intro}

The primary focus of this paper is measuring colour gradients in
galaxies, which provide important clues to the distribution of stellar
populations and assembly history. While spatially resolved
spectroscopy has provided kinematic and radial measurements for both
low-redshift \citep{spindler2017, Rowlands_2018, Wang_2018} and
high-redshift \citep{belli2017} galaxies, sample sizes remain
limited. Current and upcoming high-resolution imaging from
\textit{Euclid} \citep{mellier2025}, JWST \citep{gardner2006}, Nancy
Grace Roman Telescope \citep{spergel2015}, and ground-based adaptive optics can potentially provide such measurements for billions of galaxies, albeit with larger uncertainties \citep{liu2016}. However, high dust content at the peak of cosmic star formation ($z\sim$2), coupled with age-dust degeneracy, makes interpreting radial colour gradients challenging even with NIR observations \citep{forster-schreiber2020}.

In weak lensing, colour gradients are a systematic bias source for space-based shear measurements, particularly when using single broad-band filters with diffraction-limited PSFs. The bias arises from wavelength-dependent PSFs and spatially varying SEDs, resulting in incorrect PSF assignment \citep{voigt2012}. For \textit{Euclid}, multiplicative shear bias from colour gradients must be known to better than 1 part in 10$^4$ \citep{cropper2013}. One calibration approach uses emulated HST data \citep{semboloni2013, er2017}; alternatively, measured colour gradients from HST can be forward-modelled to quantify shear bias. In this work, we measure colour gradient distributions in $riz$ bands---the primary shear measurement passbands for \textit{Euclid} and \textit{Rubin Observatory}---and their evolution with redshift and galaxy properties.

Precise galaxy morphology estimates are crucial for studying size-mass
relation evolution \citep{vanderwel2014} and informing galaxy
formation models. They also provide essential priors for weak lensing
measurements where galaxy size correlates with ellipticity
\citep{kannawadi2019}. In particular, accurate distributions of bulge
and disc structural parameters from deep HST imaging---including
bulge-to-disc size ratios and S\'ersic indices---are essential inputs
for cosmic shear calibration in current and upcoming weak lensing
surveys. HST structural parameter distributions serve as inputs to
Bayesian shape estimation \citep{miller2013, pujol2017, congedo2024} and
calibration simulations \citep{hoekstra2017}. Ellipticity gradients from bulge-disc components with different inclination-dependent ellipticities are a potential systematic \citep{bernstein2010}, motivating 3D models that include these effects.

Previous structural parameter constraints from HST optical and near-infrared imaging (e.g., \cite{griffith_2012, vanderwel2014}) typically employed single S\'ersic fits \citep{sersic1963, margalef2016}, two-component bulge-disc models \citep{simard2002}, or machine learning approaches \citep{tuccillo2017}. These studies have fundamental limitations. First, they assume 2D intrinsic shapes rather than projections of inclined 3D light distributions, impacting measurement precision. Second, they largely neglect the connection between morphology and SEDs \citep{Lee_2013}, particularly for the bulge and disc components. For $z > 1.5$ galaxies, optical imaging samples rest-frame ultraviolet emission, making multi-wavelength constraints across the 4000\,\AA\ break essential \citep{haussler2013, Dimauro_2018}. Third, extinction effects---which preferentially suppress central light at bluer wavelengths and higher inclinations \citep{graham2008}---are typically ignored, biasing structural parameters.

In this paper, we address these limitations by: (1) defining 3D
analytical models for bulge and disc surface brightness; (2) modelling
extinction effects as functions of inclination angle, wavelength, and
dust opacity; (3) jointly fitting models (single S\'ersic and
two-component, with/without extinction) to HST CANDELS data; (4)
computing component structural parameters, ellipticities and colour
gradients---and their distributions; (5) measuring empirical relations
relevant for weak lensing and colour gradient evolution with redshift
and galaxy properties. We discuss implications for galaxy evolution
studies and weak lensing measurements.

Throughout, we adopt a $\Lambda$CDM cosmology with $H_0=70\,\mathrm{km}\,\mathrm{s}^{-1}\mathrm{Mpc^{-1}}$, $\Omega_{\mathrm{M}} = 0.27$, and $\Omega_{\Lambda} = 0.73$.

\section{GALAXY MODELS}
\label{sec:models}

The galaxy models are rendered as a dense collection of
Monte Carlo (MC) points that are drawn from an analytic function that
describes the surface brightness distribution for each component. This
follows the approach of \cite{bridle2010} for the GREAT08 Challenge
where galaxy images are created via photon shooting. The premise is
that if we draw random samples (photons) from an analytical function
that defines the light profile of the galaxy and then count the number of photons falling in each
pixel, the result is an image of the galaxy which satisfies the light
profile. The MC technique we use for shooting photons is inverse
transform sampling from a 1-D cumulative distribution function.
Table~\ref{tab:table_parameters} summarises the parameters of the models.

\begin{table}
\centering
\begin{tabular}{ | c | c | } 
\hline
 Symbol & Definition (units)\\ \hline
  
$\theta_d$ &  Inclination angle of disk ($\degree$)     \\
$\theta_b$ &  Inclination angle of bulge ($\degree$)     \\
$\theta_g$ &  Inclination angle of the galaxy ($\degree$)    \\
$\phi_d$ &     Position angle of disk ($\degree$)  \\
$\phi_b$ &     Position angle of bulge ($\degree$)  \\
$\phi_g$ &     Position angle of galaxy ($\degree$) \\
$h_d$ & disk scale length ($\arcsec$)\\
$R_e$ & bulge effective radius ($\arcsec$)\\
$R_e^g$ & Effective radius of galaxy ($\arcsec$) \\  
$z_0$ & vertical scale length of disk in $z$ direction ($\arcsec$)\\  
$n$  & bulge S\'ersic index \\
$n_g$  & S\'ersic index of galaxy from one-component fitting \\
$e_b$  &  Ellipticity of the bulge  \\
$e_d$  & Ellipticity of the disk \\
$e_g$  & Ellipticity of the summed bulge and disk model \\  
\end{tabular}
\caption{Table of galaxy parameters}
\label{tab:table_parameters}
\end{table}

\subsection{3D Disk}
\label{models:disk}

The disk assumes cylindrical geometry. The surface brightness profile
of the disk at radius $r$ is described in terms of the disk scale
length, $h_d$, of the intrinsic 3D shape such that 

\begin{equation}
I(r) = I(0) \,e^{-r/h_{\rm d}},
\label{eq:sb_disk_radial}
\end{equation}
\noindent
where $I(0)$ is the surface brightness at $r=$0. The disk model
therefore does not explicitly capture Type II or Type III disk breaks
(e.g.\ \citealt{pohlen2006}). The major axis of the
3D disk which lies along one of the axes of symmetry is defined as
$h_d$. We sample points in the radial direction from this
analytic profile and compute the cumulative probability density,
$F(r)$.

We draw 10,000 radial points from equally spaced intervals in $F(r)$
in order to generate a 3D map of the disk. We choose
the mid-point of each interval, $i$, and find the radial point, $r_i$
corresponding to that mid-point. We evaluate $F(r)$ at neighbouring
points $r=r_i$ and $r=r_i - dr$ where $dr=10^{-6} r_{\mathrm{max}}$ and
interpolate linearly between them in order to generate a new set of
radii, $r^{\mathrm{'} }_{\mathrm{i}}$. At each of these radial points,
we also generate the inclination angle, $\theta_{d,i}$, randomly from $0$ to
$2\pi \text{ rad}$. This means that some shot-noise is present in the
generated models. The coordinates of the projected model in $x-y$ plane are given by $x_i = r^{\mathrm{'}
}_{\mathrm{i}}\,\cos(\theta_{d,i})$, $y_i = r^{\mathrm{'}
}_{\mathrm{i}}\,\sin(\theta_{d,i})$. 

Next, we consider the intensity in the $z$ direction, perpendicular to
the plane of the disk. The surface brightness profile in the $z$
direction is given by \cite{vanderkruit1981a, vanderkruit1981b,
  vanderkruit1982a, vanderkruit1982b}
\begin{equation}
 I(z) = I(z_0) \sech^{2} (z/z_0).
\label{eq:sb_disk_z1}
\end{equation}
\noindent

For the disk, we generate 10,000 points in the $z$ direction, both
above and below the plane of the disk, out to a
truncation length $z_{max}$ by randomly sampling from a cumulative function
$F(z)$ such that the $z$ value of each sampled point is given by
\begin{equation}
z=z_0\tanh^{-1}(F(z)).
\label{eq:sb_disk_z2}
\end{equation}
\noindent

\subsection{Prolate bulge}
\label{models:bulge}

We model the 2D projection of the bulge assuming a 3D intrinsic
shape of a spheroid (ellipticity $\epsilon=0$) which is then flattened to make it prolate. For the bulge in the two-component fits, we assume a S\'ersic profile
\begin{equation}
I(r) = I(R_e) \, e^{-b_n [ (r/R_e)^{1/n} -1 ]} ,
\label{eq:sersic_sb_1}
\end{equation}
\noindent
where $I(R_e)$ is the surface brightness at the effective (half-light)
radius, $R_e$, which is equivalent to the length of the major axis in
our formalism, $n$ is the bulge S\'ersic index and $b_n\simeq 1.9992n -
0.3271$ for $0.5 < n < 10$ \citep{Capaccioli1989}. $R_e$ is fixed to 1
(arbitrary units) in the
model generation and is then scaled to each galaxy in the subsequent
fitting to the HST data. Points are generated
radially as for the disk except $a$ is now given by
\begin{equation}
a = b_n \left(\frac{r}{R_e}\right)^{1/n} .
\label{eq:sersic_sb_2}
\end{equation}
\noindent
Unlike the disk, the truncation radius of the bulge is set to a large
value ($r_{max}=20R_e$) when drawing points from the radial cumulative
function. At each radial point, we generate the inclination
angle, $\theta_b$, randomly (just as for $\theta_d$ for the disk) and then derive the $x_i$ and
$y_i$ coordinates of each point of the projected model from the
spherical coordinates, $r$, and angle $\theta$. We generate bulge
models in this way for each value of $n$ going from 0.5 to 4 in steps of 0.5.

The observed axis ratio for the projected 2D model and for a
given inclination angle, $\theta$, is given by
\begin{equation}
b/a = \sqrt{\cos^2(\theta_b)  + \beta_0^2 \,\sin^2(\theta_b)} ,   
\label{eq:method_two_comp_1}
\end{equation}
\noindent

where $\beta_0$ is the intrinsic flattening ratio of the bulge.


\begin{table}
\centering
\begin{tabular}{ | c | c | } 
\hline
Band&Wavelength[\AA]\\ [0.5ex] \\ \hline
UV09&912.0\\ 
UV13&1350.0\\ 
UV15&1500.0\\ 
UV16&1650.0\\ 
UV20&2000.0\\ 
UV22&2200.0\\ 
UV25&2500.0\\ 
UV28&2800.0\\ 
UV36&3650.0\\ 
b&4430.0\\ 
v&5640.0\\ 
i&8090.0\\ 
j&12590.0\\ 
k&22000.0\\
\end{tabular}
\caption{Wavelengths used in the extinction modelling
    described in Sect.~\ref{method:extinction_model}}
\label{tab:popescu_wavelengths}
\end{table}

\subsection{Frozen Parameters}
\label{models:frozen_params}

To reduce the number of free parameters and prevent over-fitting in
the modelling, we decided to freeze the following parameters:

\begin{itemize}
    \item $\textbf{$\beta_0$}$: We set $\beta_0=0.6$ which is within the range observed for bulges in extragalactic disk galaxies \citep{bertola_1991, mosenkov_2010}. Furthermore, \cite{mendezabreu2008} found no correlation between the intrinsic shapes of bulges and other properties such as bulge luminosity. We investigate the sensitivity of the galaxy parameter distributions on the choice of $\beta_0$ (see Appendix~\ref{app:priors_test}) and also find little sensitivity. This suggests that the choice of $\beta_0$ has little impact on the results, making it a suitable candidate for fixing.
    \item $\textbf{$z0/h_d$}$: We set $z0/h_d=0.17$ for the disk which is motivated by the consensus among similar studies of surface photometry of disk galaxies and simplified 3D modelling of the light distribution \citep[e.g.,][]{vanderkruit1982a, bizyaev2007}. In addition, in Appendix~\ref{app:priors_test} we show there is little sensitivity of the galaxy parameter distributions to the choice of $z_0/h_d$. 
    \item $\textbf{$r_{max}$}$, $\textbf{$z_{max}$}$: We adopt cutoff radii $r_{max}=4.5h_d$ and $z_{max}=5.0z_0$ for the disk to simplify the modelling process without introducing the significant bias and avoid over-fitting. For truncation lengths (radial and vertical), we adopt cutoff radii for the disk. This is based on empirical observations of nearby edge-on galaxies as detailed by \cite{vanderkruit1982a}. The evidence for evolution of the truncation radii with redshift, environment and galaxy type is mixed. \cite{kregel2004} found quite consistent values in nearby galaxies while \cite{pohlen2006} found some dependence on Hubble type but not on environment. \cite{trujillo2005} and \cite{azzollini2008} found a small to moderate evolution between ($z\sim$0) and ($z\sim$1).
  \end{itemize}

These choices still allow us to maintain a sufficient level of model complexity to capture the features of interest that related to structure and colour gradients in galaxies.

\subsection{Modelling the effect of extinction on surface brightness}
\label{method:extinction_model}

There are a few astrophysical effects that can potentially impact the
observed structural parameters and colour gradients in galaxies:
\begin{enumerate}
\item Dust extinction which preferentially extincts the light in the
  central region at bluer wavelengths and for more inclined disks.
  This can bias both the observed structural parameters, such as $h_d$, and the measured colour gradients. 
\item  Age and metallicity gradients which lead to different SEDs of
  the bulge and disk components \citep{sanchez-blazquez2014, roig2015,
    Sch_nrich2017}.
\item Contribution from patchy structure e.g., star-forming clumps
  which are not included in the modelling.
\end{enumerate}

Regarding point (1), previous corrections for extinction such as those
done by \cite{graham2008} use an analytic formula for $h_d$ as a function of inclination angle and
wavelength. This is the apparent scale length of the disk which increases with extinction because the central surface brightness is
more strongly suppressed than the outer one, leading to a larger
fitted $h_d$. It is more principled to explicitly account for
the change in the entire radial surface brightness profile due to central
extinction. In this work, we model it as a function of wavelength, 
inclination and dust opacity. The latter is important because the dust
amplitude can vary among the HST galaxies.
We therefore use synthetic images of axisymmetric dusty
disk galaxies, which were used to calculate the integrated attenuated SEDs of disk
galaxies by \cite{Popescu_2011} (referred to as P11) and
\cite{Tuffs_2004} (referred to as T04). They make use of a full radiative transfer
calculation of the propagation of starlight in the disk. For
simplicity, we will refer to the models provided to us by those authors as P11. The model images
were generated according to a prescription for the geometrical
distributions of stellar emissivity and dust opacity which is described in those papers.
The P11 models consist of an intrinsically dustless bulge, a thick disk of older stellar
populations and an associated dusty disk, and a thin disk
consisting of younger stellar populations and an associated dust
disk. The bulge light can still be extincted by the dust present
in either of the dust disks. Several studies have
  applied the models to correct for dust attenuation in large samples
  of local galaxies \citep{pastrav2013, battisti2017} while a few
  others have employed them at higher redshifts \citep{grootes2017,
    leslie2018}.

As the majority of the HST galaxies are
at $z \sim 1$, we expect
the thin disk to dominate the disk light and therefore neglect the thick disk (and
its associated dust disk) in this work. Out to $z\sim1$, galaxies
  have begun to show evolutionary differences from local samples but
  still largely maintain disk-dominated morphologies where the
  axisymmetric assumptions are reasonable \citep{sargent2007}. The P11 bulge is constrained to emit only in the rest-frame optical and Near Infrared
  ($bviJK$ bands) and assumes $n=$4 for the bulge, consistent with
  observations of low-redshift galaxies. This is in contrast to our
  modelling (Sect.~\ref{models:bulge}) where $n$ is allowed to vary
  and bulge stellar emissivity can also extend into the rest-frame
  UV. These limitations in the P11 models could introduce some
  noise and bias in the fitted structural parameters, particularly for $z>1$
  galaxies. Despite this, they provide a robust framework for incorporating extinction
  effects on surface brightness profiles and galaxy light spectra (and
  thereby colour gradients which we obtain from fitting the MC models
  to HST data and stratify in redshift bins). At higher redshifts, global
  modelling often requires handling diverse dust distributions\citep{buat2018}. The P11 models serve as a self-consistent and practical starting point for incorporating
  extinction effects at higher redshifts without fully resolving all
  dust distribution complexities. Using the P11 models in this way also helps maintain simplicity by not
requiring additional model assumptions or parameters to be introduced
in our own modelling, making the model-fitting more comptationally feasible.
  
The young stellar disk emits in these wavelengths as well but importantly, it contributes towards
all the stellar emissivity in the ultraviolet (shortwards of
the $b$ band). The precise monochromatic wavelengths corresponding to each band
are listed in Table~\ref{tab:popescu_wavelengths}. 
All images were sampled by the P11 authors at the same linear resolution of 34.5\,pc.
The images for the thin disk have dimensions
1399 $\times$ 1400 pixels with the galaxy centre at pixel
699.0,699.5 (the first pixel being 0-offset). The bulge images have dimension 439 $\times$ 440 pixels,
with the galaxy centre at pixel $(220.0, 219.5)$.

The P11 models were generated for a range of inclination angles sampled uniformly in
$\cos(\theta_d)$ (for the disk) and $\cos(\theta_b)$ (for the bulge) as well as wavelength $\lambda$ and total central face-on B-band
opacity $\tau$. The latter is the integral of dust opacity passing through
the centre of the galaxy along the $z$
direction from $-\infty$ and $+\infty$. Twenty one values of inclination angle
were used, corresponding to a uniform sampling of $1-\cos(\theta)$
from 0.01 to 1.0 in steps of 0.05. The values of $\tau$ were 0.0, 0.1,
0.3, 0.5, 1.0, 2.0, 4.0 and 8.0.

We then use the P11 models to apply extinction to the surface brightness profiles of the 3D MC points
of the bulge and the disk (described in Sections~\ref{models:bulge}
and ~\ref{models:disk}) such that the surface brightness and its
variation with inclination angle and wavelength represent those of observed HST galaxies. We follow these steps:  

\begin{enumerate}
\item We generate MC points with the same parameters as the
  P11 bulge and disk models for every value of $\theta_d$ and $\theta_b$.
\item We modulate the surface brightess profile of the MC points by
  applying a weight for extinction that is derived from the P11
  models, to each MC point. After binning the points into image pixels (see
  Sect.~\ref{method:single_sersic_fitting} for more details about
  gridding the points), we obtain a weight, $w_i$ per MC point, $i$,
  as the ratio of the pixel value that the particle falls into in the
  dusty P11 map to the corresponding pixel value in the dust-free P11 map, for every value of $\lambda$, $\tau$ and $\theta$. 
\item Since the models generated from the MC points will be fitted to the HST
  galaxies which have a distribution of inclination angles and 
  redshifts (and therefore correspond to different rest-frame
  wavelengths, see Sec.~\ref{method:fit_two_comp}), we obtain $w_i$
  for a given $\theta_{b,i}$ and rest-frame wavelength for the bulge, and similarly for the disk. For simplicity, instead of a two variable interpolation, we interpolate $w_i$ across the $\theta_{b,i}$ (and separately for $\theta_{d,i}$) values first for each MC point, and
  then determine the $w_i$ for a particular rest-frame
  wavelength when the models are fitted to the data. Thus, for each MC
  point, we first fit a cubic polynomial of $\cos(\theta_b)$ for the
  bulge (or cos($\theta_d$) for the disk) to the weight
  values such that $w_i(x)= a_0 + a_1x + a_2x^2 + a_3x^3$ where
  $x=\cos(\theta_b)$ (or $x=\cos(\theta_d)$). In Sect.~\ref{method:fit_two_comp}, we describe how we use the
coefficients $a_0$, $a_1$, $a_2$ and $a_3$ to obtain the correct surface
brightness for a particular rest-frame wavelength of the galaxy.
\end{enumerate}

\section{DATA}
\label{sec:data}


\begin{figure}
\centering
\includegraphics[width=0.9\columnwidth]{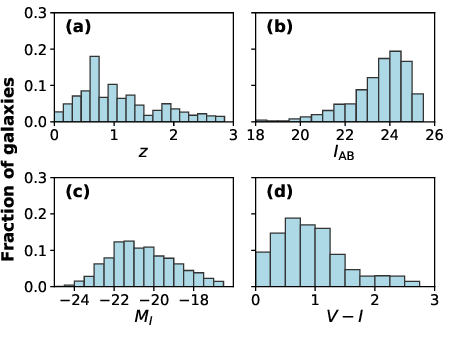}
\caption{Distributions for the magnitude-limited sample: (a) redshift,
  (b) magnitude $\rm I_{AB}$, (c) absolute magnitude $M_I$ and (d) $V-I$ colour. In this work, we examine the dependence of the colour gradient (measured by a model) on each of these quantities.}
\label{fig:sample_distributions}
\end{figure}


\begin{figure}
\centering
\includegraphics[width=0.9\columnwidth]{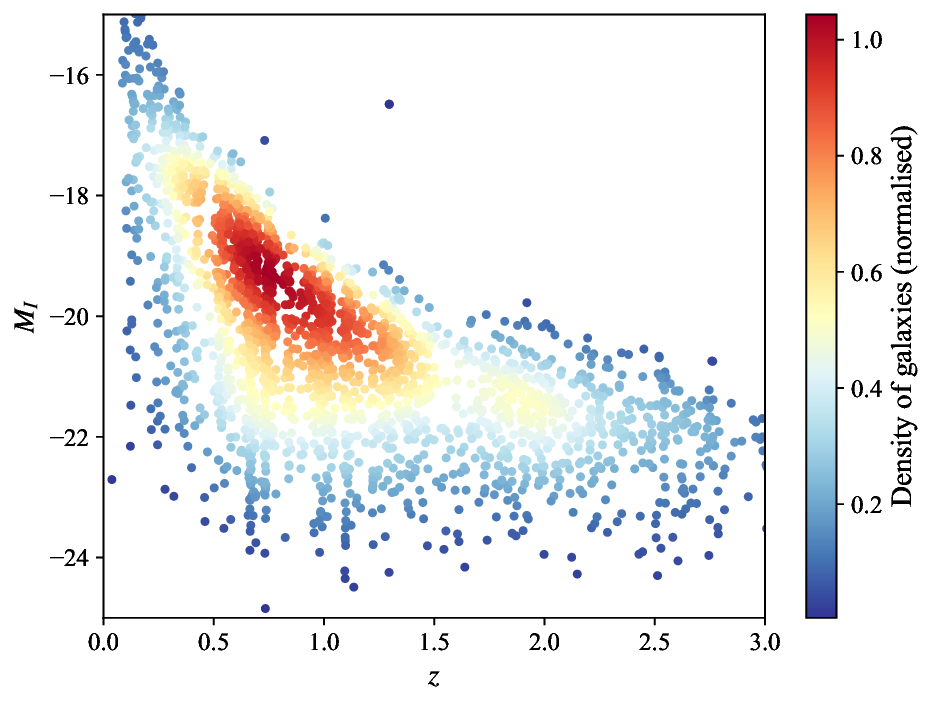}
\caption{The distribution of the galaxy sample in the $z$-$M_I$
  plane. The colour scale indicates the normalised density of galaxies.} 
\label{fig:sample_selection}
\end{figure}

\subsection{Sample selection and description}
\label{subsec:data_sample}

We start with a magnitude-limited ($I_{AB} < 25.5$) sample of 2505
galaxies. These are derived from a
photometric catalogue for the GOODS-South field from the Multiwavelength
Survey by Yale-Chile (MUSYC, \cite{cardamone2010}). The GOODS-South field  is particularly suitable for studying
resolved properties of galaxies (like colour gradients) due to its extraordinary depth of
multi-wavelength HST imaging (from UV through near-infrared) and lower
Galactic extinction compared to other CANDELS fields, combined with extensive complementary data from other observatories.
The catalog is matched on positions with
an HST Advanced Camera for Surveys (ACS) $z$-band detection catalog for the same field.\footnote{\tiny{\url{https://archive.stsci.edu/pub/hlsp/goods/catalog_r2/}}} The photometric
redshift catalog has a depth of $R_{AB}=25.3$ and is generated from deep
medium-band optical imaging from the Subaru telescope and 32-band
photometry. Grism spectroscopic redshifts from 3D-HST
\citep{brammer2012} are available for 637 galaxies in the sample and
we use photometric redshifts from \cite{cardamone2010} for the remainder.
Fig.~\ref{fig:sample_distributions} shows the redshift, absolute
magnitude, apparent magnitude and colour distribution of the
sample. The median redshift of the sample is 0.78 and the median
$I_{AB}=$23.6. Fig.~\ref{fig:sample_selection} shows how the galaxies
populate the absolute magnitude-redshift plane and also the
redshift-magnitude-colour space.
Our sample primarily consists of galaxies with  $0 < z < 2.5$,
$-22.0 < M_I < -17.0$ and $V-I < 1.5$ (observed frame), capturing the diversity of galaxy populations during the peak epoch of cosmic star formation history.

The ACS images and weight maps are version 2.0 of the reduced, calibrated, stacked mosaics
\footnote{\tiny{\url{https://archive.stsci.edu/pub/hlsp/goods/v2/h_goods_v2.0_rdm.html}}}
in the F435W ($b$), F606W ($v$), F775W ($i$) and F850LP ($z$)
bands \citep{giavalisco2004}. The ACS stacked and drizzled images in the GOODS-South field are organised
into sections each of $8192\times8192$ pixels. The pixel scale of the images is
0.03\arcsec. The Wide Field Camera 3 (WFC3\footnote{\tiny{\url{https://archive.stsci.edu/pub/hlsp/candels/goods-s/gs-tot/v1.0/hlsp_candels_hst_acs-wfc3_gs-tot_readme_v1.0.pdf}}}) images
in F105W ($Y$), F125W ($J$) and F160W ($H$) bands are obtained from
\textit{CANDELS} data in the GOODS-South field \citep{grogin2011,
  koekemoer2011}. We make postage stamp images in all
the bands centred on the coordinates of each target galaxy given in the catalogue. In the ACS
bands, the stamp dimensions are 340$\times$340 pixels. The
pixel scale of the WFC3 images 0.06\arcsec. In both ACS and WFC3,
the weight maps contain the inverse variance in each pixel arising
from background and read-out noise. We extract $168\times 160$ pixel
postage stamps in each WFC3 band.

Observations of the morphologies in the postage
  stamps suggest
  that a significant fraction of HST galaxies are edge-on. They also show
  central obscuration in the $v$ and $i$ images but a clear bulge in
  the near-infrared bands. We therefore hypothesise that the
  model-inferred colour gradients and other structural parameters of these galaxies would be biased if
extinction, particularly in the central regions of the galaxy, is
neglected. We refer to the subsample of HST galaxies with these
features (edge-on, no bulge in the $v$ or $i$ band but a bulge present
in the $H$ band) as \textit{Dusty} ($\sim$26\% of the sample) and therefore needing to be fitted with an extinction model, and describe the remainder of the galaxies as \textit{Dustless}.

\begin{figure}
\centering
\includegraphics[width=0.9\columnwidth]{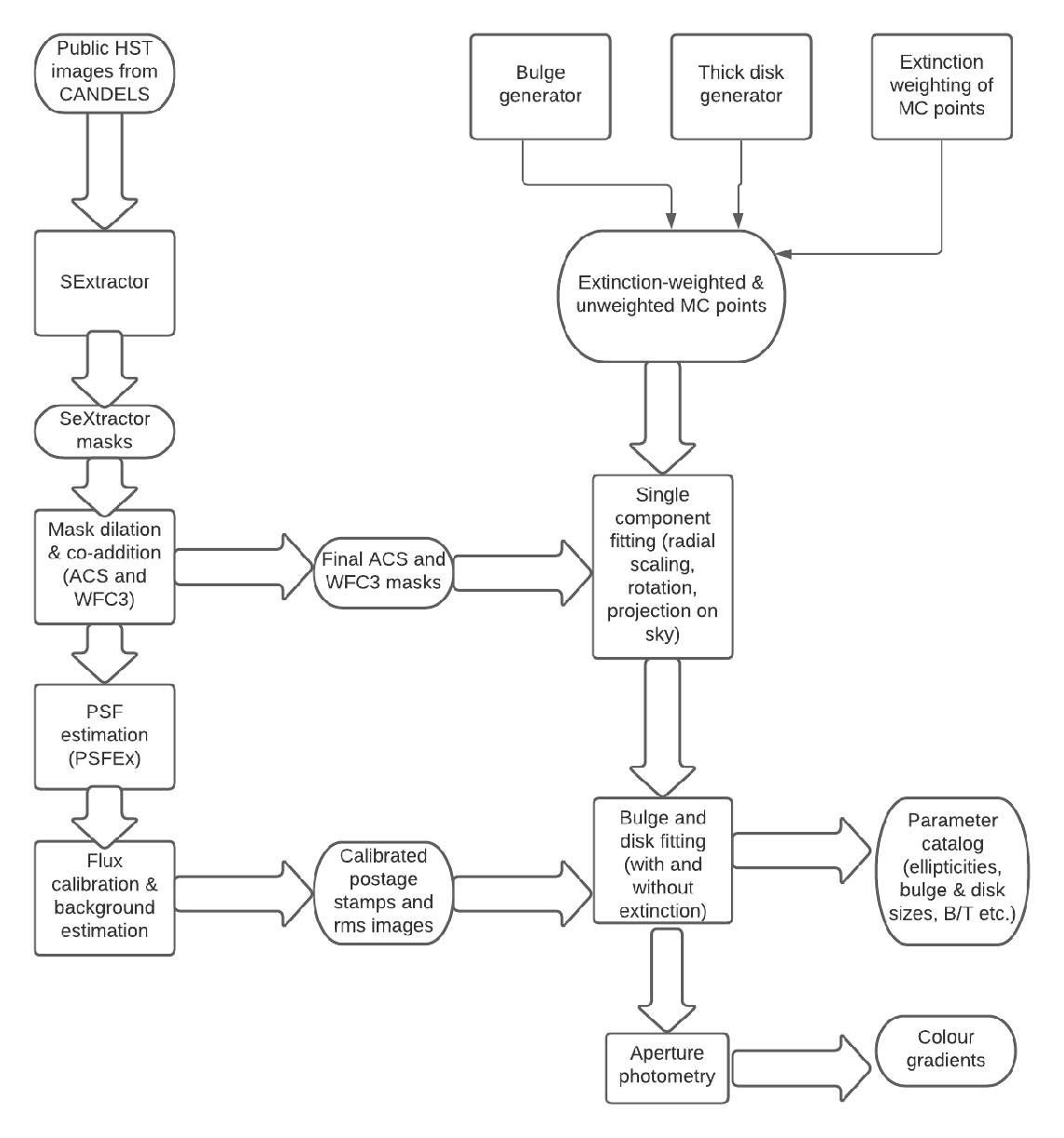}
\caption{Data flow diagram showing the processes that are followed to
  translate the public HST images through to the output galaxy catalogs.}
\label{fig:data_flow}
\end{figure}

\subsection{Pre-processing: Masks and PSF models}
\label{subsec:data_masking}

The data flow and steps in the modelling pipeline are summarised in Fig.~\ref{fig:data_flow}.
In order to mask out contaminating background galaxies, 
we first run \texttt{SExtractor} \citep{bertin1996} on the ACS and
WFC3 mosaics using a 5$\sigma$ detection threshold (\texttt{DETECT\_THRESH}), 
32 sub-thresholds (\texttt{DEBLEND\_THRESH}) and set the minimum
contrast parameter \texttt{DEBLEND\_MINCONT} to 0.0008 for the deblending. We also mask zero value
pixels in the image. Within a postage stamp, the light from the extended wings of
neighbouring galaxies could bias the model-fitting to the target
galaxy. To deal with this, we increase the area around background
galaxies by dilating the segmentation maps using a $5\times\,5$ pixel
box i.e, each masked pixel is dilated by a factor of 5 in both $x$ and $y$
directions. We co-add the individual ACS (and similarly for the WFC3)
masks to make a final ACS and WFC3 mask for each galaxy that is more 
conservative and comprehensive than the original\texttt{SExtractor}
maps.

For the results quoted in this work, we estimate the PSF in each passband using the
\texttt{PSFEx} software \citep{bertin2011} which models the PSF as a linear combination of
basis vectors. \texttt{PSFEx} requires a list of input stars generated
by running \texttt{SExtractor} on all the mosaics. In each section of
the GOODS-S field, we select the
stars from the magnitude-size plane where both quantities are measured by \texttt{SExtractor}. 
Although \texttt{PSFEx} can do its own pre-selection of stars, this procedure is important because in regions of the magnitude-size
plane where stars and galaxies coincide, we find that \texttt{PSFEx} itself cannot perform a clean distinction between them.  
The final stellar catalog is then input to \texttt{PSFEx} and we verify the
residual images generated after subtracting the local PSF model from each star. 
We obtain on average 30 stars that are usable for the PSF modelling in each
section.  Fitting a complex model of a spatially varying PSF to the
images is sub-optimal given the relatively low number of stars. 
We therefore assume the PSF is constant across the field and
generate a $35\times\,35$ pixel PSF image in each band. 
\texttt{PSFEx} can model a smooth spatial variation of the PSF
using a low-order polynomial so we tested what effect our assumption
of a constant PSF had on our measurements by allowing a polynomial 
variation of degrees 1 and 2 ($PSFVAR\_DEGREES=$ 1 and 2
respectively). We find no significant effects on the measured distributions of structural parameters
and colour gradients from the assumption of a constant PSF. We also repeated our
measurements with a Gaussian PSF (using the $\mathrm{FWHM}$ values quoted in the
ACS and WFC3 handbooks) and again find no significant differences.

\subsection{Flux calibration and background estimation}
\label{subsec:data_hst_flux_err}

The science images in all bands are count rates ($C$) in units of $e^{-}/s$. We convert to
AB flux densities in units of nano-Janskys (nJy) using the AB zero-point in that passband.


The value in each pixel of the weight maps is defined as $w=1/\sigma_{\mathrm{weight}}^2$ where $\sigma_{\mathrm{weight}}$ is
the uncertainty from the background and read-out noise in each pixel.
However, we also need to take into account the fact that when the images are drizzled, they are convolved with a
smoothing kernel which means that the output pixels in the final 
drizzled images are correlated \citep{casertano2000, guo2013}.
We do not include the smoothing due to drizzling in the models but
rather make an approximate allowance for the suppression of the noise in
the images due to correlated noise. We compute the
auto-correlation function (ACF) on the ACS and WFC3 images using
the software \texttt{astroRMS.py}\footnote{\tiny{\url{https://github.com/mmechtley/astroRMS/}}}
which is a python module that replaces the `MEDRMS' IRAF task for calculating pixel-to-pixel RMS noise in
astronomical images. The software takes into account autocorrelation and noise variations in
the images. We use only regions free of objects in the image
for the computation of the ACF by providing the masks of the detected
objects (described above) to the software. We compute the ACF in the drizzled science
image in each band within an aperture radius of 5 pixels (0.15$\arcsec$). The tool computes
(1) the peak value $\psi_0$ of the ACF; (2) the background level $\phi_b$ in the ACF
image; (3) the integral of the ACF $\phi_p$ which gives the total power within
the aperture radius. After subtracting off the background level from (1) and (3), 
we compute the RMS suppression factor $f=\phi_p/\phi_0$ due to pixel
correlations. The mean values of $f$ we find in each band are: 2.00 ($b$), 1.79 ($v$), 1.91
($i$), 1.74 ($z$), 2.50 ($Y$), 2.44 ($J$), 2.43 ($H$). The weight maps
provide the expected noise per pixel that the images would have in the
absence of the above correlations. The RMS obtained from the weight
maps is therefore divided by the value of $f$ in each band. We
then fit a 1-component disk model (see Sec.~\ref{method:single_sersic_fitting}) to each galaxy image (excluding
masked regions) in order to estimate the Poisson noise from the galaxy
itself in each band The final RMS in each pixel is obtained by adding
the Poisson noise and the suppressed noise from the
weight maps in quadrature.



%


The drizzled science images (and hence the postage stamps) have been background-subtracted but are
found to contain some residual background variations. These residuals
can bias the fitted quantities leading to a small outlier population of
galaxies with very large effective radii $R_{\mathrm{eff}} > 2.0\arcsec$. We find a local estimate of the
background to be more optimal than a global estimate when dealing with background
fluctuations across the image. We employed different annulus radii based on iterative testing to
optimise signal-to-noise across all bands. For $b$-band images, we
found that positioning annuli at $ 2.0\arcsec < r < 3.0\arcsec$ (the
galaxy centre is obtained from the model fits) yielded optimal results
compared to the other bands where the annulus was defined as $ 2.75\arcsec < r <
3.5\arcsec$. This is consistent with findings from CANDELS that
galaxies typically exhibit more compact light profiles at bluer
wavelengths \citep[e.g.,][]{wuyts2012}. Our iterative testing focused
on minimising contamination from source flux while being close enough
to capture the local background variations around each galaxy and
having sufficient pixels for statistical
robustness. Since a fixed angular background annulus
  is used for all sources, some residual contamination from extended
  galaxy light cannot be excluded for the largest or most extended galaxies in the
  sample. We subtract the median background level in the annulus from the flux density of every pixel in
the image.


\section{FITTING TO THE HST DATA}
\label{method:fitting}

\subsection{Single S\'ersic fitting to galaxy}
\label{method:single_sersic_fitting}

The single S\'ersic fitting described in
Section~\ref{method:single_sersic_fitting} provides starting points for the
two component fits. We fit jointly to all the bands for each galaxy with a
one-component model obtained from Eq.~(\ref{eq:sersic_sb_1})
with $R_e=R_e^g$, $n=n_g$ and $b_n=b_{ng}$,

\begin{equation}
I(r) = I(R_e^g) \, e^{-b_n^g \left[ \left(\frac{r}{R_e^g}\right)^{1/n_g} - 1 \right]}
\label{eq:sersic_sb_1}
\end{equation}
\noindent
and no extinction, as the goal is to
obtain preliminary estimates of galaxy parameters that will then be
used as initial guesses for the two component fitting. 
We adopt a S\'ersic index $n_g$ that varies from 0.5 to 6.0 in steps
of 0.5. \cite{shibuya2015} find that $n_g$ values above 6.0 for a single component 
are largely disfavoured by the HST data. They find a best-fit $n_g\sim1.5$ for one-component fitting to
star-forming galaxies at $z\,$$\sim$0$-$6 in CANDELS data. We fit a
S\'ersic model (characterised by $n_g$) with
the following free parameters: (1) effective (half-light) radius
$R_{e}^{g}$ of the galaxy; (2) pseudo-ellipticities $e_1'$ and $e_2'$ which are the ellipticities that
an infinitely thin disk would have if viewed at the same inclination
angle $\theta_g$; (3) centroid position $x_0$, $y_0$ and (4) flux densities in all bands
which are computed analytically (see below). The inclination angle of
the galaxy, $\theta_g$, and its
position angle, $\phi_g$, are related to $e_1'$ and $e_2'$ by
\begin{equation}
  \theta_g = \sin^{-1}(|e'|) ,
  \,\,\,\,\,;   {\rm and }  \,\,\,\,\,
  \phi_g  =  0.5 \tan^{-1}(e_2'/e_1')
\label{eq:method_one_comp_1}
\end{equation}
\noindent
where $|e'|^2=e_1'^2 + e_2'^2$. We find the pseudo-ellipticities $e_1'$ and $e_2'$ are a more robust set of
parameters to optimise than $\theta_g$ and $\phi_g$. We apply the
prior $|e'| < 1$ in the fitting. The equations for
computing $\chi^2$  and the flux density of the galaxy in each band
are given in Appendix~\ref{app:single_sersic_fits}.

\subsection{Two component fitting}
\label{method:fit_two_comp}

The results in this work come primarily from the joint bulge and disk
fitting to the HST data. In the two-component fitting, we assume the bulge and disk are co-centered and co-aligned
($\theta_b=\theta_d=\theta_g$ and $\phi_b=\phi_d=\phi_g$), with the ratio of their scale lengths
being a parameter in the fit.

\subsubsection{Bulge}
\label{method:fit_bulge}

For the bulge, we fit a S\'ersic model characterised by a $n$ (over
the range $0.5-4.0$) where the free parameters in the fit are: (1) the bulge effective radius, $R_{e}$; (2) the inclination
and position-angle dependent quantities $e_1$ and $e_2$ defined in Sect.~\ref{method:single_sersic_fitting}; (3) the
pixel coordinates in the 2D image, $x_0$ and $y_0$; and (4) flux densities
in all bands.  The bulge is treated as a spheroid which is flattened
to make it prolate, as described in Sect.~\ref{models:bulge}. We apply
the following operations to the MC points generated from the bulge model:

\begin{enumerate}
\item Compute the extinction weights for the bulge as described in Sect.~\ref{method:extinction_model}. The coefficients
  $a_0$, $a_1$, $a_2$ and $a_3$ in Sect.~\ref{method:extinction_model} are used to compute
$w_i(\theta_b)$ for each particle $i$ at an inclination angle $\theta_b$ (or equivalently
$e_1^{\prime}$ and $e_2^{\prime}$ according to Eq.~(\ref{eq:method_one_comp_1}) that is
fitted. We interpolate the $w_i$ across the 5 rest-frame wavelengths
$\lambda_j$ that the P11 models are generated for in order to find the weight for each HST
passband, for that $\theta_b$. Dust-free models (corresponding to $\tau=0$)
have $w_i=1$.
\item Apply a radial scaling to all points in the model: $x'=R_e\,x$,
  $y'=R_e\,y$  (where $R_e$ is set to 1 at
  the start).
\item Compute the observed minor-to-major axis ratio $b/a$ of the bulge for a given $\theta_b$ (derived from $e_1$ and
$e_2$) assuming an intrinsic 3D spheroid (see Sect.~\ref{models:bulge}). The
new coordinates in the plane of the sky, $x''$ and $y''$, are then
defined as $x^{\prime\prime} = x^{\prime}$ and $y^{\prime\prime} = \frac{b}{a} y^{\prime}$. 

\item Rotate the model about the new $z'$ axis (in the plane of the
  sky) by the position angle $\phi_b$ to get the final
  projected coordinates on the image, $x_f$, $y_f$.
\item Pixelate the model (using the ACS or WFC3 pixel size) by gridding up particles that fall within each
  pixel (an order $N_{p}$ operation if $N_{p}$ is the total number of pixels in
  the image). Given the weights computed in step (i), the total intensity in each pixel is given by
  $\sum_{i}{w_i}$ which accounts for the effect of any extinction on the
  surface brightness profile. 
\item Convolve the bulge model with the PSF model for that band using a Fast Fourier
  Transform (performed with the \texttt{fftw3} library in C).
\item Normalise the PSF-convolved model flux density to 1.
  
\end{enumerate}

The final projected coordinates on the image, $x_f$, $y_f$, in step
(v), after applying the final rotation defined by the position angle $\phi_b$, are
\begin{align}
  x_f = x^{\prime\prime} \cos(\phi_b) - y^{\prime\prime} \sin(\phi_b)
  \,\,\, {\rm and } \,\,\,
  y_f = x^{\prime\prime} \sin(\phi_b) +  y^{\prime\prime} \cos(\phi_b) .
 \label{eq:method_two_comp_3}
\end{align}

The component flux densities are computed
analytically in each band as described in
Appendix~\ref{app:amplitudes} and are for the stellar
components in the absence of dust.

\subsubsection{Disk}
\label{method:fit_disk}

The bulge is linked to the disk through the ratio of bulge effective
radius, $R_e$, to disk scale length, $h_d$. Since individual fits to $n$
and $R_e/h_d$ in galaxies are noisy \citep{haussler2013}, we want to fit $n$ and
$R_e/h_d$ globally i.e., for the full sample. We
arrange this by fitting on a grid for each galaxy so that we can then
combine measurements from all galaxies to get the global best values.
We then use these global values to find the best-fit model
colour gradients. We allow $R_e/h_d$ to vary from 0.05 to 1.0 in steps
of 0.05. The free parameters in the two-component fitting
are then: $h$, $e_1$, $e_2$, $x_0$, $y_0$ and $F_k$ (flux density through the
$k$th filter).

We use the MC points generated from the disk model described
Sect.~\ref{models:disk} and apply the same operations as for the bulge
except that we apply the radial scaling to all $x$, $y$ and
$z$ points in the disk (since the disk is 3D) and skip step (iii). The coordinates of the inclined model $x''$ and $y''$ are now given by
\begin{equation}
  x'' =x'
\,\,\,\, {\rm and} \,\,\,\,
y'' = y' \cos(\theta) - z'' \sin(\theta).
\label{eq:method_two_comp_3}
\end{equation}
\noindent
We then follow steps (iv)-(vii) as for the
bulge. Local values of $w_i>1$ can occur in the P11
  models that include multiple anisotropic scattering
  \citep{Tuffs_2004}, typically exceeding unity by $\sim1$--$3$ per
  cent at intermediate-to-outer disk radii; this is consistent with
  redistribution of light by scattering, and the resulting disk
  surface-brightness distribution can therefore depart slightly from a
  pure exponential. We optimise the $\chi^2$ of the joint fit of the bulge and disk to the
\textit{bvizYJH} bands. We computed the flux densities of the unextincted stellar components
as for the bulge (Appendix~\ref{app:amplitudes}). In order to measure colour
gradients for galaxies fitted by an extincted model, we multiply the
flux density of the unextincted stellar component by the best-fit extincted
model. So the total flux density in a particular pixel will be given by

\begin{equation}
f_{t} = a_bm_b + a_dm_d ,
\end{equation}
\noindent
where $a_b$ and $a_d$ are the best-fit unextincted stellar amplitudes
and $m_b$ and $m_d$ are the best-fit model flux density of the bulge
and disk respectively (either extincted or
unextincted depending on which is selected as the best fit) which are normalised to unity.

The model parameters $n$,  $R_e/h_d$ and $\tau$ are defined on a discrete grid in this work. In addition,
these bulge and disk parameters can be degenerate.
Therefore, we first find optimal values of $n$ and $R_e/h_d$ for the
sample. We assume no extinction ($\tau=$0) in all galaxies and stack the $\chi^2$ in the
$[n,R_e/h_d]$ space. We then apply extinction modelling to those
galaxies which have $f_{bulge,o}^v=0$ and $f_{bulge,o}^H > 0$ i.e.,
where the dust-free models yield zero bulge amplitude in the optical
and a positive bulge amplitude in the near-infrared image, consistent with our observations of the \textit{Dusty} HST
subsample (Sect~\ref{subsec:data_sample}). With $[n, \tau, R_e/h_d]$ thus fixed for both the
\textit{Dustless} or \textit{Dusty} samples, we fit for the remaining
bulge and disk parameters. Model parameter uncertainties are computed according to
Appendix~\ref{app:errors}.

Occasionally, the Downhill Simplex method converges onto a
local, instead of a global, minimum. We implement a few techniques to help ensure a
global minimum is found:
\begin{itemize}
\item The best-fit parameter values obtained from one-component
  fitting (see Section~\ref{method:single_sersic_fitting}) are used as
  starting points for the two-component fitting in the no extinction
  ($\tau=$0) case. When fitting the $\tau > $0 models to the HST data, we use the
  $\tau=$0 best fits as the starting points. This ensures the two-component fitting does not start a
  long way from the global minimum. 
\item Once the two-component fitting converges to a solution, we explore
  the $\chi^2$ around this minimum by evaluating it at equally
  spaced steps in each parameter using a step size of 5$\%$ of the
  parameter value at the minima. If a lower $\chi^2$ is found during
  this search, then that is adopted as the global minimum instead. 
\end{itemize}

\begin{figure}
\centering
\includegraphics[width=0.9\columnwidth]{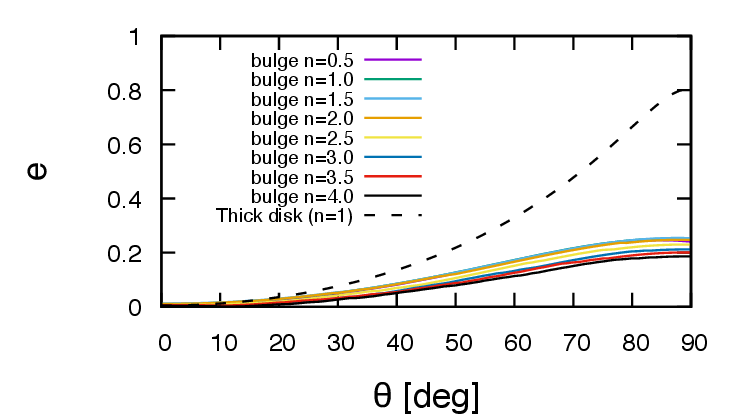}
\caption{Relation between inclination angle and ellipticity for
  different galaxy components, i.e., prolate bulges with different $n$ and unextincted disk models. The ellipticity
  is computed on the noise-free and PSF-free models for each component
  using the quadrupole moments described in Sect.~\ref{subsec:method_ellipticity}.
}
\label{fig:theta-ellipticity}
\end{figure}

\subsection{Computing ellipticities from the best-fit models}
\label{subsec:method_ellipticity}

The calibration of the shear measurement in upcoming weak lensing surveys will rely on accurate distributions of galaxy
priors. One important relation is the one between galaxy ellipticity and other
galaxy parameters such as disk scale length \citep{kannawadi2019}.
The ellipticity for the bulge, disk and galaxy (summed bulge and disk) are derived from the
quadrupole moments $Q_{xx}$, $Q_{xy}$ and $Q_{yy}$
\citep{Schneider_2006}. The real and imaginary components of the
ellipticity are defined as

\begin{equation}
 e_1 =   \frac{Q_{xx} - Q_{yy}}{ Q_{xx} + Q_{yy} + 2\sqrt{Q_{xx}Q_{yy}
     - Q_{xy}^2} }    ,  
 \label{eq:ellipticity_e1}
\end{equation}
\noindent
and-
\begin{equation}
  e_2 = \frac{2Q_{xy}}{ Q_{xx} + Q_{yy} + 2\sqrt{ Q_{xx}Q_{yy} - Q_{xy}^2}    }.
  \label{eq:ellipticity_e2} 
\end{equation}
\noindent
The ellipticity (which is quoted in the catalog) is then given as

\begin{equation}       
 |e| = \sqrt{e_1^2 + e_2^2} .
 \label{eq:ellipticity_e}
\end{equation}
\noindent
The ellipticities of the bulge, disk and galaxy (summed intensities of
the bulge and disk models) are denoted by $e_b$, $e_d$ and $e_g$
respectively. In Fig.~\ref{fig:theta-ellipticity}, we show the
dependence on the inclination angle of the ellipticity of a prolate bulge of varying
$n$ as well as an unextincted disk. The relation is obtained
by generating mock bulge and disk components at many values of
$\theta_b$ and $\theta_d$ respectively.
It is worth noting that since we fix the bulge flattening parameter
$\beta_0$ in the fitting to HST galaxies, the range of observed
bulge ellipticities will be constrained as a
result.


\section{RESULTS}

The catalogue generated by both the one-component and two-component
fitting is summarised in Table~\ref{table:catalog}. Here we describe the results from each type of fit to the HST galaxies.

\begin{figure}
\centering
\includegraphics[width=0.45\columnwidth]{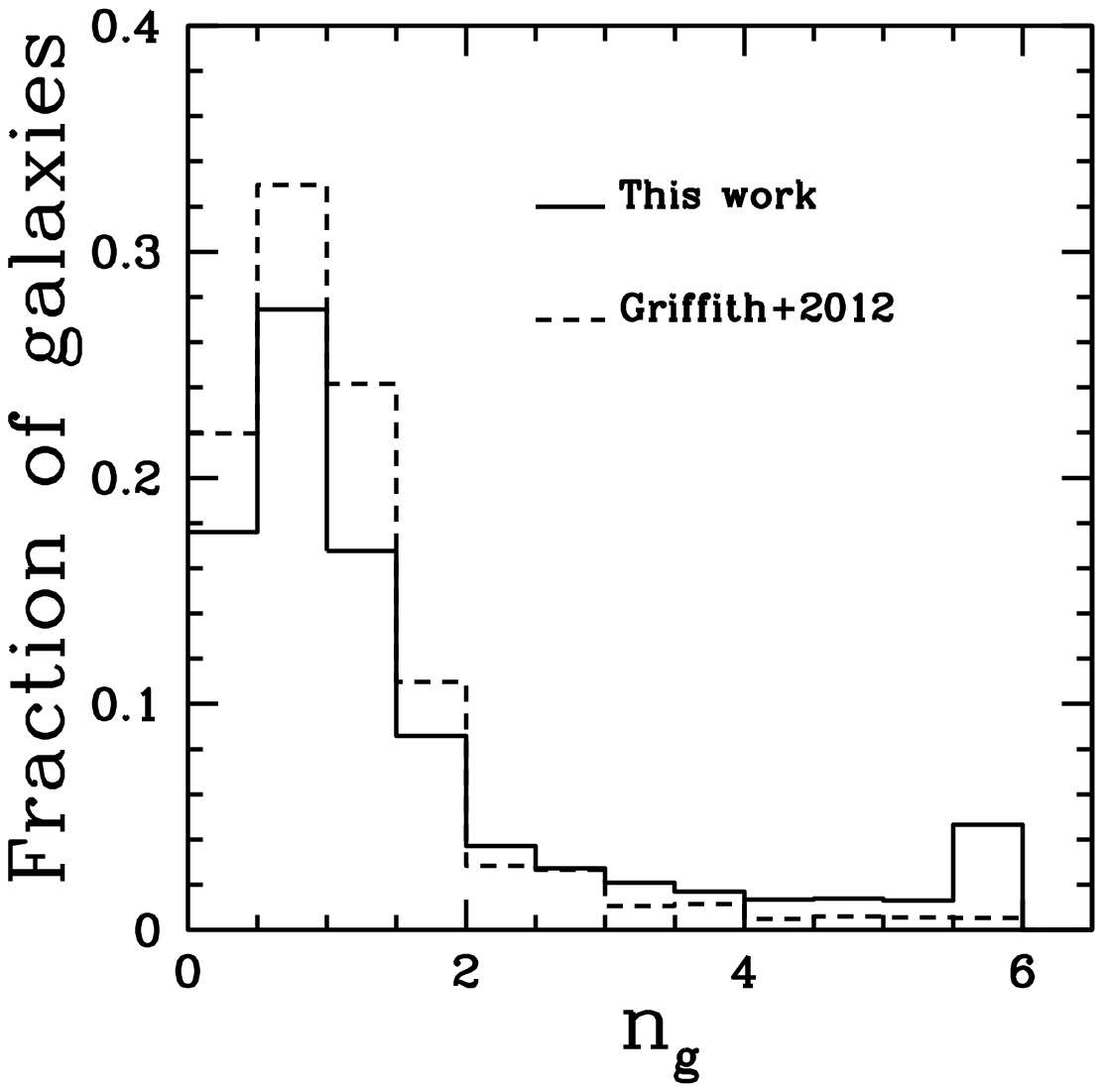}
\hfill
\includegraphics[width=0.45\columnwidth]{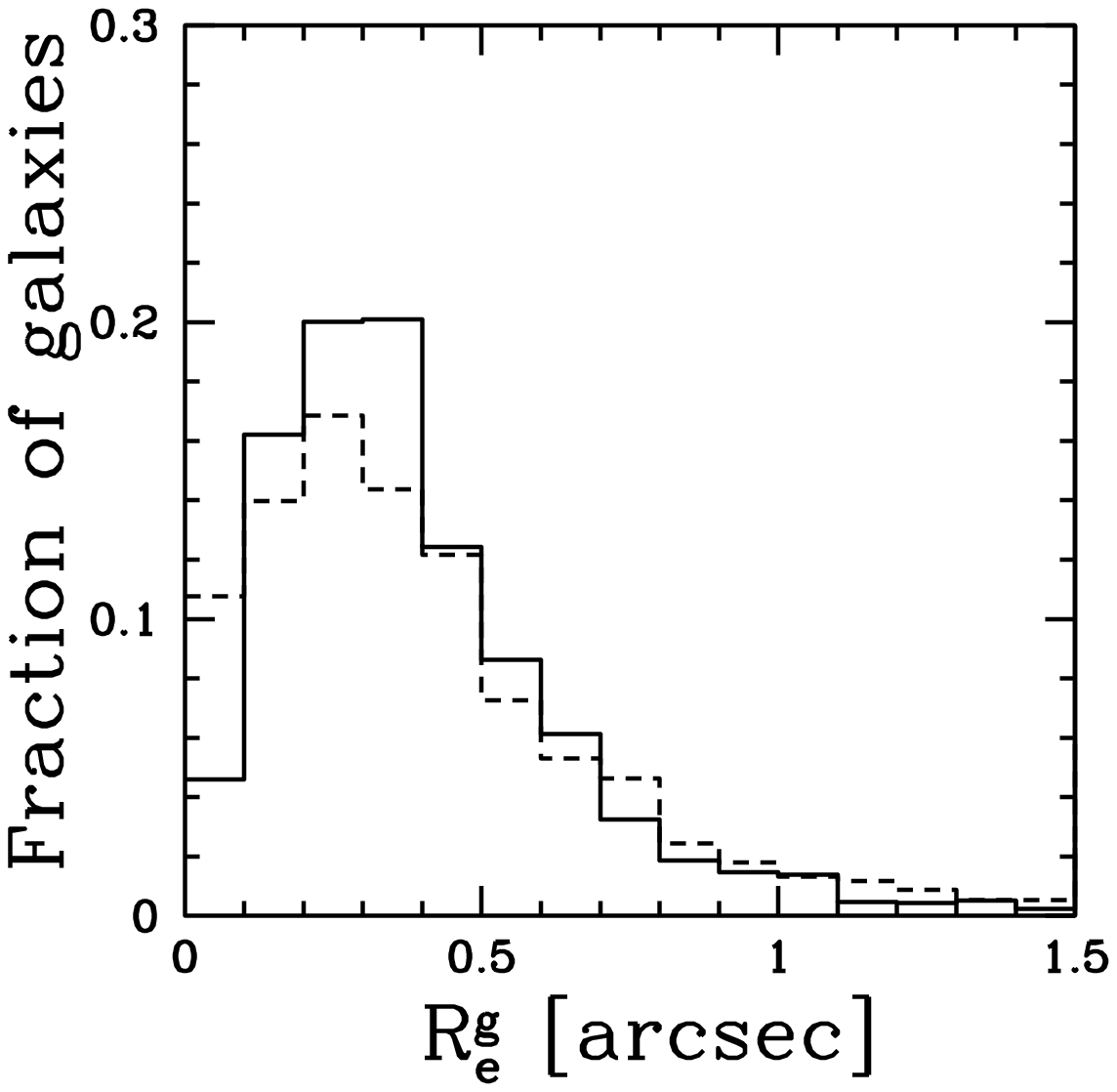}
\caption[Distributions of $n_{\rm g}$ and $R_e^{\rm g}$]{Distributions of key galaxy structural parameters, S\'ersic
  index $n_{\rm g}$ (left) and half-light radius of galaxy $R_e^{\rm g}$
  (right), from two methods: single component 3D multi-band S\'ersic fits (this work) and single-band S\'ersic fits
  with GALFIT by \protect\citet{griffith_2012} for $I_{AB} < 25.5$ (see
  Sec.~\protect\ref{method:single_sersic_fitting}). The results of the single
  component fits are also used as starting points for the two-component fits.}%
\label{fig:structural_params_1comp}
\end{figure}

\subsection{Single S\'ersic fits without extinction}

We start first with multi-band one-component S\'ersic fits (no
extinction applied) which are used as starting points for the
two-component fitting. These fits also give us the
first insights into the sizes and surface brightness profiles of this
sample with these newly constructed 3D models to a depth $I_{AB} <
25.5$. In Fig.~\ref{fig:structural_params_1comp}, we show the
distribution of $n_g$ and $R_e^g$ for the one-component model. We find a median $n_g$ of 1.5, implying the galaxies in the
sample have moderately concentrated central regions (slightly more concentrated
than a purely exponential profile i.e., with $n=1$). These would
normally correspond to intermediate-type galaxies on the Hubble
Diagram e.g., barred spirals or ellipticals with disky isophotes. In terms of the sizes,
the median $R_e^g$ as measured in the $v$ band is 0.35$\arcsec$ which
is approximately 12 times larger than the ACS pixel size and around 4 times larger than the WFC3
pixel size. $R_e^g$ in Eq.~(\ref{eq:sersic_sb_1}) is also approximately 6 times larger than the PSF
HWHM for ACS and 4 times larger than the HWHM for the WFC3 camera.
This means that most of the galaxies in the sample are well resolved.
We investigate this picture of galaxy structure further with 3D bulge and disk decomposition.

\begin{figure*}
\centering
\scriptsize

\begin{tabular}{@{}c c c c c c@{}}
  & \makebox[0.18\textwidth]{\centering \textbf{Data}} 
  & \makebox[0.18\textwidth]{\centering \textbf{Bulge}} 
  & \makebox[0.18\textwidth]{\centering \textbf{Disk}} 
  & \makebox[0.18\textwidth]{\centering \textbf{Summed}} 
  & \makebox[0.18\textwidth]{\centering \textbf{Residual}} \\[0.2cm]

\raisebox{0.4\height}{\rotatebox{90}{\textbf{F435W
  (B)}}} &
\includegraphics[width=0.18\textwidth]{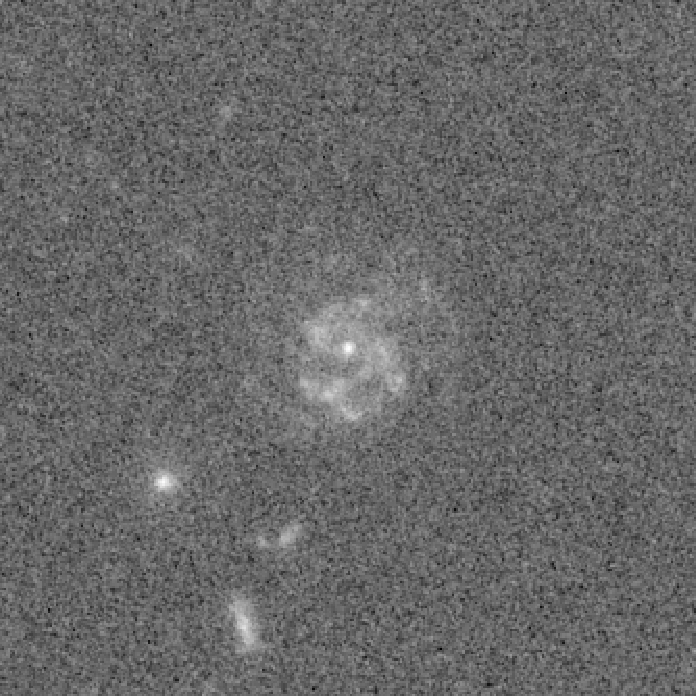} &
\includegraphics[width=0.18\textwidth]{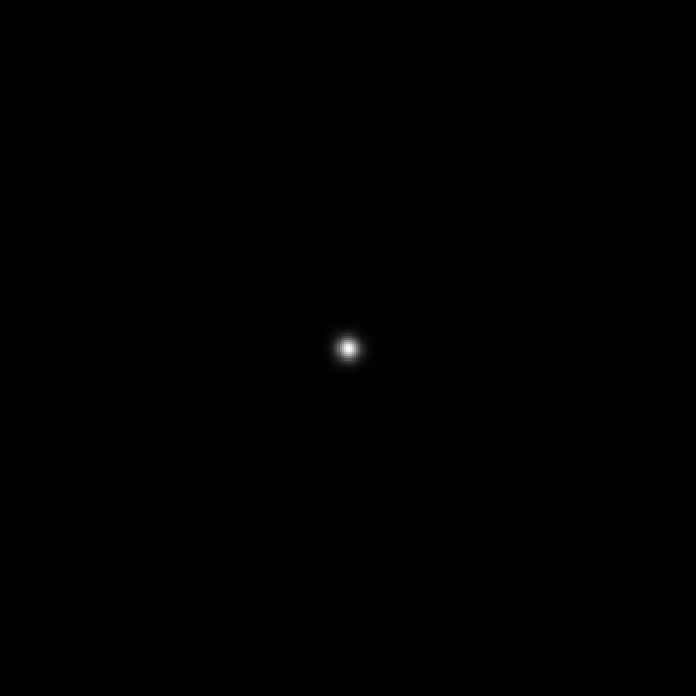} &
\includegraphics[width=0.18\textwidth]{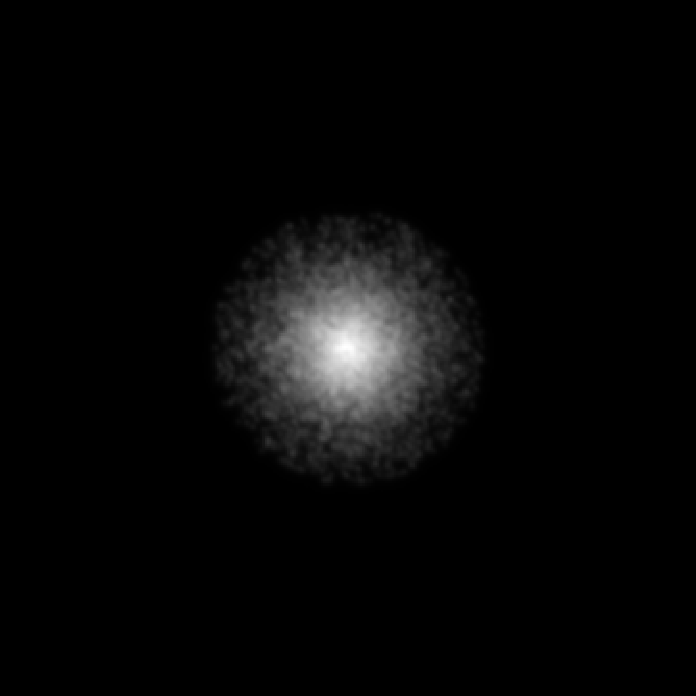} &
\includegraphics[width=0.18\textwidth]{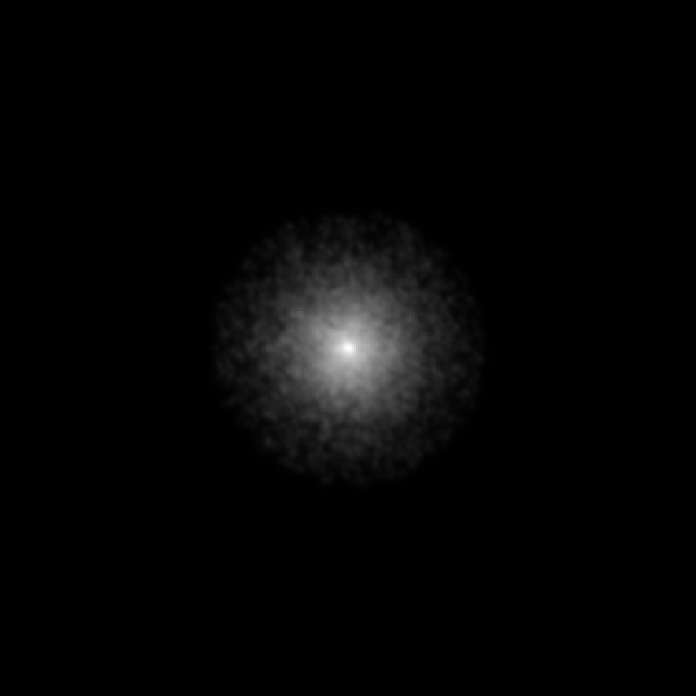} &
\includegraphics[width=0.18\textwidth]{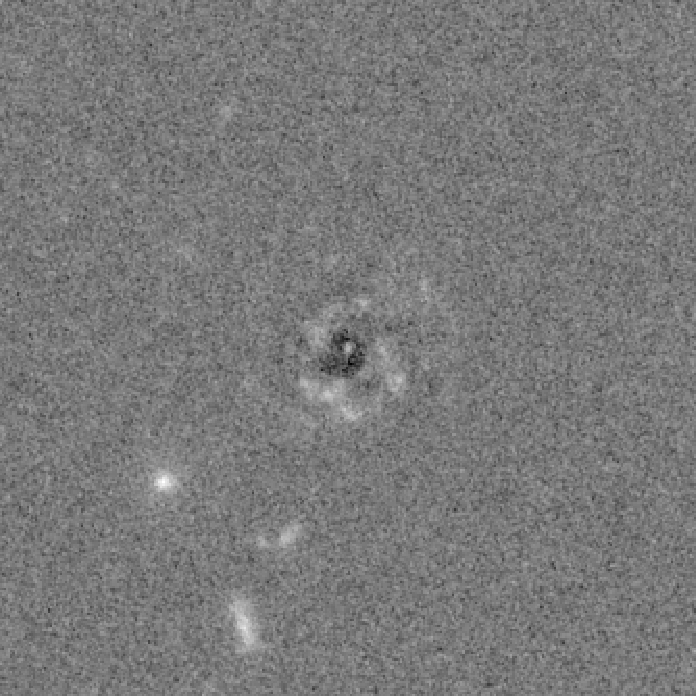} \\[0.2cm]

\raisebox{0.4\height}{\rotatebox{90}{\textbf{F606W
  (V)}}} &
\includegraphics[width=0.18\textwidth]{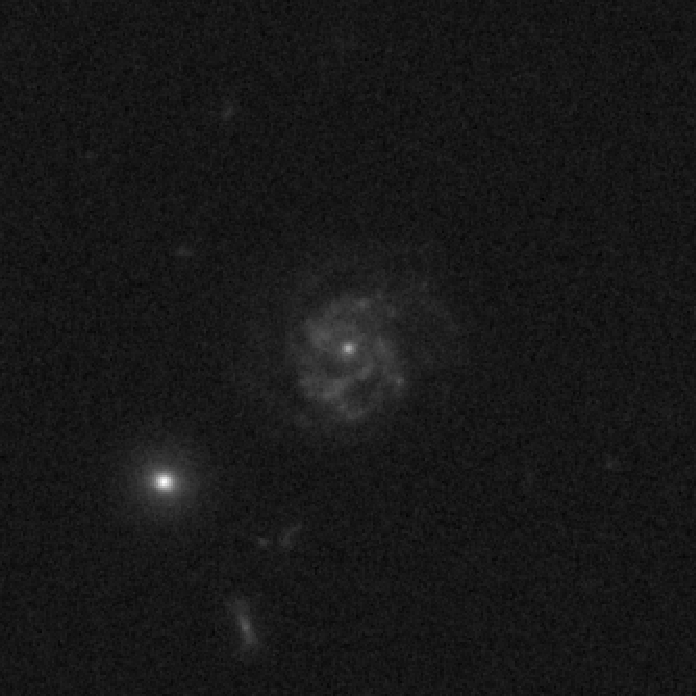} &
\includegraphics[width=0.18\textwidth]{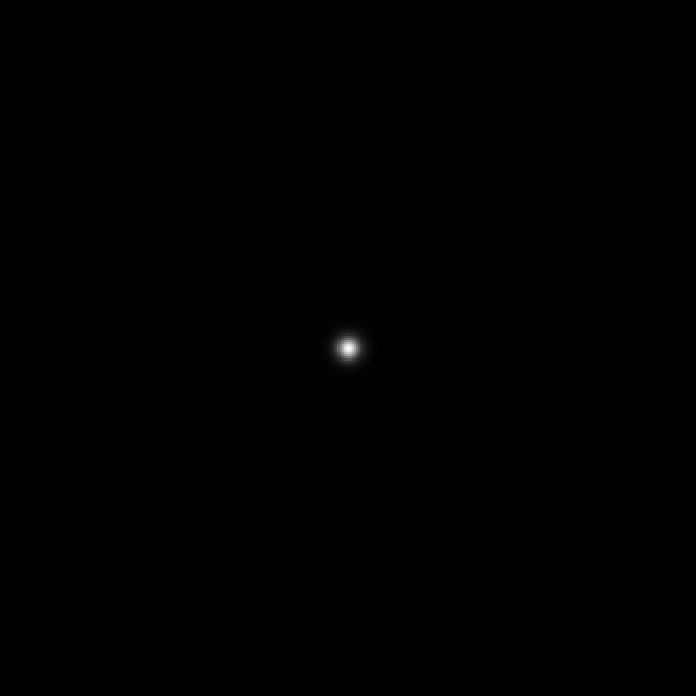} &
\includegraphics[width=0.18\textwidth]{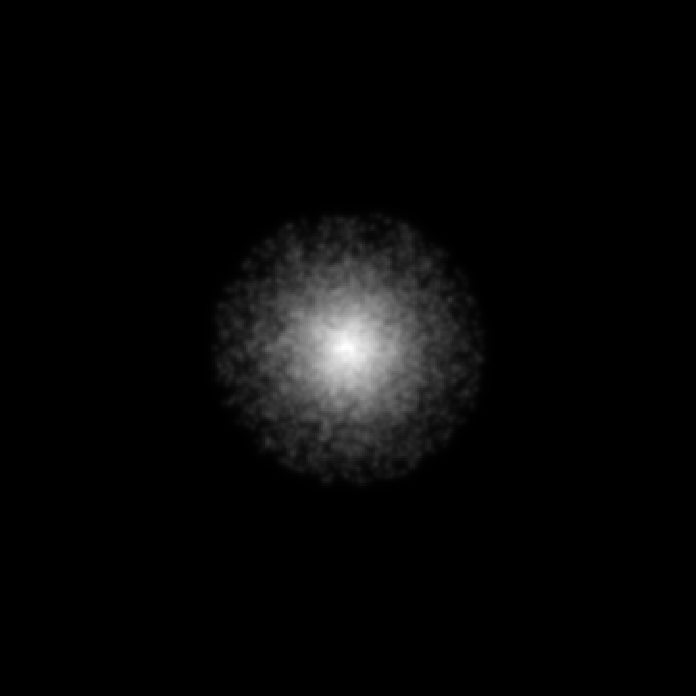} &
\includegraphics[width=0.18\textwidth]{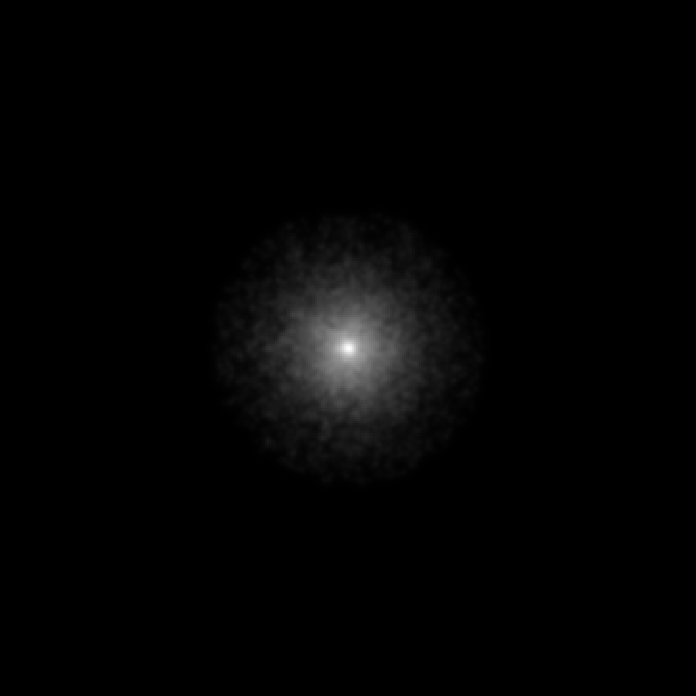} &
\includegraphics[width=0.18\textwidth]{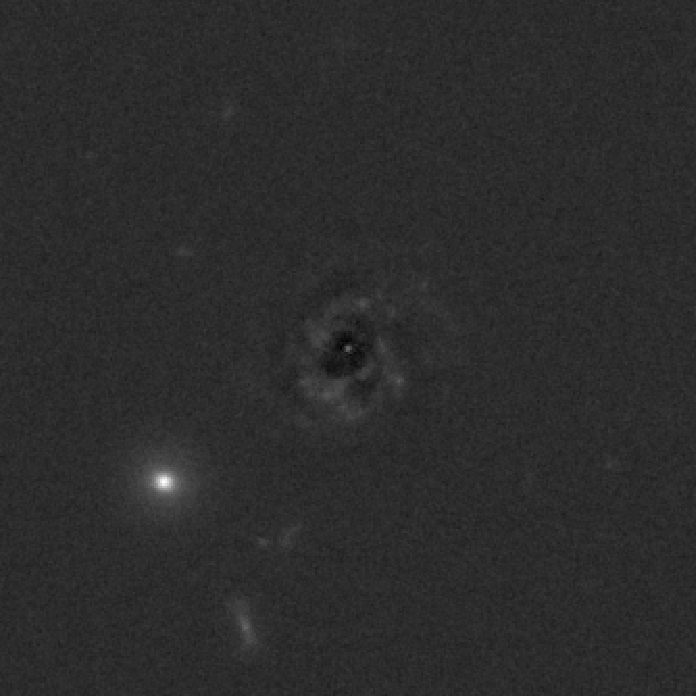} \\[0.2cm]

\raisebox{0.4\height}{\rotatebox{90}{\textbf{F160W
  (H)}}} &
\includegraphics[width=0.18\textwidth]{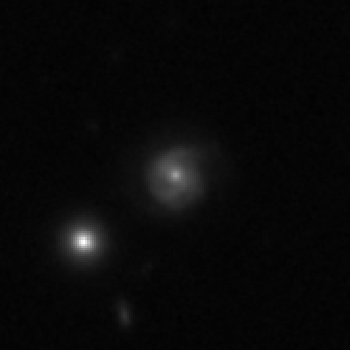} &
\includegraphics[width=0.18\textwidth]{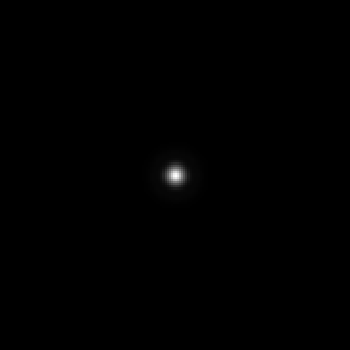} &
\includegraphics[width=0.18\textwidth]{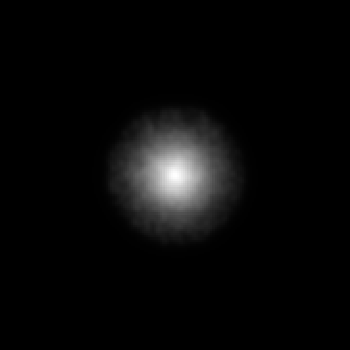} &
\includegraphics[width=0.18\textwidth]{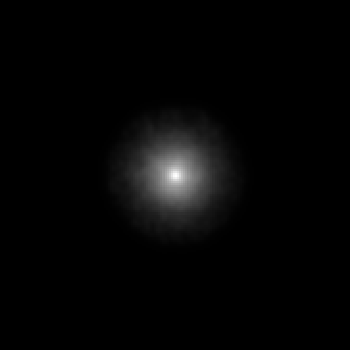} &
\includegraphics[width=0.18\textwidth]{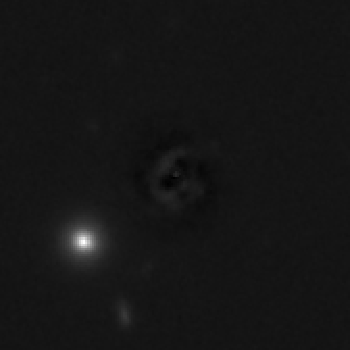} \\

\end{tabular}

\vspace{0.3cm}

\caption{\small Fit for a face-on \textit{Dustless} galaxy $(I_{\rm AB}=22.2)$ at $z=0.62$ with a visual spheroidal and disk component. Rows correspond to filters F435W ($b$), F606W ($v$), and F160W ($H$). Columns show: (1) data; (2) best-fit PSF-convolved bulge model; (3) disk model; (4) summed model; and (5) residual image (in $\sigma$). A linear greyscale is used throughout. The model shown is the individually best-fitting model for this galaxy, with $\tau=0$ assumed.}
\label{fig:images_dustless_faceon_bd}
\end{figure*}

\begin{figure*}
\centering
\scriptsize

\begin{tabular}{@{}c c c c c c@{}}
  & \makebox[0.18\textwidth]{\centering \textbf{Data}} 
  & \makebox[0.18\textwidth]{\centering \textbf{Bulge}} 
  & \makebox[0.18\textwidth]{\centering \textbf{Disk}} 
  & \makebox[0.18\textwidth]{\centering \textbf{Summed}} 
  & \makebox[0.18\textwidth]{\centering \textbf{Residual}} \\[0.2cm]

\raisebox{0.4\height}{\rotatebox{90}{\textbf{F435W
  (B)}}} &
\includegraphics[width=0.18\textwidth]{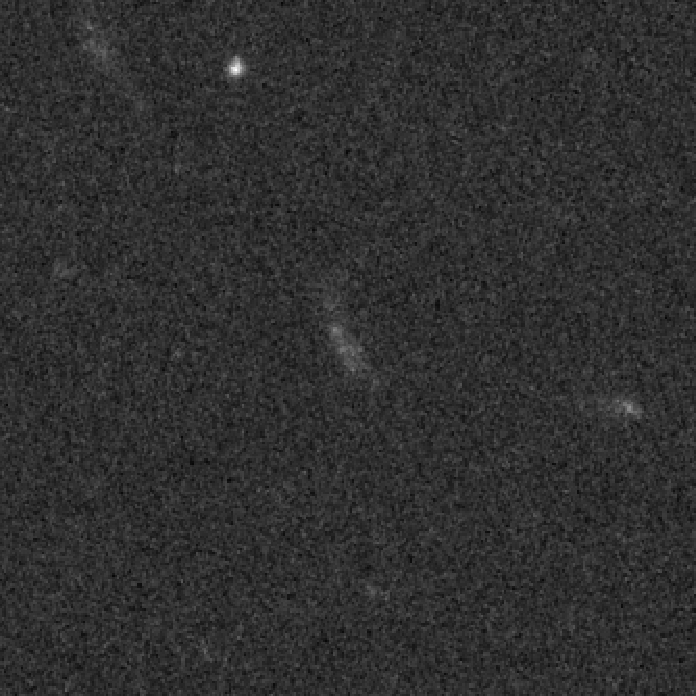} &
\includegraphics[width=0.18\textwidth]{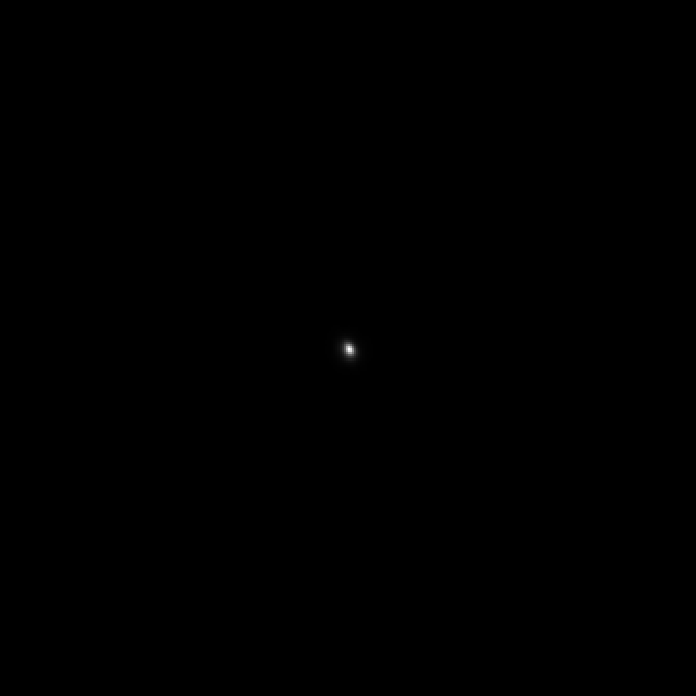} &
\includegraphics[width=0.18\textwidth]{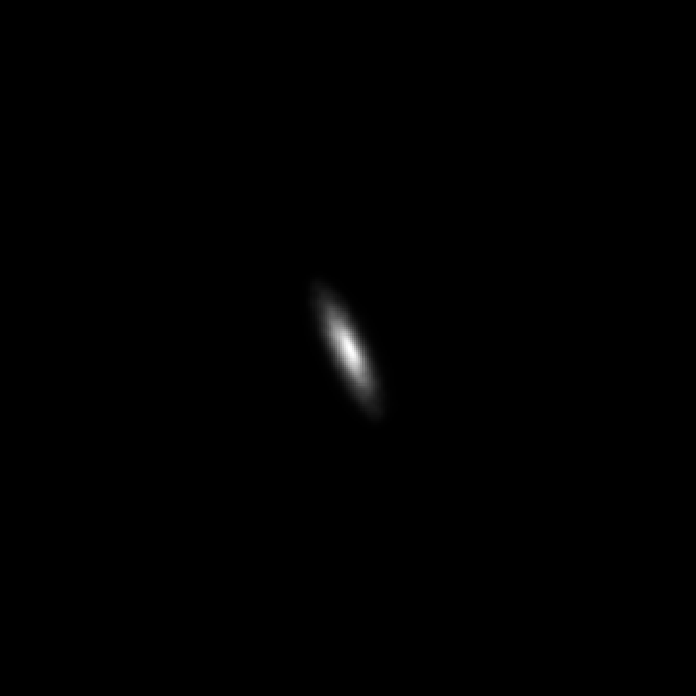} &
\includegraphics[width=0.18\textwidth]{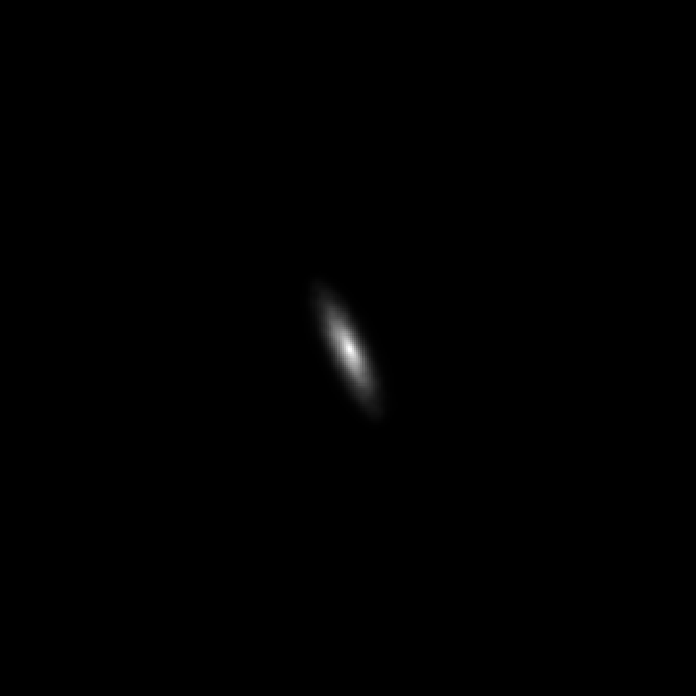} &
\includegraphics[width=0.18\textwidth]{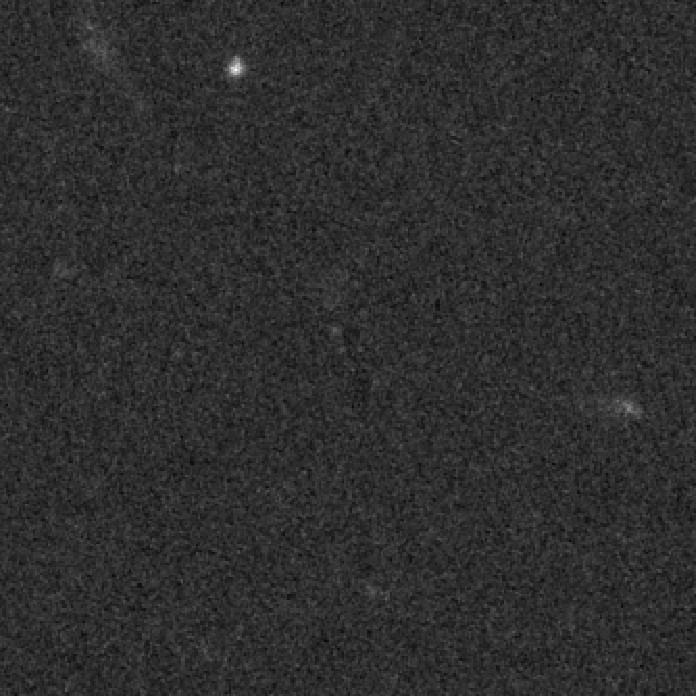} \\[0.2cm]

\raisebox{0.4\height}{\rotatebox{90}{\textbf{F606W
  (V)}}} &
\includegraphics[width=0.18\textwidth]{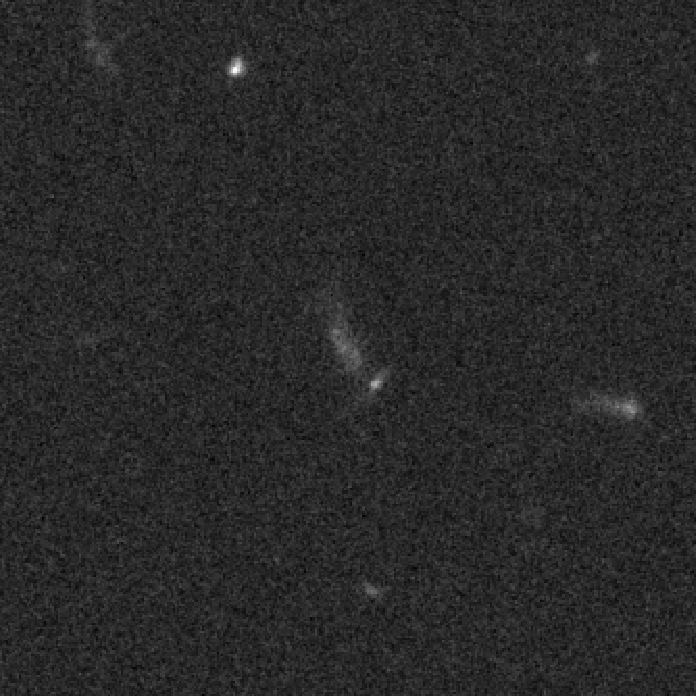} &
\includegraphics[width=0.18\textwidth]{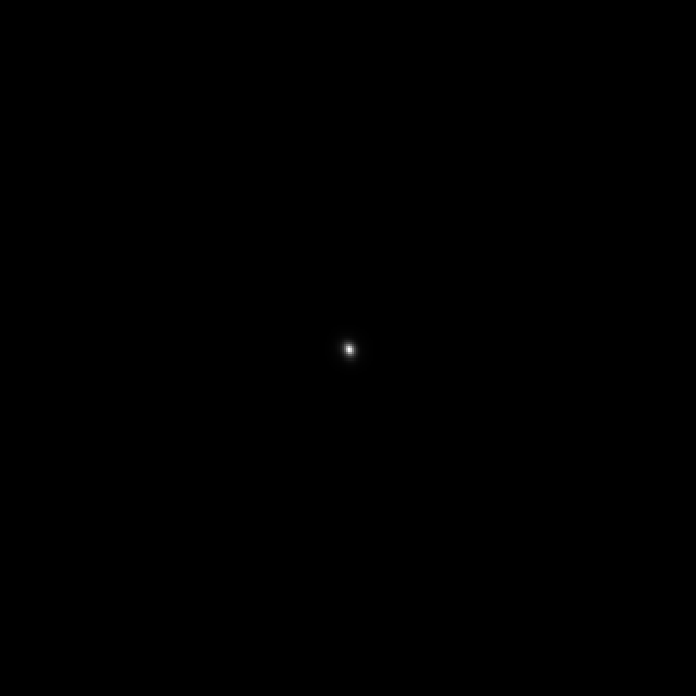} &
\includegraphics[width=0.18\textwidth]{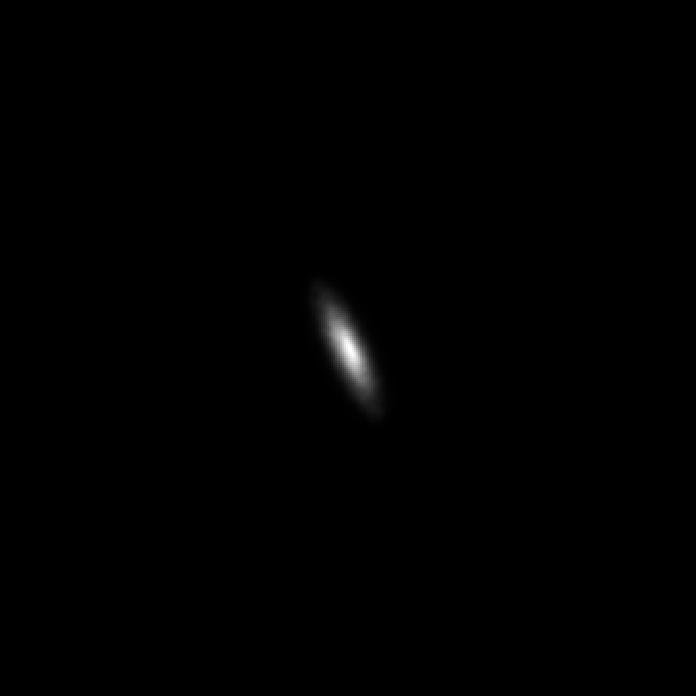} &
\includegraphics[width=0.18\textwidth]{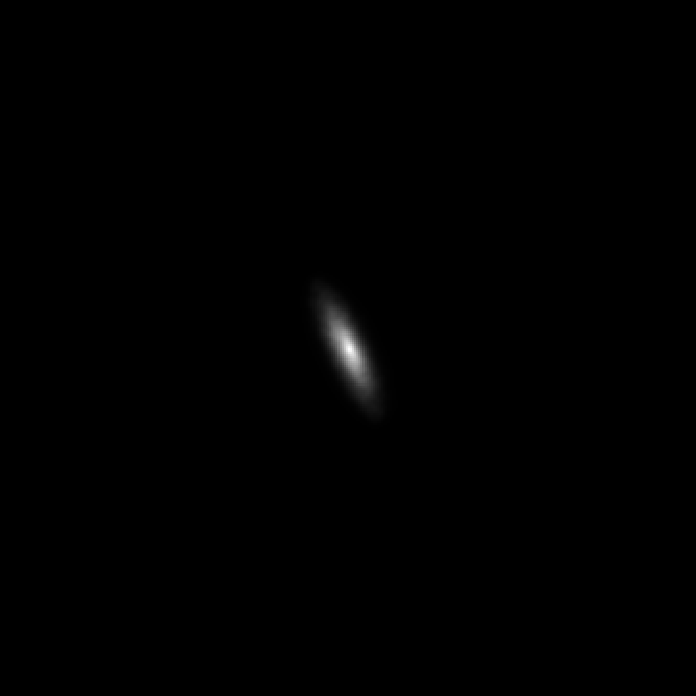} &
\includegraphics[width=0.18\textwidth]{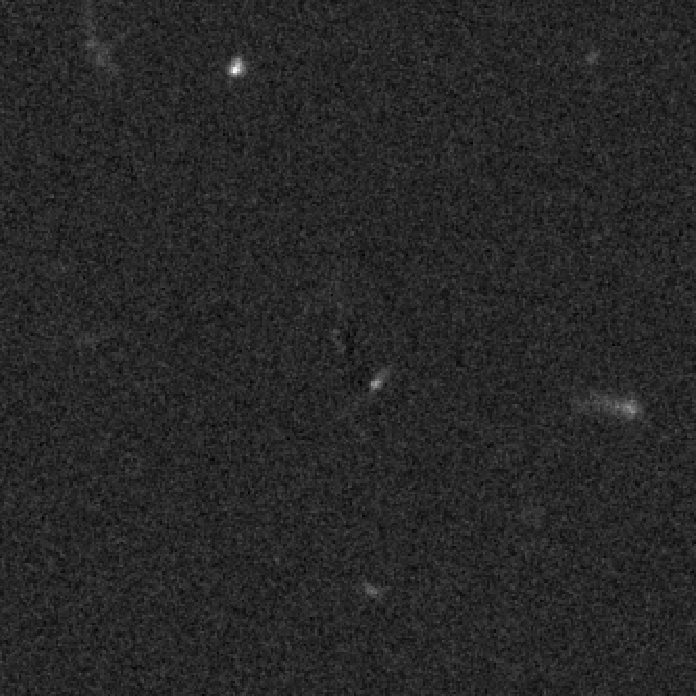} \\[0.2cm]

\raisebox{0.4\height}{\rotatebox{90}{\textbf{F160W
  (H)}}} &
\includegraphics[width=0.18\textwidth]{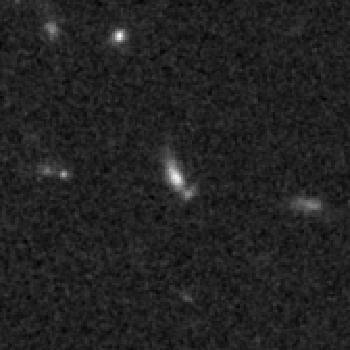} &
\includegraphics[width=0.18\textwidth]{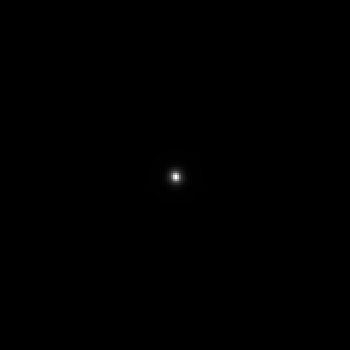} &
\includegraphics[width=0.18\textwidth]{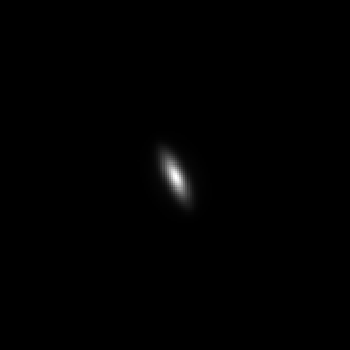} &
\includegraphics[width=0.18\textwidth]{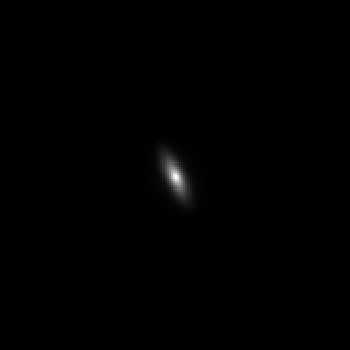} &
\includegraphics[width=0.18\textwidth]{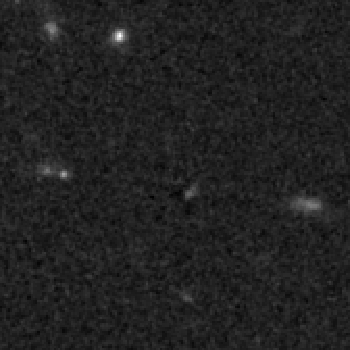} \\

\end{tabular}

\vspace{0.3cm}

\caption{\small Fits to an edge-on \textit{Dusty} galaxy $(I_{\rm AB}=24.9)$ at $z=2.1$ with a visual spheroidal and disk component. The model shown is the individually best-fitting extincted model for this galaxy selected from the full relevant parameter grid, and yields $\tau=2$.}
\label{fig:images_dusty_edgeon_bd}
\end{figure*}

\begin{figure*}
\centering
\scriptsize

\begin{tabular}{@{}c c c c c c@{}}
  & \makebox[0.18\textwidth]{\centering \textbf{Data}} 
  & \makebox[0.18\textwidth]{\centering \textbf{Bulge}} 
  & \makebox[0.18\textwidth]{\centering \textbf{Disk}} 
  & \makebox[0.18\textwidth]{\centering \textbf{Summed}} 
  & \makebox[0.18\textwidth]{\centering \textbf{Residual}} \\[0.2cm]

\raisebox{0.4\height}{\rotatebox{90}{\textbf{F435W
  (B)}}} &
\includegraphics[width=0.18\textwidth]{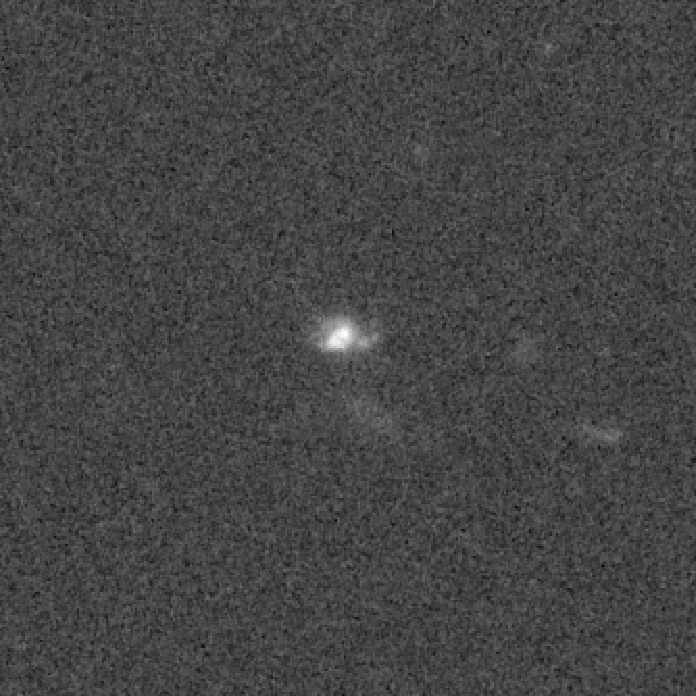} &
\includegraphics[width=0.18\textwidth]{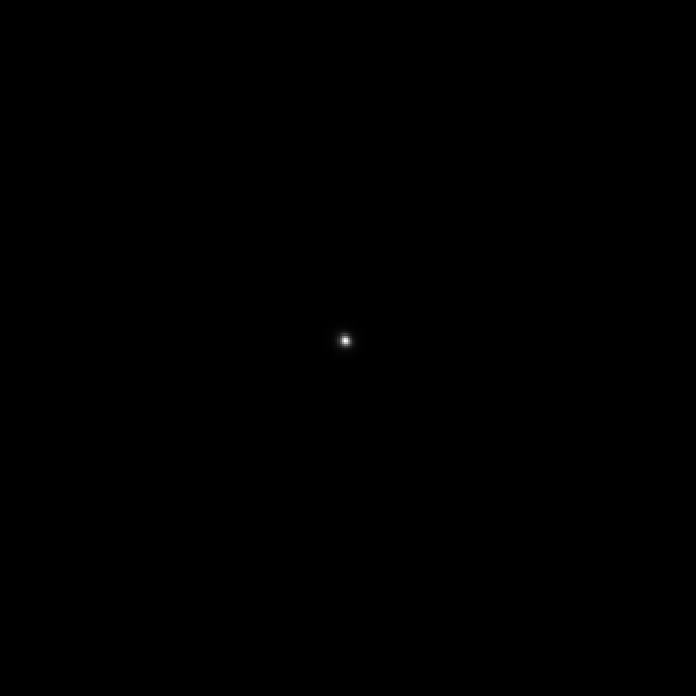} &
\includegraphics[width=0.18\textwidth]{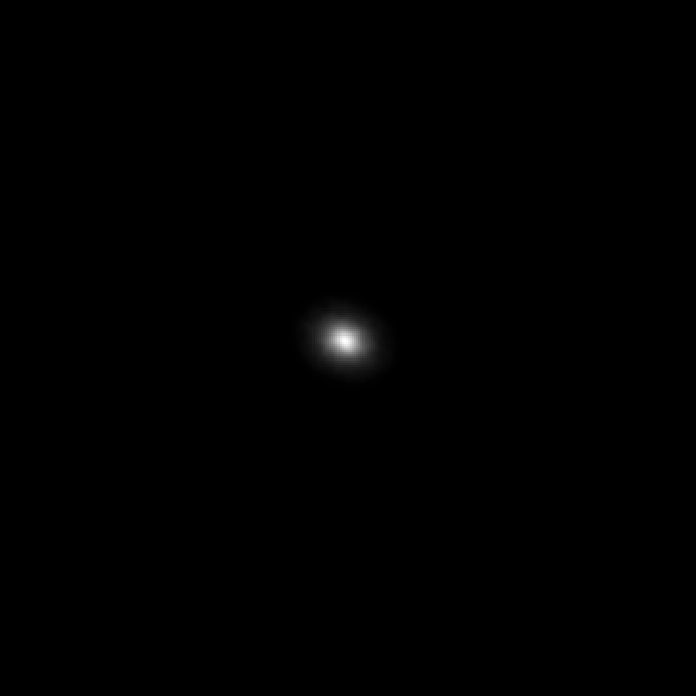} &
\includegraphics[width=0.18\textwidth]{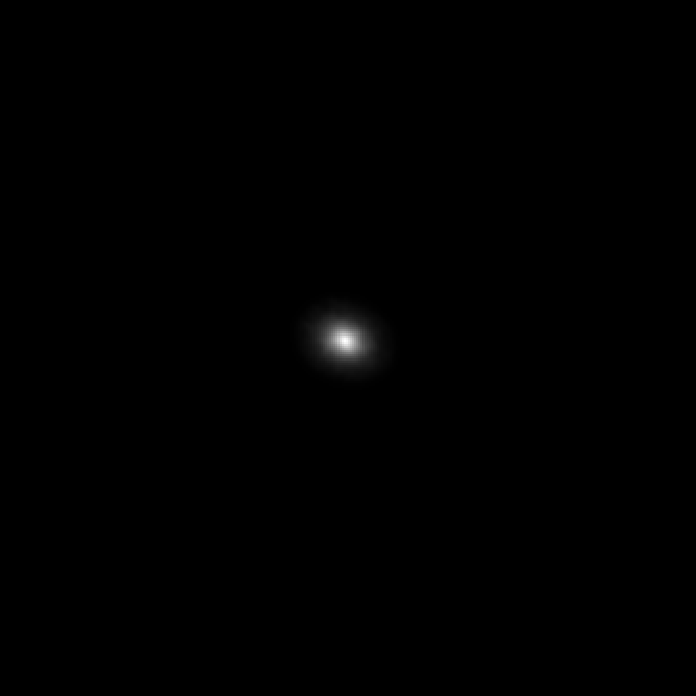} &
\includegraphics[width=0.18\textwidth]{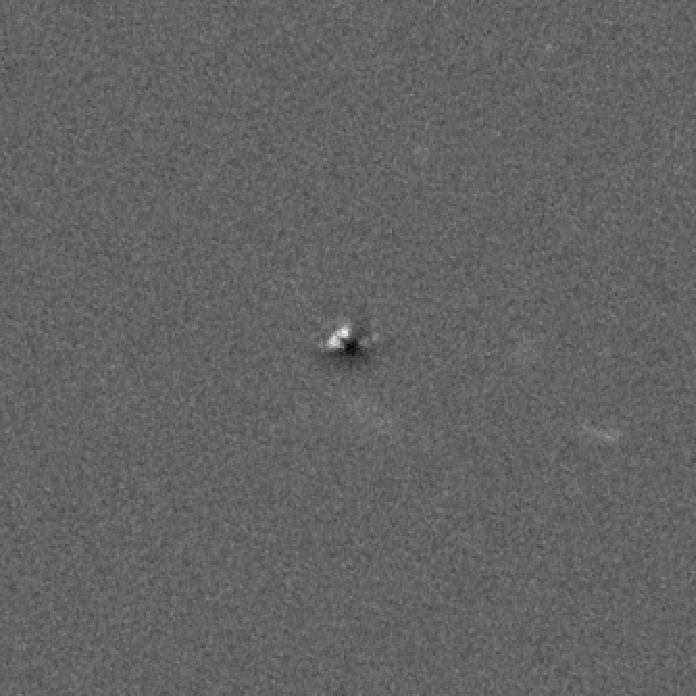} \\[0.2cm]

\raisebox{0.4\height}{\rotatebox{90}{\textbf{F606W
  (V)}}} &
\includegraphics[width=0.18\textwidth]{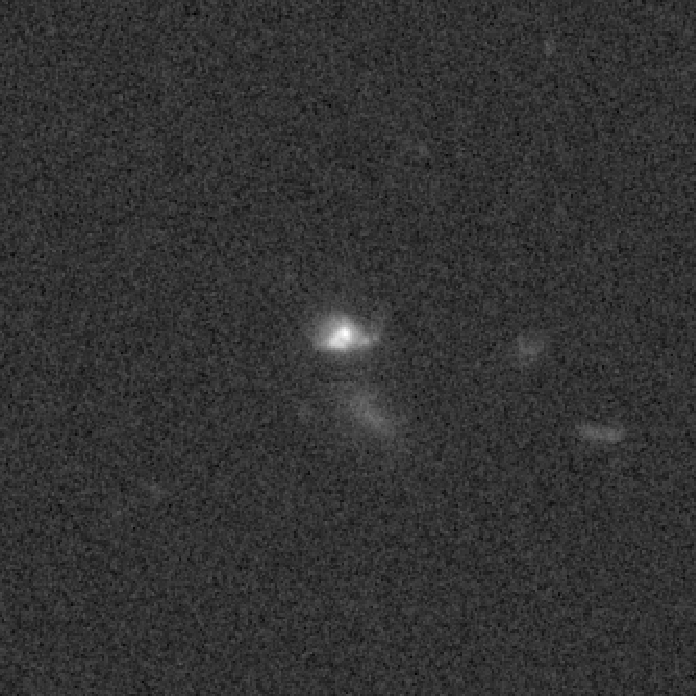} &
\includegraphics[width=0.18\textwidth]{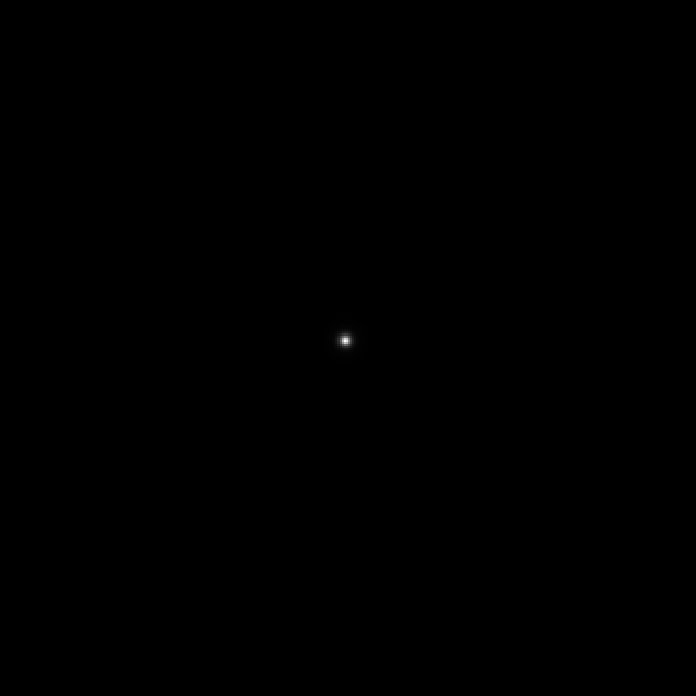} &
\includegraphics[width=0.18\textwidth]{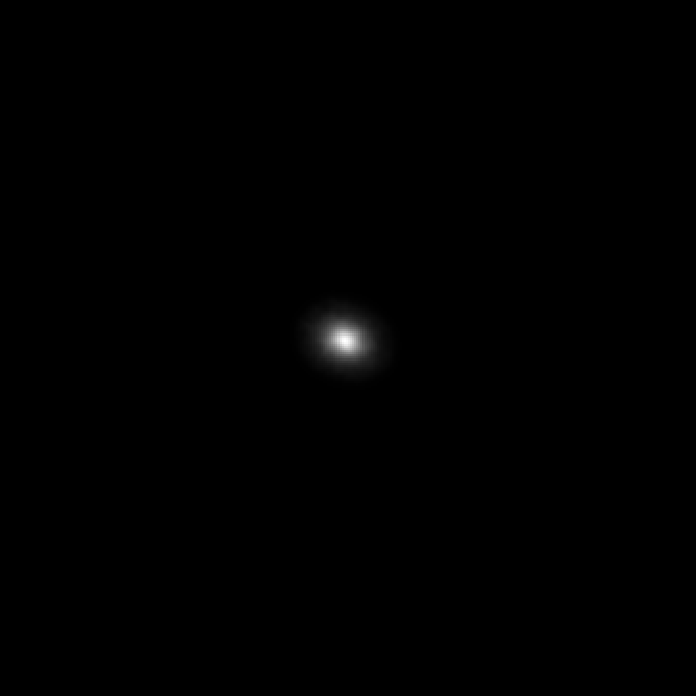} &
\includegraphics[width=0.18\textwidth]{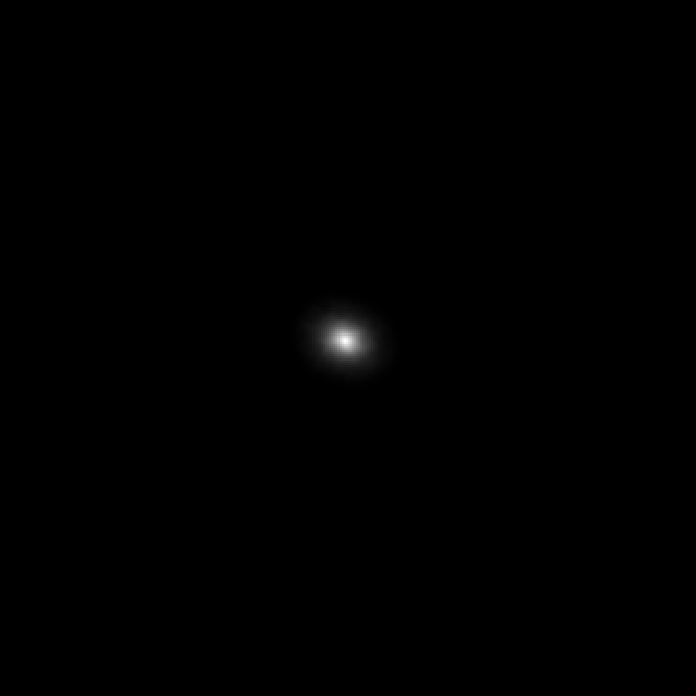} &
\includegraphics[width=0.18\textwidth]{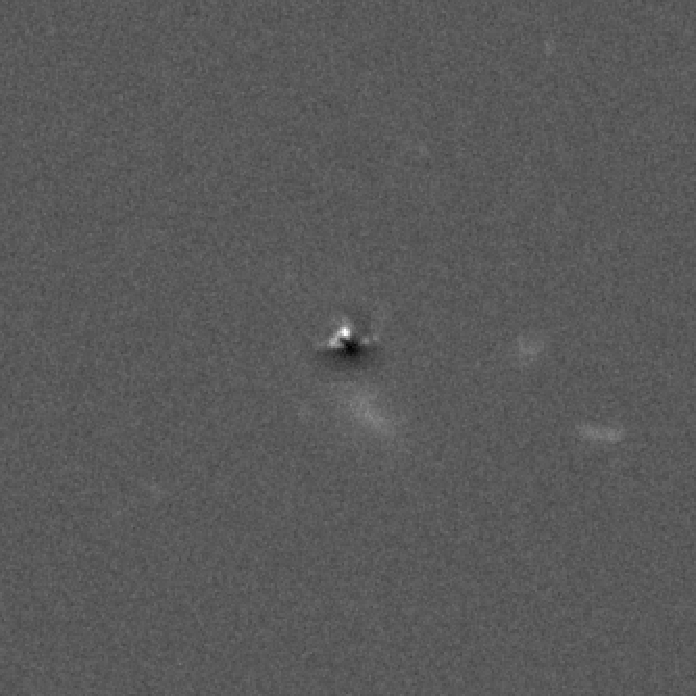} \\[0.2cm]

\raisebox{0.4\height}{\rotatebox{90}{\textbf{F160W
  (H)}}} &
\includegraphics[width=0.18\textwidth]{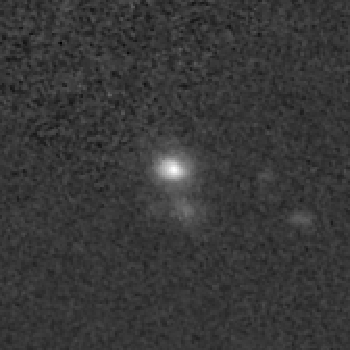} &
\includegraphics[width=0.18\textwidth]{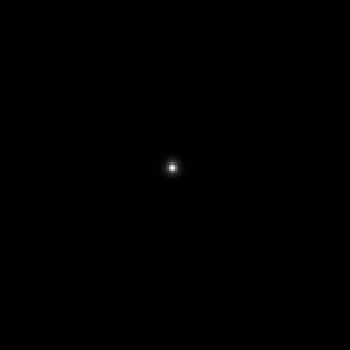} &
\includegraphics[width=0.18\textwidth]{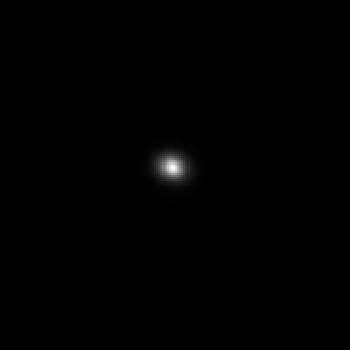} &
\includegraphics[width=0.18\textwidth]{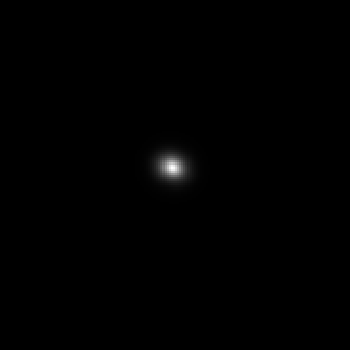} &
\includegraphics[width=0.18\textwidth]{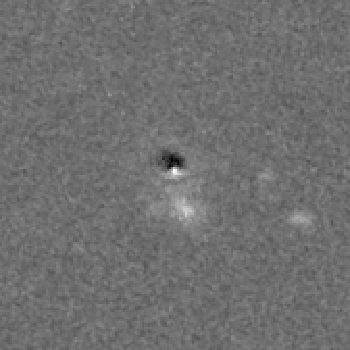} \\

\end{tabular}

\vspace{0.3cm}

\caption{\small Fit to a face-on galaxy $(I_{AB}=23.0)$
  at $z=0.85$ with an irregular morphology.}
\label{fig:images_dusty_faceon_irr}
\end{figure*}

\begin{figure}
\centering  
\includegraphics[width=.9\columnwidth]{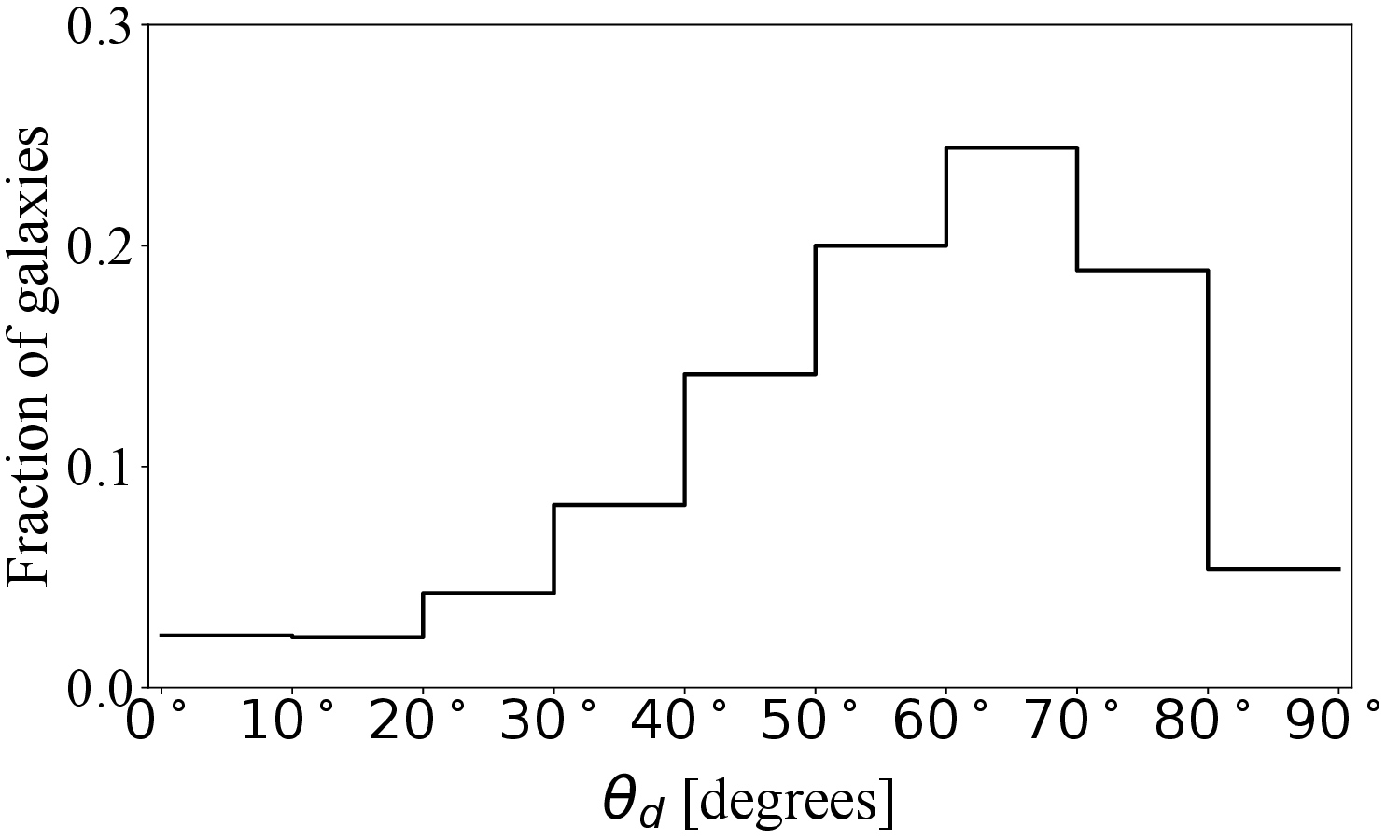}
\caption{Distributions of fitted inclination angles of the disk $\theta_d$ for the
  sample to a depth of $I_{AB}=$25.5, obtained from the two-component
  fits.}
\label{fig:theta_distrib}
\end{figure}

\begin{figure}
\centering
\includegraphics[width=0.48\columnwidth]{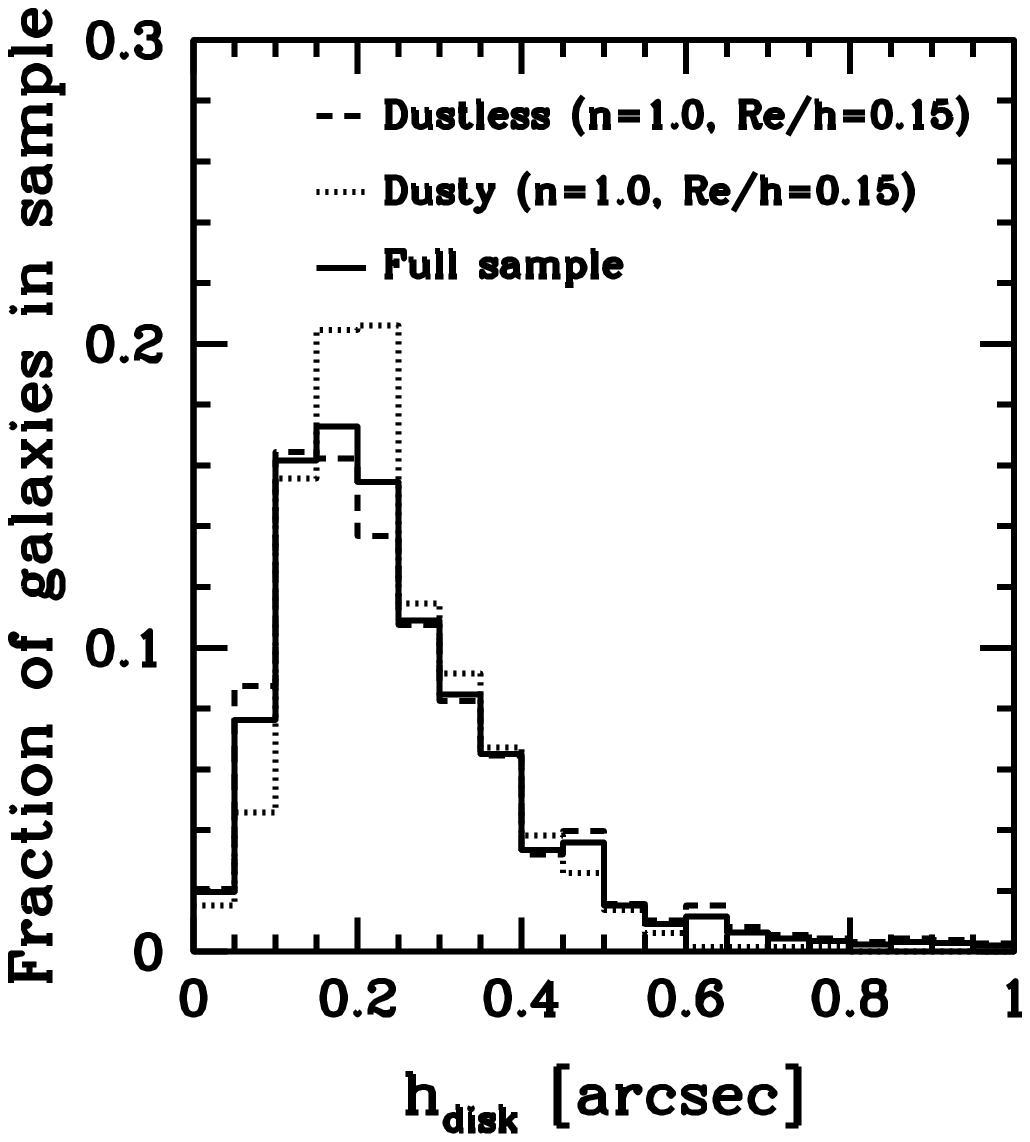}
\hfill
\includegraphics[width=0.48\columnwidth]{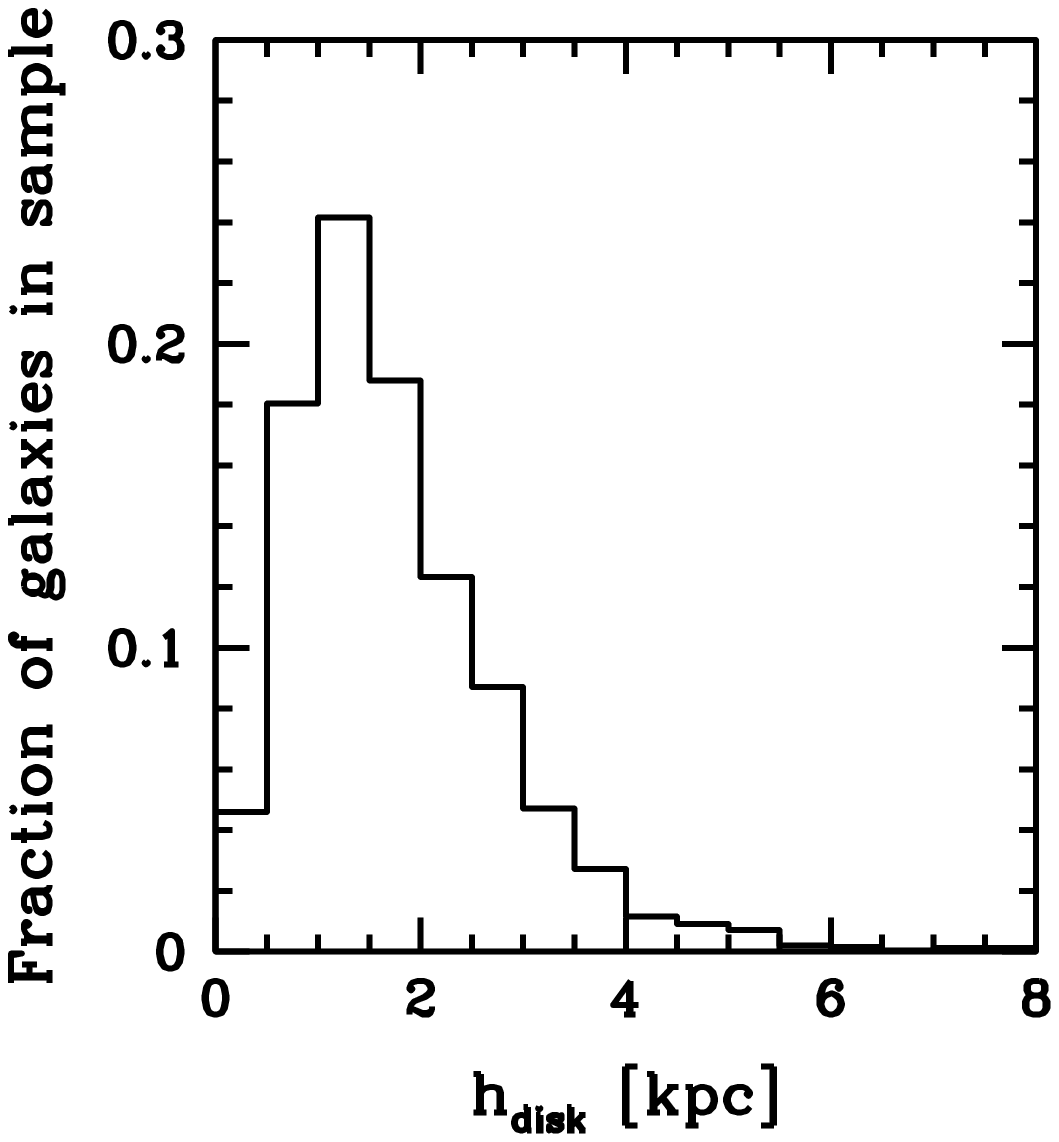}
\caption{Left panel: distribution of scale lengths for the intrinsic 3D
  disk $h_d$ for $I_{AB} < 25.5$. This quantity is wavelength-independent and is the same as the observed disk
  scale length for unextincted galaxies ($\tau=$0) in the
  \textit{Dustless} sample (the scale length is the size of the major
  axis of the disk model by construction). This is not the case for
  the \textit{Dusty} sample where the inferred observed disk scale length will vary with wavelength
  and the surface brightness profile no longer corresponds to an
  exponential disk. Right panel: the corresponding distribution of physical scale
  lengths in proper coordinates.}
  \label{fig:hdisk}
\end{figure}

\begin{figure}
\centering
\includegraphics[width=0.45\columnwidth]{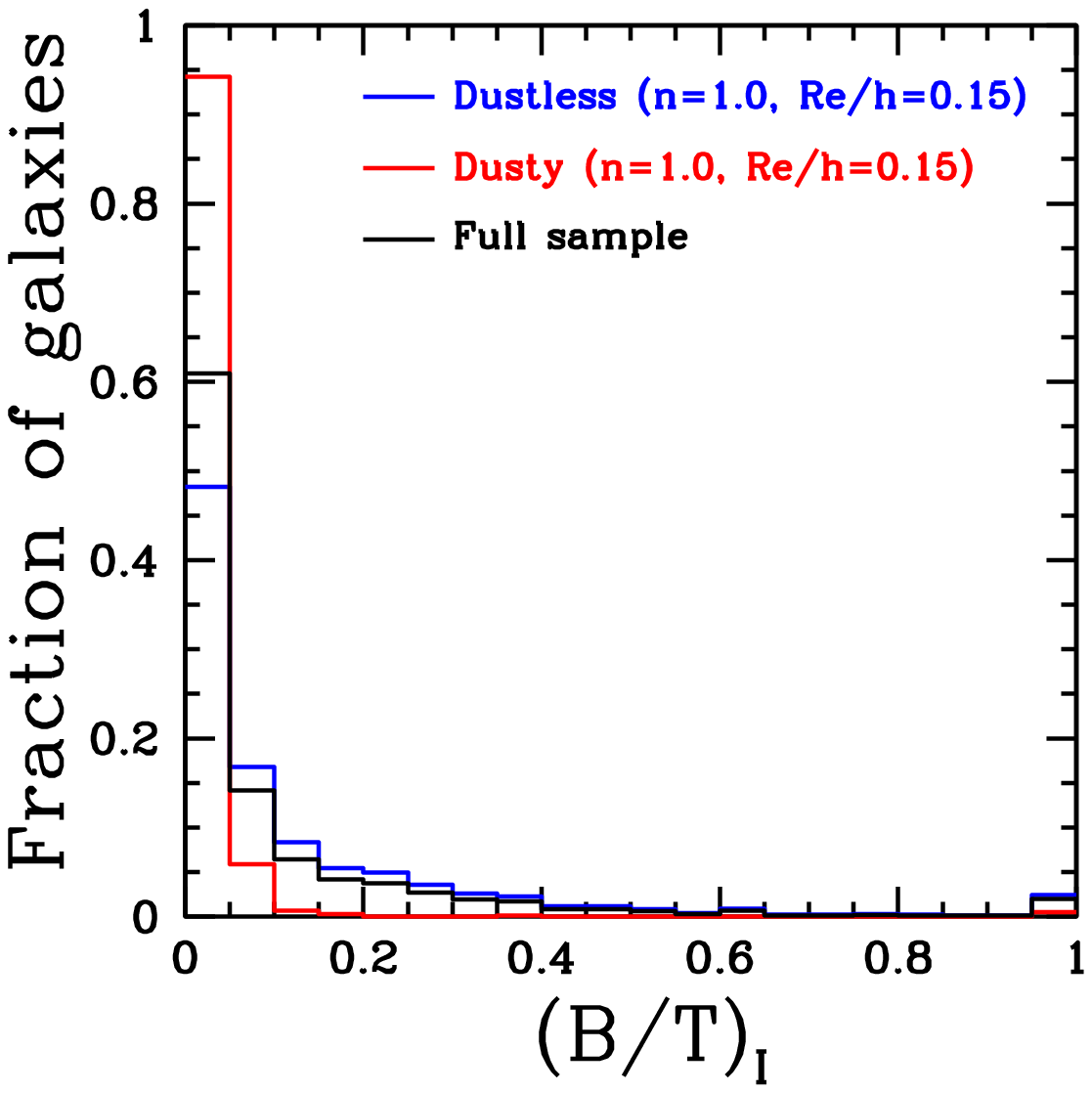}
\hfill
\includegraphics[width=0.45\columnwidth]{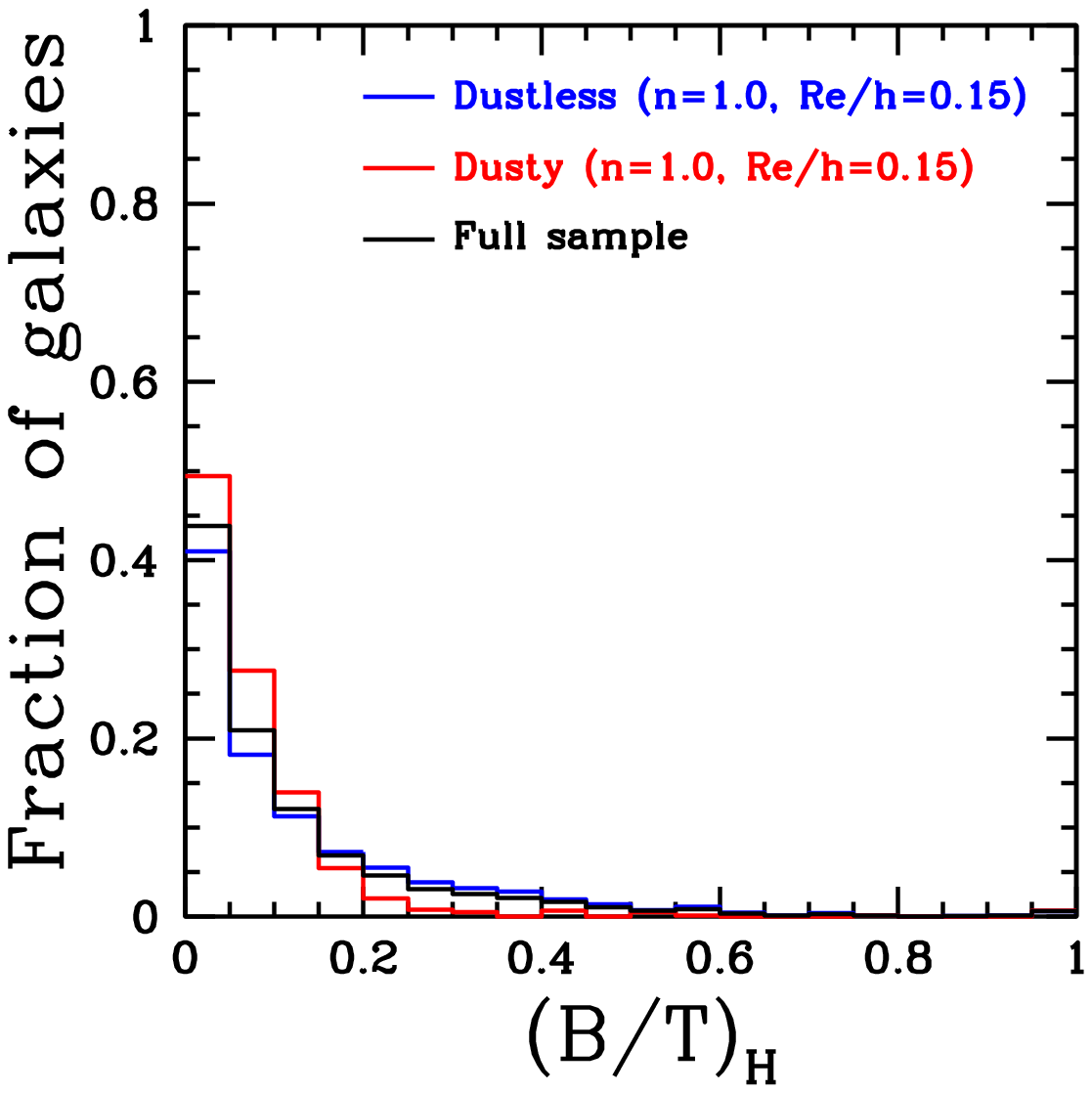}
\caption{Bulge-to-total distributions for the \textit{Dustless},
  \textit{Dusty} and full samples for $I_{AB} < 25.5$ for (a) $F775W$ ($i$) (b)
  $F160W$ ($H$).}
\label{fig:bt}
\end{figure}

\subsection{Bulge and disk fits with and without extinction}
\label{results:bulge_disk_fits}

In the two-component fitting, we first fit for all bulge and disk
quantities (including $n$) in individual galaxies.
Fig.~\ref{fig:images_dustless_faceon_bd} shows an
example of a \textit{Dustless} system, a face-on spiral galaxy which
has a clear central bulge and disk component. In the two-component fitting, the bulge is clearly visible in all the
bands while residual structure (after model subtraction) arising from the
spiral arms of the galaxy, which are not fitted well by the smooth
bulge plus disk models, are most prominent in the bluer $b$ band. By contrast,
Fig.~\ref{fig:images_dusty_edgeon_bd} is an example of an edge-on
galaxy that is classified as \textit{Dusty} and has an individually
best-fitting model yielding $\tau=2.0$. The bulge component is relatively faint compared to the disk in the optical bands but
contributes a higher fraction to the total light in the galaxy in the
near-infrared bands. Finally, Fig.~\ref{fig:images_dusty_faceon_irr} is an
example of an irregular morphology ($\theta_d=35.0$\degree)
showing several distinct clumps of star formation. The model fits
suggest some central extinction. However, fitting a bulge and disk
component to this patchy light distribution results in some over-subtraction in the residual image.


As discussed in Sect.~\ref{method:fit_disk}, individual fits to galaxies are noisy
but they indicate that the distribution is skewed to low values of $n$
(see Fig.~\ref{fig:n_individual_fits}). In addition, for such fits,
neglecting extinction would increase the inferred fraction of low-$n$
bulges (approximately 10\% at $n=0.5$). Following the approach
described in Sec.~\ref{method:fit_disk}, the maximum of the stacked
likelihood results in $n$=1.0 and
$R_e/h_d=$0.15 for the full sample when no extinction (dustless) is
applied to the models. The stacked likelihood for
the \textit{Dusty} sample yields $n=$1.0, $R_e/h_d=$0.15 and
$\tau=$4.0. We use an F-test to compare the dusty and dustless models, with the null hypothesis that the additional parameter $\tau$ does not improve the fit. For galaxies in the \textit{Dusty} sample, the dusty model is favoured ($F/F_c>1$, where $F_c$ is the critical value), so we reject the null hypothesis and assign an extincted bulge and disk model to that galaxy. With the values of $n$, $R_e/h_d$ and $\tau$ fixed for the \textit{Dustless}
and \textit{Dusty} subsamples, we obtain the
distributions of $\theta_d$, $h_d$ and $B/T$
in each HST band.

The distribution of $\theta_d$ shown in Fig.~\ref{fig:theta_distrib} is skewed to higher inclination angles
of the disk although the fraction drops sharply between 80\degree and
90\degree. This supports the requirement to model the attenuation in
the surface brightness profile which would be a function of
$\theta_d$ \citep{graham2008}. The distribution of $h_d$ in both angular and
physical units is shown in Fig.~\ref{fig:hdisk}. The corresponding distribution of
physical scale lengths in proper coordinates is also shown for the
same depth. The median $h_d=$0.22$\arcsec\pm$0.004$\arcsec$ for the
full sample for the magnitude limit $I_{AB} < 25.5$. The median
physical scale length of the disk in the sample is $(1.57 \pm
0.03)\,\text{kpc}$.

For most galaxies in the sample, the residual flux fraction after
subtracting the best-fit bulge-plus-disk model from the data in either
the $v$ or $i$ bands accounts for less than 10 percent of the total
flux density in the aperture. So although the model
accounts for most of the flux in the image, there are typically differences
between the model and the data in specific locations in the image due
to patchy structure. The $(B/T)$ distributions in the optical bands
(Fig.~\ref{fig:bt}) suggest the HST sample is disk-dominated.
The low $n$ and $B/T$ values for HST galaxies (whether or not they are affected by
extinction) are very different from S0 or Sa galaxies observed locally
or at low redshifts \citep{graham2008}. It is important to note that the $B/T$ values reported here are
observed ones and not extinction-corrected. Extinction is expected to be strongest at bluer wavelengths and
for more inclined galaxies, so the $B/T$ distribution observed in the
$bviz$ bands in the \textit{Dusty} sample is expected to be skewed to
lower values compared to the \textit{Dustless} one. In the
near-infrared bands, the difference in the $B/T$ distributions between the
two samples is much smaller as there is less attenuation of the light
at these wavelengths. We also test the effect of various assumptions in the
models. We find negligible sensitivity of the distributions on the assumed values of $\beta_0$
for the bulge and $z_0/h_d$ for the disk (see
Appendix~\ref{app:priors_test}). For example, the median $h_d$ changes
by $1 \%$ when $z_0/h_d$ changes by $20 \%$, and the median $e_g$
changes by $ 1\%$ when $\beta_0$ changes by $25 \%$.

\begin{figure}
\centering  
\includegraphics[width=.85\columnwidth]{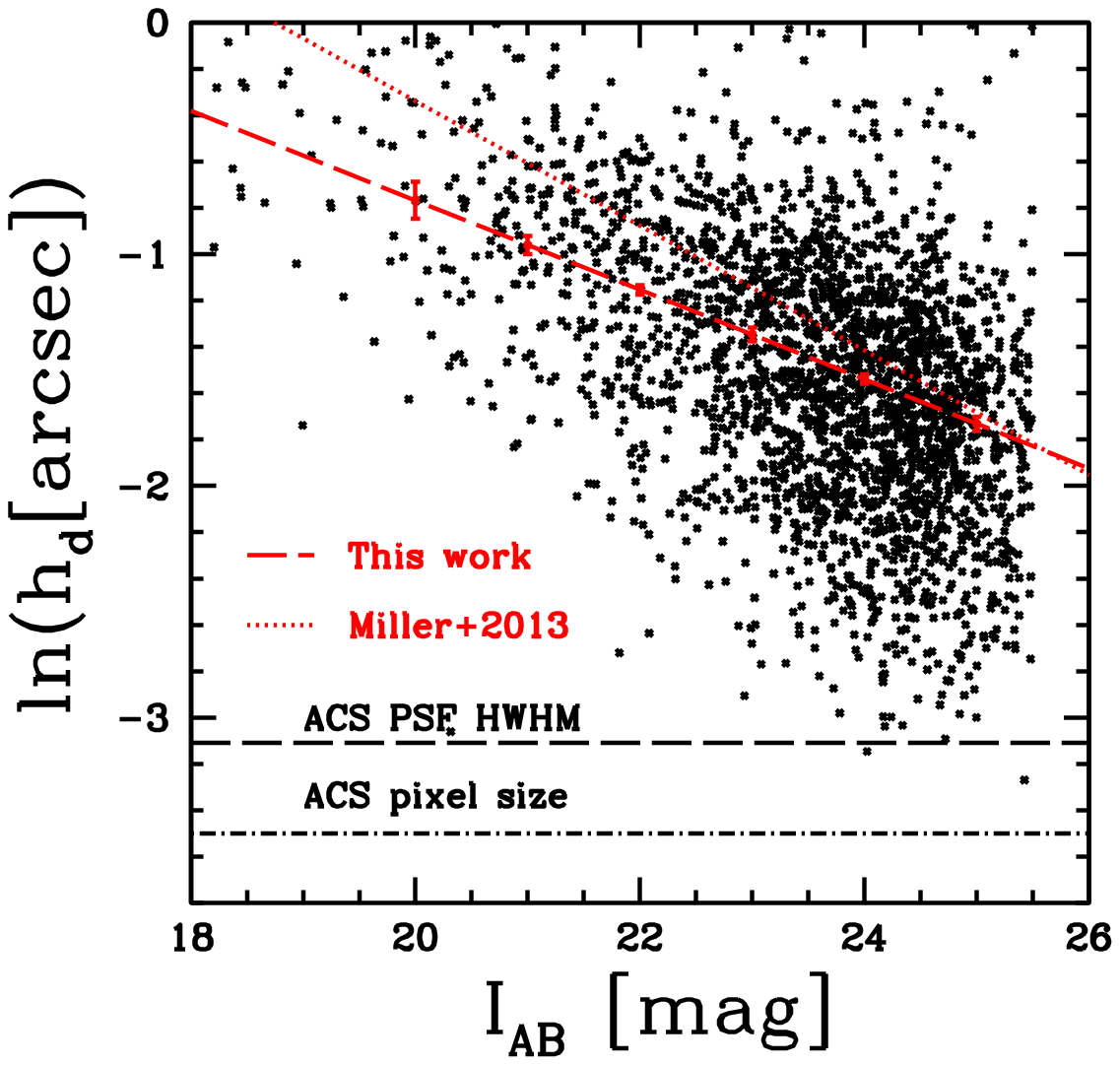}
\caption{Relation between disk scale length, $h_d$, and magnitude, $I_{AB}$ for galaxies. The long dashed red line is a least-squares fit to the median $h_d$ in 0.5 magnitude bins in this work according to Eq.~(\ref{eq:mag_hd_fit_welikala}).}
\label{fig:size-magnitude}
\end{figure}

\begin{figure}
\centering
\includegraphics[width=.9\columnwidth]{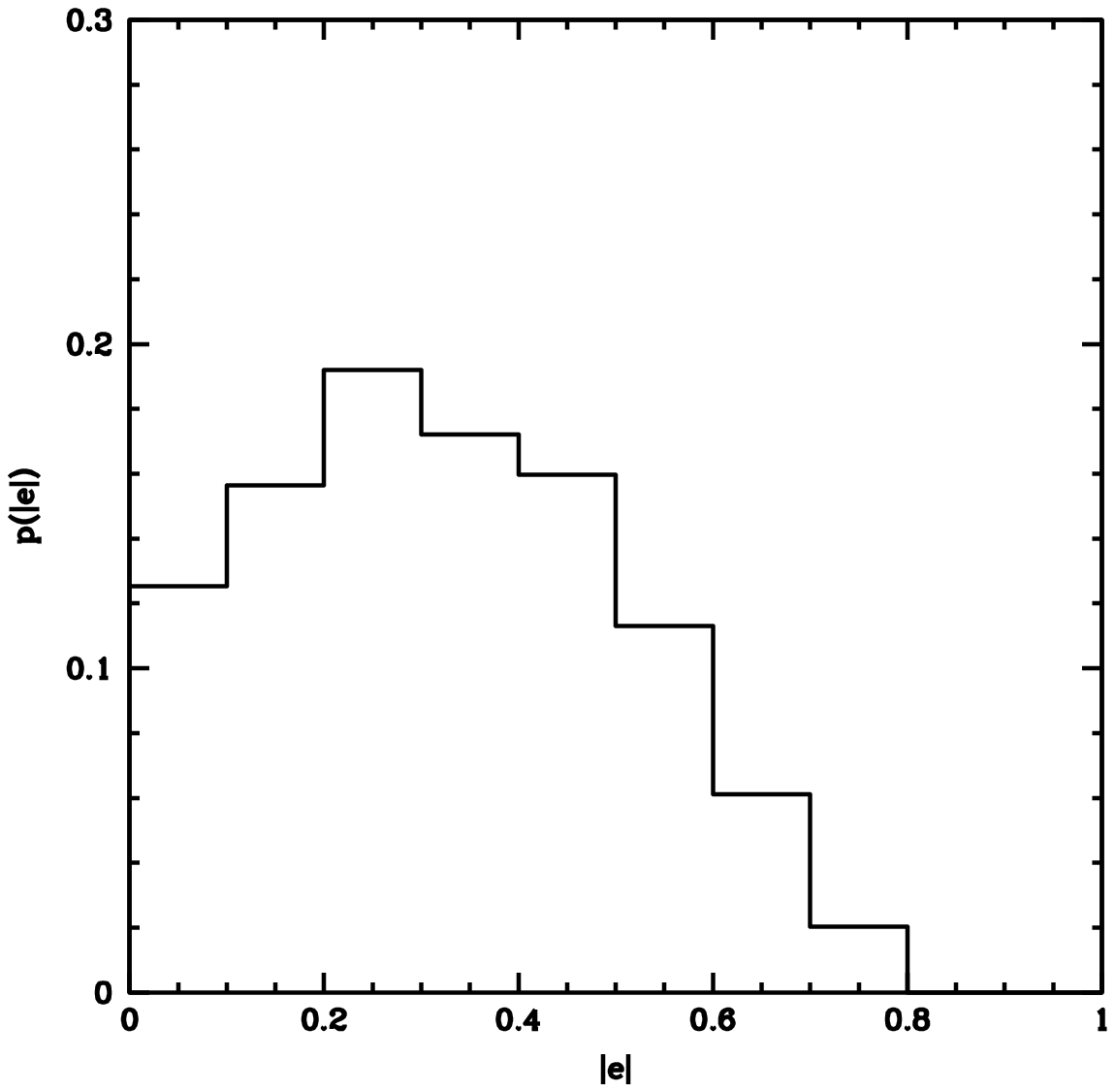}
\caption{Distribution of ellipticities computed from quadrupole
  moments on the noise-free and PSF-free best-fit model for the
  galaxy i.e., the summed bulge and disk model.} 
\label{fig:ellipticities}
\end{figure}

\begin{figure}
\centering
\includegraphics[width=0.45\columnwidth]{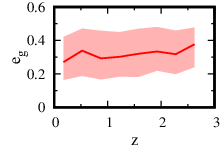}
\hfill
\includegraphics[width=0.45\columnwidth]{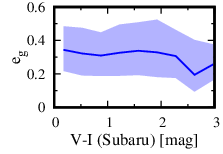}
\includegraphics[width=0.9\columnwidth]{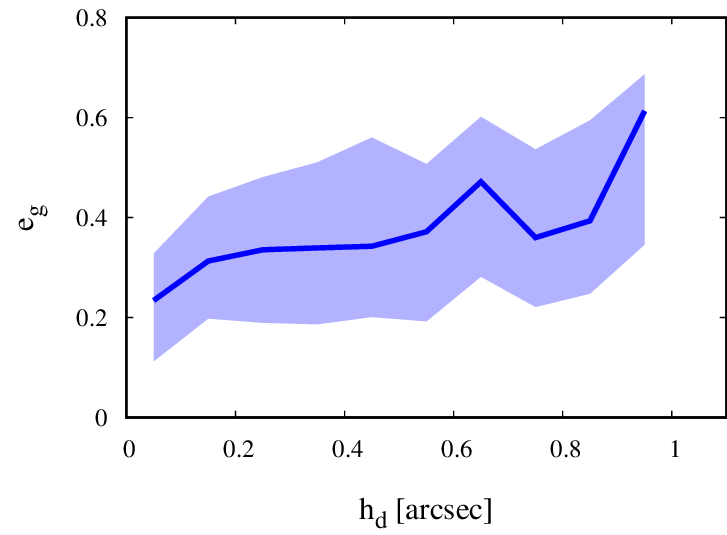}

\caption{Relation between galaxy ellipticity $e_g$ and (a) photometric
  redshift $z$ and (b) $V-I$ colour (top panel);  and (c) disk scale
  length $h_d$ (bottom). The
  galaxy ellipticity is measured using the quadrupole moments on the
  best-fit summed bulge and disk model for each galaxy, as described in
  Sect.~\ref{subsec:method_ellipticity}. The solid line corresponds to
  the median $e_g$ in each bin and the shaded region shows the 25th–75th
  percentile range of the $e_g$ distribution.}
\label{fig:ellipticity_correlations}
\end{figure}

\subsection{Prior distributions and relations for weak lensing}
\label{subsec:priors_weak_lensing}

Fig.~\ref{fig:size-magnitude} shows the relation between the disk
scale length and magnitude for this magnitude-limited
sample. \cite{miller2013} used disk and bulge scalelength priors
derived from fits to HST WFPC2 observations in the Groth strip by
\cite{simard2002} for disk-dominated galaxies and found a linear
relation between the median of the $\ln(h_d)$ and $i_{ab}$ in the magnitude
range $18.5 < i_{ab} < 25.5$ given by
\begin{equation}
\ln (h_d/{\rm arcsec}) = -1.145 - 0.269(i_{ab} - 23)  .
\label{eq:mag_hd_fit_simard}
\end{equation}
\noindent
For our measurements in GOODS-South, a least-square fit to the median values
in the same magnitude range yields
\begin{equation}
\ln (h_d/{\rm arcsec}) = -1.346 - 0.193(i_{ab} -23  ).
\label{eq:mag_hd_fit_welikala}
\end{equation}
\noindent

As well as a slower fall-off of size with magnitude, the scale lengths found in this work are
systematically smaller than the ones reported by
\cite{simard2002} given the statistical uncertainties shown in the
figure. The disagreement is most notable at the brightest magnitudes
and the sizes agree at $i=$25.5. The effects on shear measurement and
calibration schemes that use these prior distributions of
galaxy size will need to be examined in future work. Future weak
lensing measurements also depend on ultra-precise measurements of the galaxy
ellipticity distribution \citep{viola2014}. In Fig.~\ref{fig:ellipticities}, we show the distribution
of $e_g$ for the samples which are obtained from the
quadrupole moments computed on the best-fit summed bulge and disk
model, described in
Sect.~\ref{subsec:method_ellipticity}. We note the caveat that the
distribution of $e_g$ is set by the construction of the model and
limited by the assumption of the disk thickness. For most galaxies in the
sample, the ellipticity of the galaxy is determined by the disk component. We
find a cutoff value of $e_g=$0.8 which is similar to the one obtained
in the SDSS data by \cite{miller2013} using a numerical fit to their
ellipticity distribution. For the bulge, we find $e_b < 0.3$
although as noted previously, $e_b$ is constrained by imposing $\beta_0=$0.6 and we do not
fit for $e_b$ in this work. However, in the Appendix~\ref{app:priors_test}, we show
that the ellipticity distribution for the galaxy (both bulge and disk) is only weakly sensitive to
the choice of $\beta_0$ for the bulge.

Finally, an accurate knowledge of which galaxy properties are
correlated with the ellipticity is important for quantifying shear
biases. Any observed correlation between ellipticity and galaxy SED for
example suggests a correlation between shear and PSF quantities in the
case of a chromatic PSF. We examine correlations of the
galaxy ellipticity with redshift, overall galaxy colour (or SED) and
the disk scale length, which are shown in
Figs.~\ref{fig:ellipticity_correlations}.
We observe a weak dependence of the median ellipticity on the
redshift of the galaxy, with a suggestion of an increase in
ellipticity in the highest redshift bins ($2.5 < z \le 3.0$). We also observe a weak decrease in
ellipticity with the broadband $V-I$ colour in galaxies - this is most
pronounced in the colour range $2.0 < V-I \le 2.5$. The strongest correlation is between
ellipticity and the disk scale length i.e., galaxies with the largest
angular size have the highest ellipticities. Our result is qualitatively
consistent with that of \cite{kannawadi2019} who do not correct for
central extinction in the galaxies. In this work, we observe a correlation between $h_d$ and $e_g$ even after
accounting for central extinction so the reason for the
correlation needs to be further investigated. The ellipticity-colour
relation in particular shows a sizeable asymmetric intrinsic scatter
which is much larger than the noise associated with measuring
quadrupole moments on the HST images.

\begin{figure*}
\centering
\includegraphics[width=.3\textwidth]{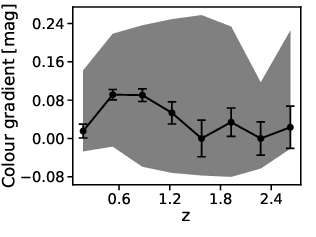}
\includegraphics[width=.3\textwidth]{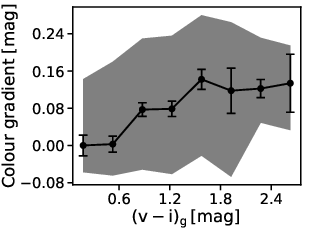}
\includegraphics[width=.3\textwidth]{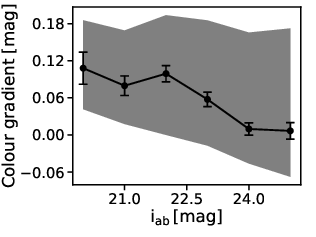}
\includegraphics[width=.3\textwidth]{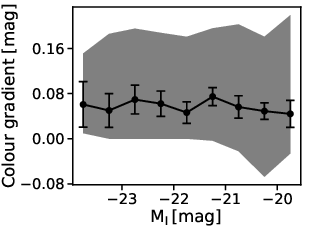}
\includegraphics[width=.3\textwidth]{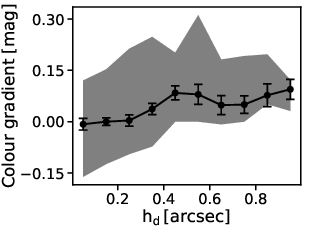}
\includegraphics[width=.3\textwidth]{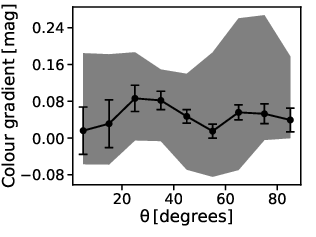}
\includegraphics[width=.3\textwidth]{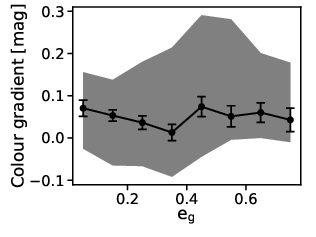}
\caption{Observed aperture colour gradients obtained from the galaxy model as a
  function of quantities independent of the model: (a) redshift; (b)
  $V-I$ colour; (c) apparent magnitude $i_{ab}$; (d) $M_I$; (e) $h_d$;
  (f) $\theta$ and (g) $e_g$. The colour gradient is obtained from the best-fit summed
  bulge and disk model  and is defined as the difference in colour
  between two apertures, $r < 0.5\,\mathrm{FWHM}$ and $0.5\,\mathrm{FWHM} < r < 4\,\mathrm{FWHM}$,
  where the $\mathrm{FWHM}$ is the full-width at half maximum of the PSF in the V
  and I bands. The median colour gradient is shown by the solid black line while the
  error bars in each bin show the Standard Error of the Mean (SEM). The shaded region
  shows the scatter in the colour gradient distribution and is bounded by the 25th and 75th 
percentiles in each bin.}
\label{fig:cg_trends}
\end{figure*}

\begin{figure}
\centering
\includegraphics[width=0.45\columnwidth]{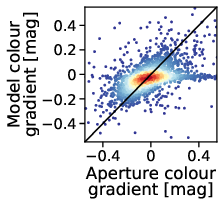}
\hfill
\includegraphics[width=0.45\columnwidth]{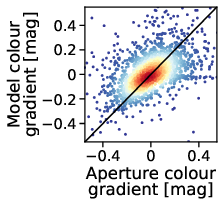}
\caption{Scatter plots comparing the aperture photometry estimate of
  the colour gradient with the best-fit PSF-convolved model for: (a)
  $\tau=$0 (no extinction) applied to all galaxies (left panel) and (b) the best
  value of $\tau=$4.0 applied to the \textit{Dusty} galaxies (around
  26$\%$ of the sample, right panel). A best $n=$1.0 and $R_e/h_d=$0.15 is
  applied to all galaxies when we fit to the bulge and disk flux
  densities and other bulge and disk parameters.}
\label{fig:cg_data_model_comparison}
\end{figure}

\begin{figure}
\centering
\includegraphics[width=0.85\columnwidth]{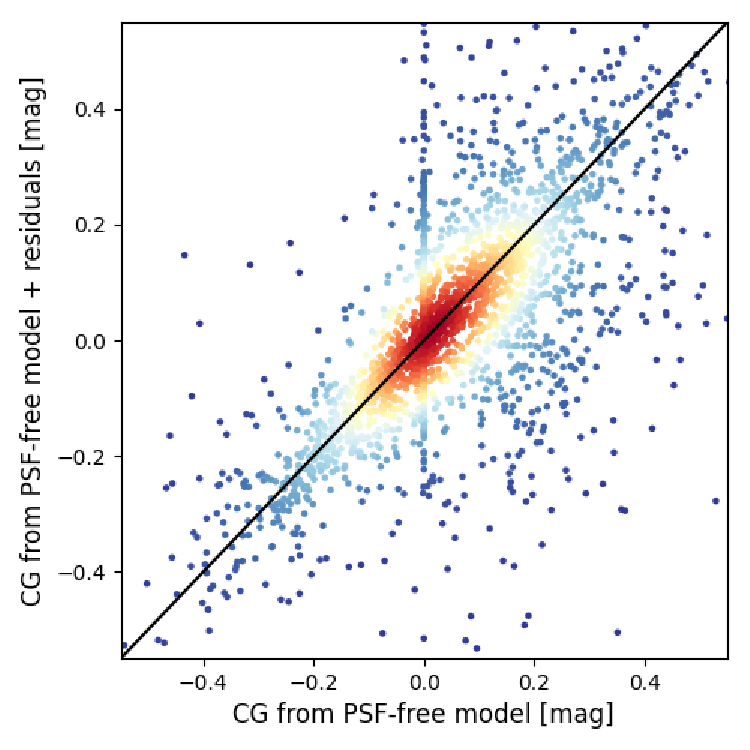}
\caption{Test of complex morphology on colour gradients: comparison of colour gradients measured on the PSF-free
  models in both apertures in the $V$ and $I$ bands with the colour
  gradient remeasured in the same apertures after adding the residual
  flux from model subtraction back into the apertures. The vertical concentration of points near model colour gradient $\approx0$ contains 348 galaxies (13.9$\%$ of the sample) and arises predominantly from objects with near-zero fitted bulge amplitudes in both the $v$ and $i$ bands (86.2$\%$ of this population).}
\label{fig:cg_residual_test}
\end{figure}

\subsection{Colour gradients}
\label{results:colour_gradients}

Colour gradients measured on PSF-free models enable us to track the
evolution of the underlying stellar populations in the bulge and disk
components of the galaxy. PSF-convolved models are useful for testing
the model fitting by comparing with the data. Here, we show (a) the PSF-removed colour gradients measured in the
observed frame and show how they correlate with various galaxy
properties; (b) the robustness of the colour gradients
by comparing gradients measured in the data with those measured using
the PSF-convolved models; (c) the effect of residual
light on the PSF-removed colour gradients in order to understand the
impact of patchy galaxy morphologies on the colour gradients. In each
case, we measure
colour gradients using the best-fit bulge and disk flux
densities in each band using the appropriate model. The colour
gradient reported in this work is the difference in observed $v-i$
colour between two radial apertures: $r < 0.5\,\mathrm{FWHM}$ and
$0.5\,\mathrm{FWHM} < r < 4\,\mathrm{FWHM}$ centred on the
fitted model centroid. The sizes of these apertures are therefore on
the scale of the HST ACS PSF. Error bars are propagated from model flux
uncertainties as described in Appendix~\ref{app:errors}. 

\subsubsection{PSF-removed colour gradients and correlation with galaxy properties}
\label{results:colour_gradient_results}

We show the evolution of PSF-free $V-I$ colour gradient in the
observed frame with various model-independent galaxy properties in
Fig.~\ref{fig:cg_trends}. As we noted in
Sect.~\ref{method:fit_disk}, the effect of extinction is included in
the colour gradients - they can result from a radial extinction profile even if
the best-fit unextincted stellar amplitudes of the bulge or the disk are zero.

The strongest correlation is with the
overall $V-I$ colour of the galaxy (the $V$ and $I$
magnitudes have been obtained from the Subaru data in the \citet{cardamone2010}
catalog). The bluest galaxies have close to zero
median colour gradient. Redder galaxies have more positive colour
gradients i.e., the core is redder than the outskirts. We also observe
a strong anti-correlation with the apparent magnitude of the galaxy
i.e., brighter galaxies have redder cores.
We observe a positive but declining median colour gradient (core redder than outskirts) in the
redshift bins $0.35 < \,z \le 1.4$ one
consistent with zero in the other redshift bins considered including
at $z \sim 0.5$ and at $z > 1.4$. There is a large asymmetric scatter in the colour gradient
distribution (much larger than the model uncertainties on the median
colour gradient) in any given bin of the galaxy parameter in
question.

We also observe a weaker correlation of colour gradient with $h_d$,
noting an increase beyond 0.7$\arcsec$ and an anti-correlation with
$e_g$ up to $e_g=$0.4, although the gradient then increases and
plateaus for higher ellipticities. We see a positive colour gradient for an intermediate range of inclination angles
$25 \le \,\theta \le \, 50$\degree and for edge-on galaxies
($\theta \ge 60.0$\degree).

\subsubsection{Robustness of the colour gradients}

To test the robustness of the model measurements, we measure colour gradients using (a) aperture
photometry on the $v$ and $i$ ACS images and (b) the best-fit
PSF-convolved summed bulge and disk model in the same bands.
Aperture photometry estimates done on the data are noisier than model estimates. On the other hand, a model-based colour gradient
can be systematically offset from the true colour gradient due to
incomplete modelling of effects contributing to the colour gradients
or due to incorrect assumptions in the model. We illustrate this in
Fig.~\ref{fig:cg_data_model_comparison} where we compare the colour
gradients estimated using either method for (1) no
  extinction ($\tau=0$) applied to any galaxies in the sample and (2) a
  best-fit extinction value ($\tau=4.0$) applied to the \textit{Dusty}
  sample and $\tau=$0 for the \textit{Dustless} sample. In both cases, the
  best-fit sample values of $n=$1.0 and $R_e/h_d=$0.15 are used.

When no extinction is applied to any galaxy, we observe a large bias in the model colour gradients relative to the
  aperture photometry ones. Most of these affected model colour
  gradients are biased towards zero for positive
  aperture photometry colour gradients.
  The outliers in the HST sample (scattered along the $y=0$ line in Fig.~\ref{fig:cg_data_model_comparison}) constitute 7.4$\%$ of the sample
for a depth of $i_{ab}=$25.5 and is generally magnitude independent (7.0$\%$ of the sample
for a depth of $i_{ab}=$24.5). When the
  extinction modelling which affects only the \textit{Dusty}
  sub-sample is applied,  we observe a much improved agreement between
  the two methods although there is some scatter in the
  relation. The fraction of outliers which are scattered along the
  $y=0$ line reduces now to 3$\%$. So despite the known presence
  of patchy morphologies in the HST data, we are still able to 
  recover the colour variation in the data if we include the
  extinction model which reddens the centres of the
  affected galaxies.

Finally, we assess the impact of complex morphology on the colour
gradient measurements in the full sample. We first measure the PSF-free colour gradient in the V and
I bands. Then we add the residual flux into the same apertures and
re-measure the colour gradient. Although the residual flux is
convolved with the PSF, it contains colour contributions from
non-smooth morphologies. This is therefore a direct test of any
systematic bias in the model colour gradients from
non-smooth morphologies. The 7.4$\%$ of outlier galaxies are confirmed to have $B/T_v\sim0$
and $B/T_H > 0$ and are mostly face-on irregular
galaxies. We note a concentration of points in
  Fig.~\ref{fig:cg_residual_test} at model colour gradient $\sim0$,
  comprising 348 galaxies (13.9$\%$ of the sample), arising
  predominantly from galaxies with near-zero fitted bulge amplitudes in
  both the $v$ and $i$ bands (86.2$\%$ of this population). For these
  galaxies, the PSF-free model reduces to a single disk component,
  producing zero colour gradient by construction. Overall, we see in Fig.~\ref{fig:cg_residual_test}, that the
majority of galaxies show no systematic bias in the colour gradients
as a result of complex morphology which indicates that the model
captures the bulk variation in the colour gradients.

It should be noted that the extrapolation to weak lensing surveys
depends on the same selection criteria being used. The fact that the
colour gradient-redshift relation is not significantly affected when
we remove the outliers from the sample suggests that they should not
contribute significantly to the colour gradient bias. However, they can still give rise to a separate morphology bias.
The effect of these outliers on the shear bias may be investigated via
simulations and emulations.

\section{DISCUSSION}
\label{sec:discussion}

\subsection{Implications for galaxy evolution}

We first discuss structural properties of galaxies measured using the
bulge and disc models. The one-component fitting suggests a
disc-dominated Main Sequence population with disc profiles, consistent
with other studies observing disc-like characteristics from surface
brightness profiles, axial ratios and kinematics at $1 < z < 2.5$. For
the two-component fitting, the sample is best described by a discy
bulge plus a disc of finite thickness, yielding a double exponential
surface brightness profile. The distribution of $n$ from the
two-component fitting to the full sample is consistent with
pseudo-bulges whose profiles resemble exponential or near-exponential
discs rather than de Vaucouleurs profiles ($n= 4.0$) characteristic of
`classical' bulges \citep{gadotti2009}. Our findings show that central
regions of HST galaxies share similar structural properties to Freeman
Type-I late-type spiral galaxies \citep{macarthur2003} and local dwarf
galaxies \citep{herrmann2013}, also well-modelled by broken or double
exponential profiles. We also find $R_e/h_d\sim0.15$ (from the maximum
likelihood of the stacked sample), consistent with bulge-to-disk size
ratios of late-type systems dominated by pseudo-bulges. While
\cite{Dimauro_2018} report a median $\langle R_e/h_d \rangle\sim0.3$
for their two-component systems, our value lies within their broad
distribution and likely reflects high-redshift fitting uncertainties,
sample selection and the relative prominence of pseudo-bulges in our
galaxies at $z\sim1$. Accounting for extinction within
  our 3D modelling framework also has a measurable impact on the recovered structural parameters
  (see Sect.~\ref{results:bulge_disk_fits}), although this does not constitute a controlled 3D-versus-2D benchmark.

Understanding pseudo-bulge formation remains an open question. The standard picture invokes secular evolution through slow internal processes such as bar-driven gas flows, disc instabilities and stellar migration \citep{kormendy2004}, contrasting with classical bulges formed through major mergers \citep{brooks2016}. In the Auriga simulations, \cite{Gargiulo_2019} find the majority of bulges are pseudo-bulges with mostly in-situ origin (21 per cent with negligible accreted fraction), forming most of their stars centrally by $z=0$. They note that part of this growth is rapid and merger-driven while another part is more secular. Alternatively, \cite{Okamoto2013} suggest pseudo-bulge formation through high-redshift starbursts before $z \sim 2$, where central disky components formed at these epochs account for most of the final pseudo-bulge mass via angular momentum exchange with accreting material \citep{Stark2008}. Different formation mechanisms likely dominate depending on galaxy mass, and our HST results could constrain these through $B/T$ and component sizes.

Resolved HST imaging samples stellar light above and below the 4000\,\AA\ break out to $z \sim 2.5$. While this work does not provide extinction-corrected bulge and disc SEDs directly, it enables more robust component SEDs through appropriate dust opacity corrections. Future attenuation corrections (e.g. in SED-fitting) should match the dust opacity associated with our measured surface brightness distributions. We find similar bulge and disc optical colours (observed frame) at $z \ge 1.15$ and redder bulges at $0.35 < z < 1.15$; extinction corrections would yield even redder centres. This trend---redder centres towards high mass and later times---is consistent with previous studies, except at $z < 0.35$ where we observe near-zero median colour gradients, possibly due to higher fractions of low-mass galaxies in this bin.

The observed colour gradients suggest similar bulge and disc colours at $z \sim 1$, except at $z \sim 0.5$ where redder centres could indicate either older stellar populations, lower specific star formation rates in bulges, or higher central extinction. While positive, dust-free colour gradients at $z \sim 0.5$ (redder bulges than discs) would support inside-out secular formation, massive star-forming galaxies at cosmic noon ($z \sim 2$) are dusty. Our surface brightness modelling provides opacity values that will enable extinction-corrected SED-fitting of bulge and disc components in future work to derive rest-frame stellar population properties and more reliable formation scenario interpretations, complementing other optical approaches using the dust-sensitive UV slope $\beta$ and rest-frame UVI colour-colour diagrams.

\subsection{Implications for weak lensing shear measurements}

The measurements on HST deep images are important for reconstructing accurate distributions of bulges and discs that are essential in cosmic shear calibration \citep{congedo2024}. Our results suggest that in the optical-only bands, the median colour gradient is small across $0< z <3$, although we detect a slightly positive gradient. While this suggests the resulting shear bias from colour gradients could average to zero (since shear is computed over many galaxies), we observe large, asymmetric scatter in the colour gradient distribution which may not average to zero. This scatter includes both noise in the component SEDs and photometric noise. 

We observe a strong correlation between colour gradient and galaxy SED (through $V-I$ colour), with redder galaxies showing redder cores. Since galaxy colour strongly correlates with environment \citep{hogg2004, grutzbauch2011}, this produces environment-dependent shear bias. A colour or environment-dependent correction to the colour gradient bias may therefore be needed for ultra-precise dark energy measurements.

Since the model captures colour variation in galaxies in the HST
sample, forward modelling of colour gradient shear bias is
feasible. We can construct component SEDs to inform the PSF model
assigned to each component, minimising shear bias from both the total
SED and individual components in the presence of noise. For
\textit{Euclid}, which has a single broadband filter, this approach
allows statistical correction for colour gradients. Recent studies
suggest the CANDELS fields may contain just enough galaxies for colour
gradient calibration \citep{voigt2012,semboloni2013}, although these
assumed classical bulges with de Vaucouleurs profiles. The colour
gradients we observe (e.g., at $z\sim1$) are much smaller than those
assumed by \cite{semboloni2013}, suggesting less calibration data may
be needed for \textit{Euclid}. The 3D framework may
  also be useful for modelling wavelength-dependent ellipticity
  gradients and assessing their impact on shear measurement.

Regarding outlier galaxies from the model-fitting (7.0 per cent of the HST sample for $i_{\rm AB}=24.5$), extrapolation to weak lensing surveys depends on consistent selection criteria. The colour gradient-redshift relation is not significantly affected when removing outliers, suggesting they should not contribute significantly to colour gradient bias, although they may still produce separate morphology bias. The effect of these outliers on shear bias may be investigated via simulations and emulations.

\section{SUMMARY}
\label{sec:summary}

We model the surface brightness distribution of the bulge and disk
components of galaxies in 3D and perform joint fitting to images
of galaxies in multiple passbands. The simultaneous joint fitting approach
provides stronger constraints on galaxy structural parameters than fitting to a
single band. We apply the method to a sample of 2505 galaxies in the
GOODS-South field with multi-band ($bvizYJH$)
imaging from the \textit{Hubble Space Telescope} to a depth of
$i_{ab} < 25.5$. Our findings are summarised as follows:

\begin{itemize}

\item Single-S\'ersic fitting of the 3D models suggests a broadly disk-like HST population, with a median $n_g=1.5$, only slightly more centrally concentrated than a pure exponential profile.
\item With two component fitting, the HST sample is largely
  disk-dominated and is best characterised
  by $n=$1.0 for the bulge and a bulge effective radius to disk scale
  length ratio $R_e/h_d=0.15$. While the majority of the HST population can be modelled
without extinction, roughly 26 percent of the sample are mostly
edge-on galaxies with significant central obscuration. We find a
B-band rest-frame opacity corresponding to this obscuration of
$\tau\approx 4.0$. We
find that the sample is largely disk-dominated (i.e. a lower bulge
fraction).
\item When $n$, $\tau$ and $Re/h_d$ are fitted for each galaxy,
  neglecting extinction would cause us to report a higher fraction of
  low $n$ bulges (by approximately $10 \%$ for $n = 0.5$) in the HST sample and a higher fraction of galaxies where the model
underpredicts the colour gradients, compared to the colour gradients
in the data.
\item We derive new weak lensing priors with this work. The median
  $h_d= 0.22 $\arcsec to a depth of $i_{ab}= 25.5$. Disk scale lengths are systematically lower at all
  magnitudes than  those from \cite{miller2013}. We
  also observe a weak correlation between galaxy ellipticity and
  redshift and a strong correlation between galaxy ellipticity and
  disk scale length.
\item Despite the low bulge fraction, we detect a slight positive median colour
  gradient ($< 0.05$ mag) in the redshift range $0.35 < z < 1.15$ i.e, bulge redder than
  the disk, and zero median colour
  gradient at all other redshifts. This implies that the bulge and
  disk could have similar stellar populations although this needs to be confirmed
  via SED-fitting. However, the colour gradient
  distribution has a large asymmetric scatter. 
\item  We observe a weak correlation between colour gradients and
  redshift but a strong correlation between colour gradients and
  overall galaxy colour. Redder galaxies show increasingly positive colour gradients (core becomes
redder compared to the outskirts). In addition, the colour gradient
distribution shows a large asymmetric scatter across the redshift and
colour range of the sample.

\end{itemize}

\section{DATA SOURCES}

The sources of the data used in this paper are listed below:

\begin{itemize}
\item Multiwavelength Survey by Yale-Chile (MUSYC, \cite{cardamone2010}).
\item GOODS-S HST Advanced Camera for Surveys (ACS) $z$ band detection
  catalog (\url{https://archive.stsci.edu/pub/hlsp/goods/catalog_r2/})
\item Grism spectroscopic redshifts from 3D-HST
\item ACS images and weight maps are version 2.0 of the reduced,
  calibrated, stacked mosaics
  (\url{https://archive.stsci.edu/pub/hlsp/goods/v2/h_goods_v2.0_rdm.html})
\item WFC3 images from the \textit{CANDELS} data (\url{https://archive.stsci.edu/pub/hlsp/candels/goods-s/gs-tot/v1.0/hlsp_candels_hst_acs-wfc3_gs-tot_readme_v1.0.pdf})
\end{itemize}

\section*{DATA AVAILABILITY}

The data underlying this article are publicly available from the sources listed in the Data Sources section.

\section*{ACKNOWLEDGEMENTS}

NW thanks R.~Tuffs for providing data from the P11 models which
were used in this work, M.~Holliman for help with high performance
computing resources, Bryan Gillis for discussions and Shona Matthew
for editing the paper. NW and GC acknowledge support from the UK
Space Agency (UKSA). AT acknowledges support from the Science and
Technology Facilities Council (STFC) and UKSA.
This work is based on observations taken by the CANDELS Multi-Cycle
Treasury Program with the NASA/ESA HST, which is operated by the
Association of Universities for Research in Astronomy, Inc., under
NASA contract NAS5-26555.

\bibliographystyle{mnras}
\bibliography{modeling}

\appendix
\section{Appendices}
\label{appendix}

\begin{figure}
\centering  
\includegraphics[width=.9\columnwidth]{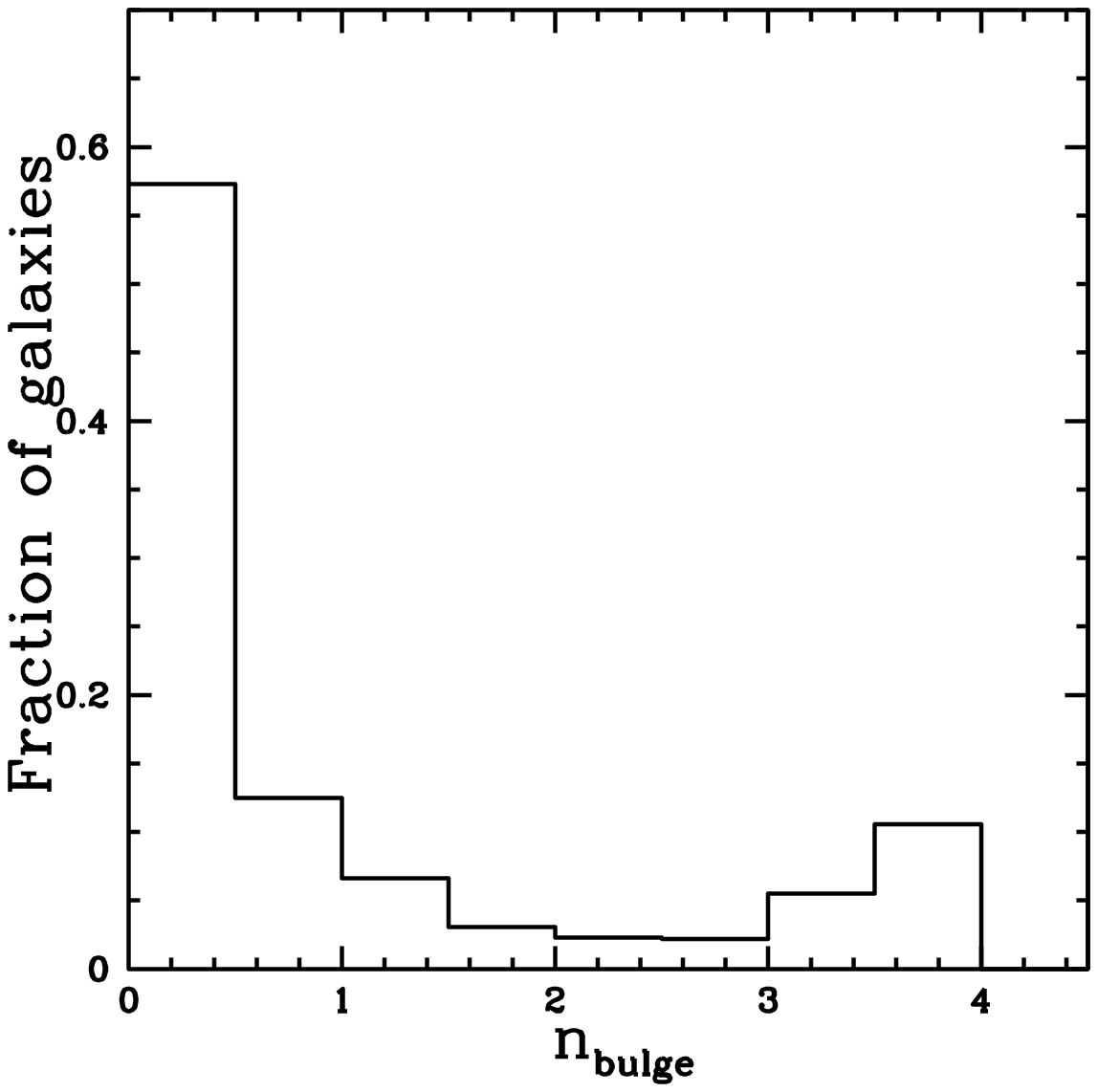}
 \caption{Distribution of bulge S\'ersic index from two component
   fitting to the galaxy sample.}
\label{fig:n_individual_fits}
\end{figure}

\begin{figure}
\includegraphics[width=0.7\columnwidth]{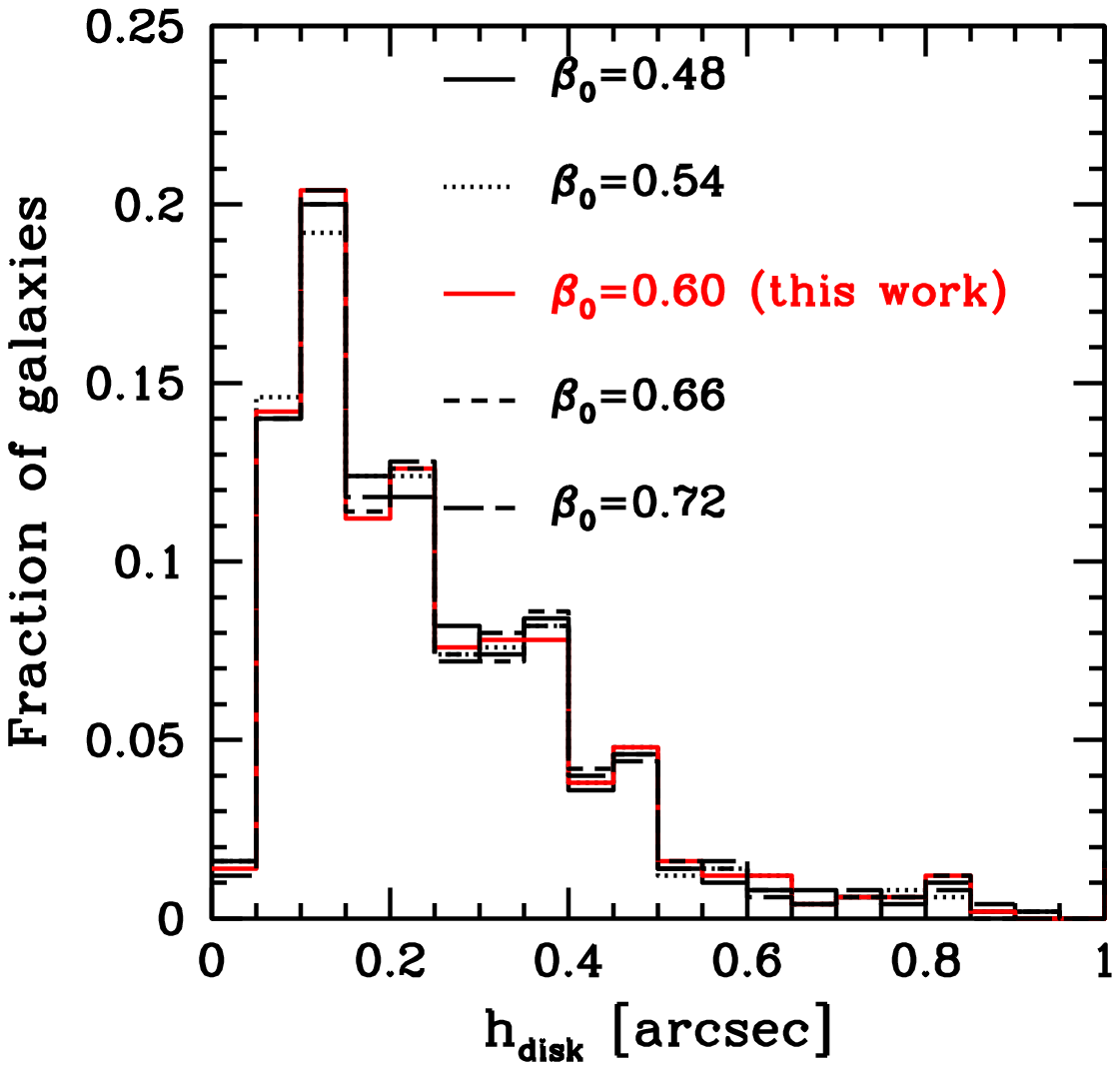}
\includegraphics[width=0.7\columnwidth]{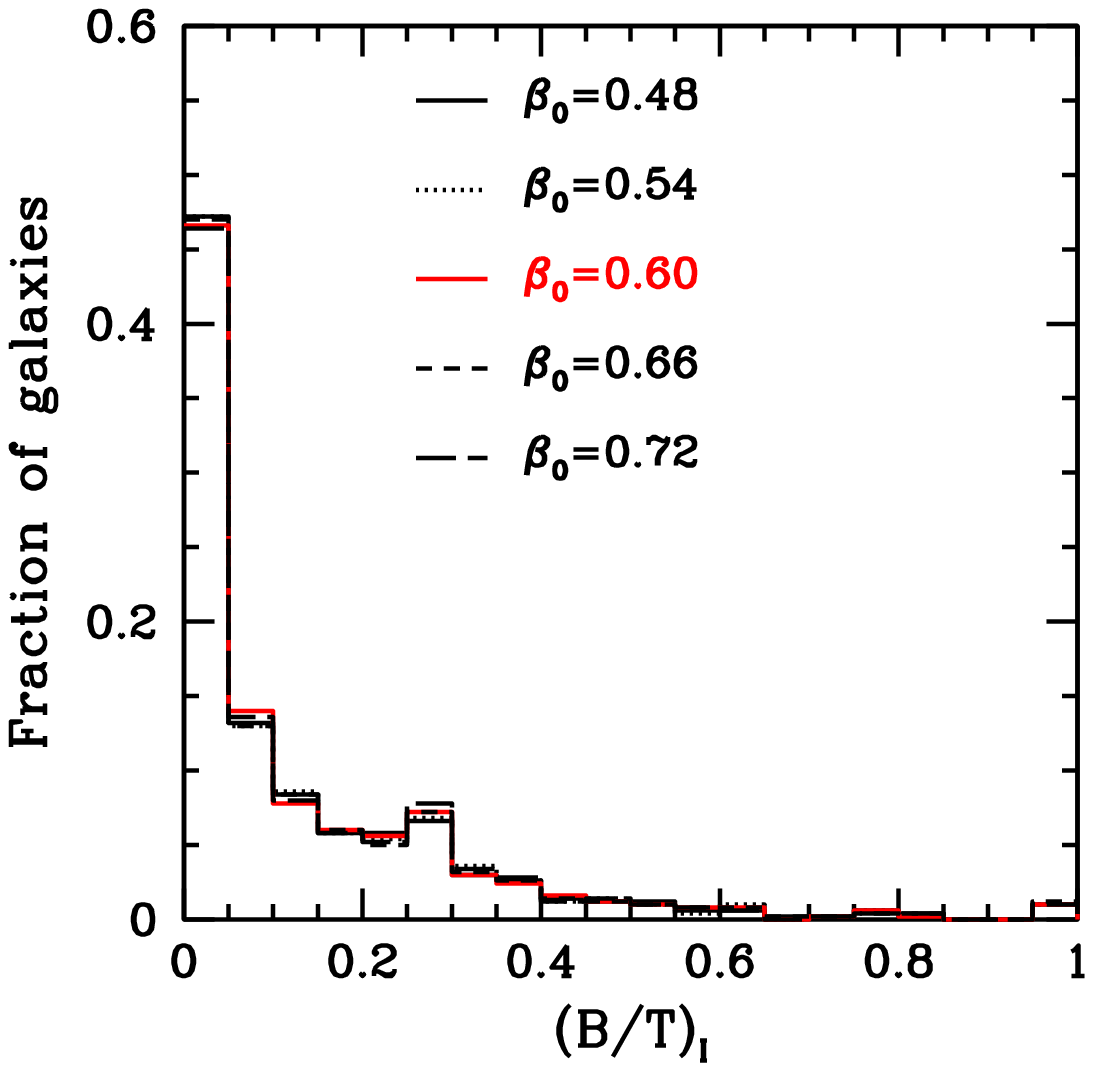}
\includegraphics[width=0.7\columnwidth]{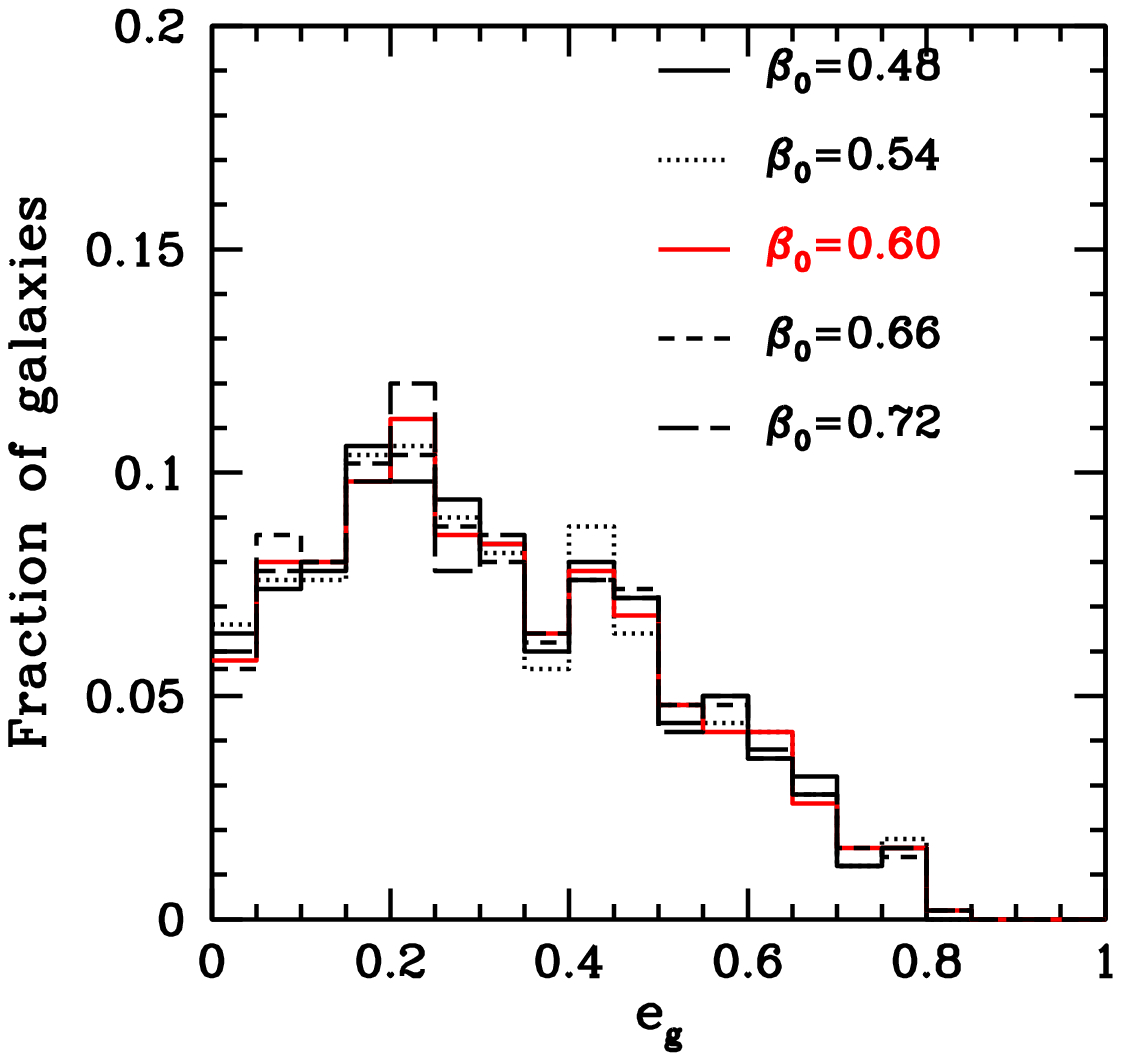}
\vspace{-4mm} 
\caption{Sensitivity of the distributions of $h_d$, $(B/T)_I$ and $e_g$ obtained from the two-component
   fitting on the assumed value for the bulge intrinsic flattening
   ratio $\beta_0$ (described in Sect.~\ref{models:bulge}). The
   same 500 randomly chosen galaxies are used in each distribution and
   we use $n=$1.0, $R_e/h_d=$0.15 and $\tau=$0 for the \textit{Dustless} sample. The red line corresponds to the fiducial value
   of $\beta_0$ used in this work and the reference distribution in the
   K-S test. The other values of $\beta_0$ are $\pm$10$\%$ and
   $\pm$20$\%$  of the fiducial value.}
\label{fig:priors_bulge}
\end{figure}

\begin{figure}
\centering
\includegraphics[width=0.45\columnwidth]{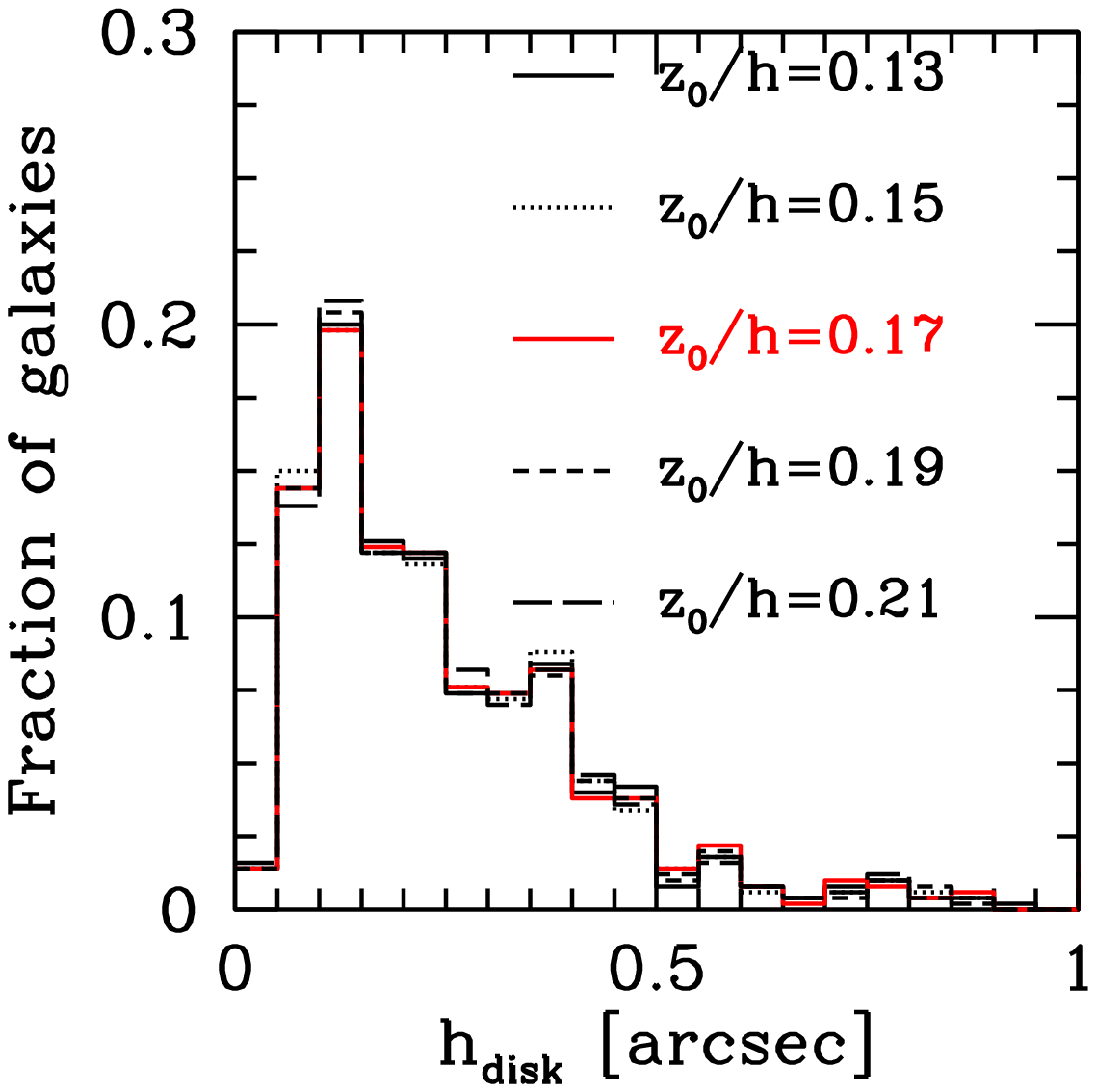}
\includegraphics[width=0.45\columnwidth]{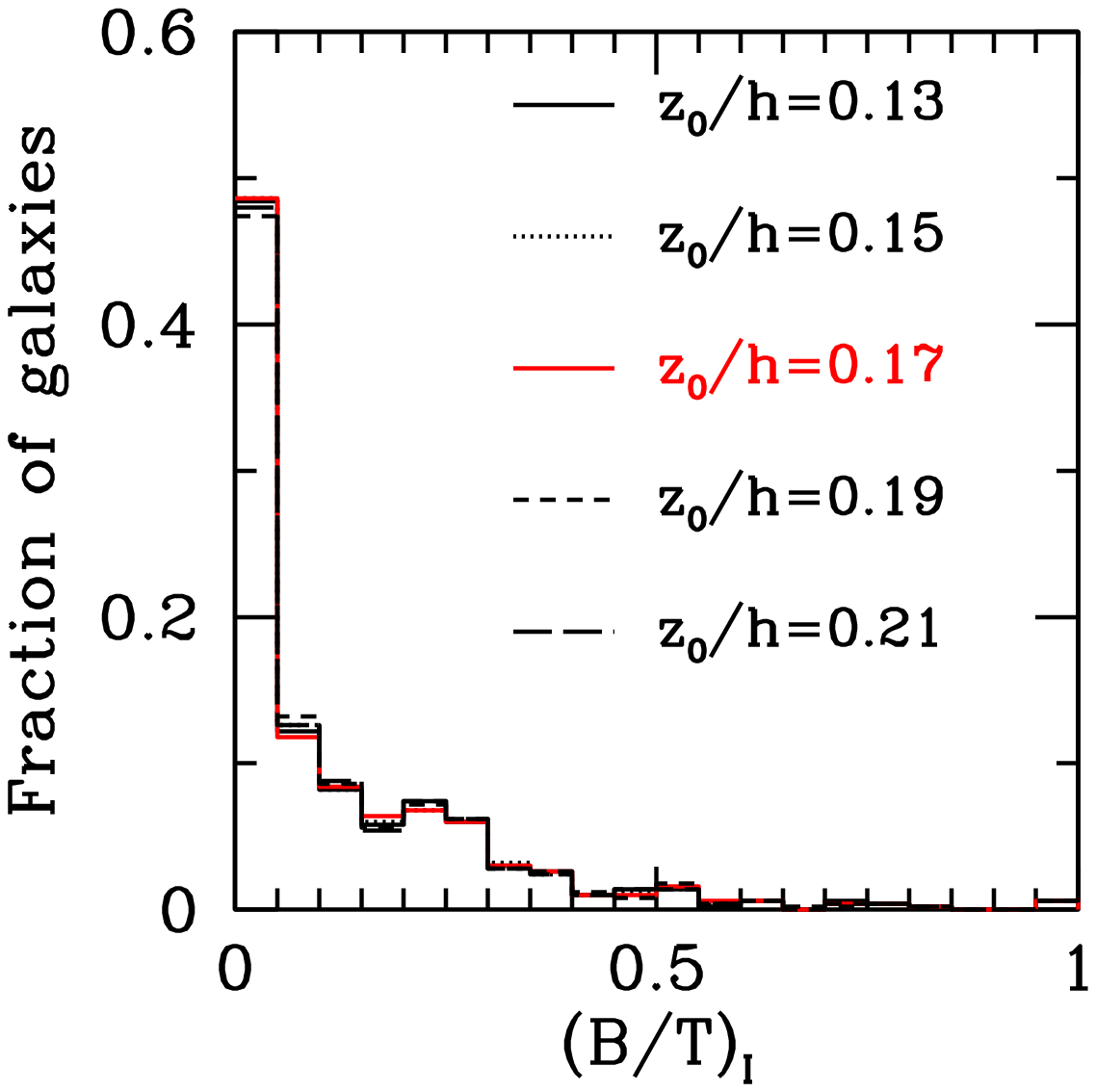}
\caption{Sensitivity of the distributions of $h_d$ 
   and $(B/T)_I$ obtained from the two-component
   fitting on the assumed value for the disk vertical-to-horizontal
   thickness ratio $z_0/h$ (see Sect.~\ref{models:disk}). The
   same 500 randomly chosen galaxies are used in each distribution and
   we use $n=$1.0, $R_e/h_d=$0.15 and $\tau=$0 corresponding to the
   \textit{Dustless} sample. The red line corresponds to the fiducial value
   of $z_0/h$ used in this work and the reference distribution in the
   K-S test. The other values of $z_0/h$ are $\pm$10$\%$ and
   $\pm$20$\%$  of the fiducial value.}
\label{fig:priors_disk}
\end{figure}

\subsection{Single S\'ersic fits}
\label{app:single_sersic_fits}

The $\chi^2$ of the Single S\'ersic fit is defined as	

\begin{equation}
\chi^2 = \sum_{i,j}{ \left(  \frac{d_{ij}  -  a_im_{ij}}{\sigma_{ij}}
  \right)^2} , 
\label{eq:method_one_comp_2}
\end{equation}
\noindent
where $d_{ij}$ is the HST flux density in pixel $i$ in the $j$th
filter, $m_{ij}$ is the model flux density in the same pixel and
filter and $\sigma_{ij}$ is the RMS error in that pixel and filter. Both
$d_{ij}$ and $\sigma_{ij}$ are obtained from the HST images (see Sect.~\ref{subsec:data_hst_flux_err}). 
The sum is over all pixels $i$ in the image which are
not masked out (see Sect.~\ref{subsec:data_masking} for details of
masking). The amplitude of the model in the $i$th band that minimises
$\chi^2$ is

\begin{equation}
a_i = \frac  {\sum_{i,j}{ d_{ij}m_{ij}}/{\sigma_{ij}^2} }
{\sum_{ij} {m_{ij}^2}/ {\sigma_{ij}^2 } }   ,  
\label{eq:method_one_comp_3}
\end{equation}
\noindent
provided the denominator (determinant of the $\chi^2$ matrix) is
non-zero. 

\subsection{Calculation of bulge and disk unextincted stellar component amplitudes}
\label{app:amplitudes}

We define the determinant of the $\chi^2$ matrix as
\begin{equation}
  D = \sum_{i,j}{ \left( \frac{m^{disk}_{ij}}{\sigma_{ij}}  \right)^2} \sum_{i,j}{ \left( \frac{m^{bulge}_{ij}}{\sigma_{ij}}  \right)^2}
   - \sum_{i,j}{ \left(
       \frac{m^{disk}_{ij}m^{bulge}_{ij}}{\sigma_{ij}^2} \right) },
\label{eq:determinant}
\end{equation}
\noindent
where $m_{i,j}$ is flux density in the $i$th pixel and $j$th filter of
the model image and $\chi^2$ is then given as

\begin{equation}
\chi^2 = \sum_{j,j}{ \left(  \frac{d_{ij}}{\sigma_{ij}}
  -  \left( a^{bulge}_i \frac{m_{ij}}{\sigma_{ij}} +  a^{disk}_i \frac{m_{ij}}{\sigma_{ij}}  \right)   \right)^2}, 
\label{eq:chi2}
\end{equation}
\noindent
where $a^{bulge}_i$ and $a^{disk}_i$ are the amplitudes of the bulge
and disk components, respectively, in the $i$th filter. 
If $D\leq0$, we revert to bulge-only fitting, setting the flux in the
disk to zero:
\begin{equation}
  a^{disk}_i  = 0,
\label{eq:method_two_comp_6}
\end{equation}
\noindent
and

\begin{equation}
a^{bulge}_i = \frac{ \sum_{i,j}{ \left(\frac{m^{bulge}_{ij}}{\sigma_{ij}}  \right)^2 }}  {
    \sum_{i,j}{\frac{d_{ij}m^{bulge}_{ij}}{\sigma^2_{ij}  }    } } .
\label{eq:method_two_comp_7}
\end{equation}

\noindent
For $D > 0$, $a^{bulge}$ and $a^{disk}$ are computed according to

\begin{multline}
a^{\mathrm{bulge}}_i =
\frac{
 \sum_{i,j}\!\left(\frac{m^{\mathrm{disk}}_{ij}}{\sigma_{ij}}\right)^2 
 \sum_{i,j}\!\frac{d_{ij}m^{\mathrm{bulge}}_{ij}}{\sigma_{ij}^2}
 - \sum_{i,j}\!\frac{m^{\mathrm{bulge}}_{ij}m^{\mathrm{disk}}_{ij}}{\sigma_{ij}^2}
 \sum_{i,j}\!\frac{d_{ij}m^{\mathrm{disk}}_{ij}}{\sigma_{ij}^2}
}{D},
\label{eq:method_two_comp_8}
\end{multline}

\noindent

and

\begin{equation}
a^{\mathrm{disk}}_i =
\frac{
  \displaystyle
  \Biggl[
    \sum_{i,j}\!\left(\frac{m^{\mathrm{bulge}}_{ij}}{\sigma_{ij}}\right)^{2}
    \sum_{i,j}\!\frac{d_{ij} m^{\mathrm{disk}}_{ij}}{\sigma_{ij}^{2}}
    -
    \sum_{i,j}\!\frac{m^{\mathrm{bulge}}_{ij} m^{\mathrm{disk}}_{ij}}{\sigma_{ij}^{2}}
    \sum_{i,j}\!\frac{d_{ij} m^{\mathrm{bulge}}_{ij}}{\sigma_{ij}^{2}}
  \Biggr]
}{
  \displaystyle D
}.
\label{method_two_comp_9}
\end{equation}

\noindent

If $D > 0$ but we encounter either a negative bulge or disk amplitude
($a^{bulge}_i < 0$ or $a^{disk}_i < 0$), we revert to one-component
fitting (see Sect.~\ref{method:single_sersic_fitting}), putting all the flux of the galaxy into the remaining
component. So in the case of a negative bulge amplitude:

\begin{equation}
  a^{bulge}_i=0,
 \label{eq:method_two_comp_10}
\end{equation}
\noindent
and

\begin{equation}
a^{\mathrm{disk}}_i =
\frac{
  \displaystyle\sum_{i,j}\!
  \left(\frac{d_{ij} m^{\mathrm{disk}}_{ij}}{\sigma_{ij}^{2}}\right)^{2}
}{
  \displaystyle\sum_{i,j}\!
  \left(\frac{m^{\mathrm{disk}}_{ij}}{\sigma_{ij}}\right)^{2}
}.
\label{eq:method_two_comp_11}
\end{equation}

\noindent

In the case of a negative disk amplitude:
\begin{equation}
 a^{disk}_i=0,
\label{eq:method_two_comp_12}
\end{equation}
\noindent
and
\begin{equation}
a^{bulge}_i= \frac{ \sum_{j}\sum_{i}{\frac{d_{ij}m^{bulge}_{ij}}{\sigma_{ij}^2} }  } {
  \sum_{j}\sum_{i}{ {\left
      (\frac{m^{bulge}_{ij}}{\sigma_{ij}}\right)^2} } } .  
 \label{eq:method_two_comp_13}
\end{equation}
\noindent

\subsection{Computing model parameter uncertainties}
\label{app:errors}

Uncertainties on model parameters are obtained by evaluating $\chi^2$ around the best-fit
minimum. We first choose the
global minimum $[p_{min},\chi^2_{min}]$ and two sampled points on either side
of this minimum $[p_{+},\chi^2_{+}]$ and $[p_{-},\chi^2_{-}]$, where
$p_{min}$, $p_{+}$ and $p_{-}$ are the parameter values at the global
minimum and the two neighbouring parameter values sampled on either
side of it ($\pm 5\%$ on either side of the best-fit parameter), and $\chi^2_{min}$, $\chi^2_{+}$ and $\chi^2_{-}$
are the corresponding $\chi^2$ values. For a given parameter $p$, $\chi^2_{-}$ and
$\chi^2_{+}$ are evaluated by fitting to the galaxy images with $p$
fixed to $p_{-}$ and $p_{+}$ respectively. In the case where $p$ is
the flux density of one component (bulge or disk) in a particular HST
passband, we evaluate the contribution to $\chi^2_{-}$ or $\chi^2 _{+}$ in that passband
by fixing that component's amplitude to $p_{-}$ or $p_{+}$ and calculating the remaining
component's amplitude in that band. In the case of $p$ being the
fitted flux density in a HST passband of the bulge, $a_{bulge}$, and we are estimating its
uncertainty, $\delta_p$, we can compute the disk amplitude $a_{disk}$
by computing the following quantities,

\begin{equation}
 D_1  = \sum_{j}\sum_{i}\frac{d_{ij} m^{disk}_{ij}}{\sigma_{ij}^2}   ,
\label{eq:method_two_comp_param_err_1}
\end{equation}
\noindent

\begin{equation}
 D_2=  \sum_{j}\sum_{i}\frac{m^{bulge}_{ij} m^{disk}_{ij}}{\sigma_{ij}^2}   ,
\label{eq:method_two_comp_param_err_2}
\end{equation}

\noindent
and

\begin{equation}
 D_3=  \sum_{j}\sum_{i}\frac{(m^{disk}_{ij})^2}{\sigma_{ij}^2}  .
\label{eq:method_two_comp_param_err_3}
\end{equation}
\noindent
In this case, since the bulge amplitude is fixed in this particular
band, $D_3$ becomes the determinant $D$ (see Eq.(~\ref{eq:determinant})). If $D$ is non-zero, then-

\begin{equation}
a_{disk} = (D_1 - a_{bulge}D_2)/D_3;
\label{eq:method_two_comp_16}
\end{equation}
\noindent
The contribution to $\chi^2$ from the remaining passbands is computed
in the same way as in Eq.~(\ref{eq:chi2}). If we define $x_0$ as $(p_+ - p_{min})$, $y_0$ as $(\chi^2_+ -
\chi^2_{min})$, $x_1$ as $(p_- - p_{min})$, $y_1$ as $(\chi^2_- -
\chi^2_{min})$, then the uncertainty in $p$ which we call $\delta_p$ is obtained using
a quadratic function for a least-squares fit,-

\begin{equation}
\delta_p = 2 \left(  \frac{ {x_0}^4 + {y_0}^4     }{  {x_0}^2y_0 +    {x_1}^2y_1    }          \right)^{1/2} .
\label{eq:least_squares_fit}
\end{equation}
\noindent
The above procedure is also applicable to finding the
uncertainty in all the other galaxy parameters. The model uncertainties in
the PSF-removed colour gradient are obtained from the bulge and disk flux
density uncertainties for each galaxy. We continue to observe a positive median colour gradient in the
second and third redshift bins $0.35 < z \le 0.70$ and
$0.70 < z \le 1.15$ to the same significance as before while the
median colour gradient is consistent with zero for all the other
redshift bins. 


\subsection{Validation of the colour gradient-redshift relation}

A non-zero colour gradient-redshift relation is potentially
important in some weak lensing surveys due to the effect of the chromatic
PSF. We conduct three tests of the colour gradient-redshift relation:
\begin{itemize}
\item In computing model uncertainties in Appendix~\ref{app:errors}, we use a least-squares fit
  using two neighbouring parameter values $\pm 5\%$ on either
  side of the global minimum. We investigate any sensitivity by using 
  points at $\pm 10\%$ from the minimum and find no significant change
  in the statistical results.
\item We compare the model uncertainties in the median colour gradient
  (i.e, the uncertainty associated with one galaxy corresponding to
  the median) with the standard error in the mean ($\sigma_i/\sqrt(N_i)$ where $\sigma_i$
  is the standard deviation of the colour gradient distribution in
  each redshift bin $i$ and $N_i$ is the number of galaxies in
  that bin). The two types of uncertainties are comparable in
  magnitude except for the $z\sim1.5$ bin. However, both types of error bars give
  a result that is consistent with a zero median colour gradient in
  those bins.
\end{itemize}

\subsection{Sensitivity to assumptions}
\label{app:priors_test}

We test the sensitivity of $(B/T)_I$, $h_d$ and $e_g$ to the assumed
value of $\beta_0$, and of $(B/T)_I$ and $h_d$ to $z_0/h$. We fix $n=1$, $R_e/h_d=0.15$,
$\tau=$0.0, corresponding to median values adopted for the \textit{Dustless}
sample. We use measurements from 500
randomly chosen galaxies in the sample. The results are shown in
Fig.~\ref{fig:priors_bulge} ($\beta_0$) and Fig.~\ref{fig:priors_disk}
($z_0/h$) for $\pm$10\% and $\pm$20\% variations about the fiducial
values. All distributions
   appear to be consistent with each other. This is also confirmed by
   a Kolmogorov-Smirnov (K-S) test performed on (a) each distribution
   corresponding to a different $\beta_0$ (or $z_0/h$) and (b) the
   reference distribution produced by two-component fitting with the
   adopted fiducial values.





   
\begin{table*}
\centering     
\begin{tabular}{ | p{5.5cm} | p{7cm} | }  
\hline
Column name & Column description \\ 
\hline\hline
\texttt{OBJID} & Object ID for HST galaxy\\
\texttt{RA} & Right Ascension (J2000)\\
\texttt{DEC} & Declination (J2000)\\
\texttt{ZPHOT} & photometric redshift from MUSYC catalog\\
\texttt{ZSPEC} & spectroscopic redshift (-1 if unavailable)\\
\texttt{IAB} & I band AB magnitude (MUSYC) [mag]\\
\texttt{VAB} & V band AB magnitude (MUSYC) [mag]\\
\texttt{M\_}I & Absolute magnitude in I band (MUSYC) [mag]\\
\texttt{CG\_VI\_DATA\_APERTURE} & Aperture $\delta_{V-I}$ [AB mag]\\
\texttt{CG\_VI\_DATA\_APERTURE\_{ERR}} & Uncertainty in 9\\   
\texttt{CG\_VI\_CONV\_APERTURE} & Aperture $\delta_{V-I}$ best-fit convolved model [AB mag]\\
\texttt{CG\_VI\_CONV\_APERTURE\_{ERR}} & Uncertainty in 11 [AB magnitudes]\\
\texttt{CG\_VI\_UNCONV\_APERTURE} & Aperture $\delta_{V-I}$ best-fit PSF-free model [AB mag]\\
\texttt{CG\_VI\_UNCONV\_APERTURE\_{ERR}} & Uncertainty in 13 [AB mag]\\   
\texttt{CG\_VI\_UNCONV\_BULGE\_DISK} & Bulge-disk colour difference for PSF-free models [mag]\\
\texttt{CG\_VI\_UNCONV\_BULGE\_DISK\_{ERR}} & Uncertainty in 12 [AB mag]\\
\texttt{NBULGE} & bulge S\'ersic index (fixed to sample mean 1.0)\\
\texttt{NBULGE\_{ERR}} & Uncertainty in NBULGE\\
\texttt{TAU} & B band face-on optical depth (no extinction in models)\\
\texttt{TAU\_{ERR}} & Uncertainty in TAU\\
\texttt{R\_E/H} & Fixed to sample mean 0.15\\
\texttt{R\_E/H\_ERR} & Uncertainty in \texttt{R\_E/H}\\
\texttt{H\_DISK} & $h_d$ [arcseconds]\\
\texttt{H\_DISK\_{ERR}} & Uncertainty in H\_DISK\_{ERR} [arcseconds]\\
\texttt{XCENTRE} & x coordinate of centroid \\
\texttt{YCENTRE} & y coordinate of centroid \\
\texttt{FLUX\_BULGE\_B} & Bulge F435W (B) flux density [nJy]\\
\texttt{FLUX\_BULGE\_B\_{ERR}} & Uncertainty in FLUX\_BULGE\_B [nJy]\\
\texttt{FLUX\_DISK\_B} & Disk F435W (B) flux density [nJy]\\
\texttt{FLUX\_DISK\_B\_{ERR}} & Uncertainty in FLUX\_DISK\_B [nJy]\\
\texttt{FLUX\_BULGE\_V} & Bulge F606W (V) flux density [nJy]\\
\texttt{FLUX\_BULGE\_V\_{ERR}} & Uncertainty in FLUX\_BULGE\_V \\
\texttt{FLUX\_DISK\_V} & Disk F606W (V) flux density [nJy]\\
\texttt{FLUX\_DISK\_V\_{ERR}} & Uncertainty in FLUX\_DISK\_V [nJy]\\
\texttt{FLUX\_BULGE\_I} & Bulge F775W (I) flux density [nJy]\\
\texttt{FLUX\_BULGE\_I\_{ERR}} & Uncertainty in FLUX\_BULGE\_I [nJy]\\
\texttt{FLUX\_DISK\_I} & Disk F775W (I) flux density [nJy]\\
\texttt{FLUX\_DISK\_I\_{ERR}} & Uncertainty in FLUX\_DISK\_I [nJy]\\
\texttt{FLUX\_BULGE\_Z} & Bulge F850LP (Z) flux density [nJy]\\
\texttt{FLUX\_BULGE\_Z\_{ERR}} & Uncertainty in FLUX\_BULGE\_Z \\
\texttt{FLUX\_DISK\_Z} & Disk F850LP (Z) flux density [nJy]\\
\texttt{FLUX\_DISK\_Z\_{ERR}} & Uncertainty in FLUX\_DISK\_Z [nJy]\\
\texttt{FLUX\_BULGE\_Y} & Bulge F105W (Y) flux density [nJy]\\
\texttt{FLUX\_BULGE\_Y\_{ERR}} & Uncertainty in FLUX\_BULGE\_Y\\
\texttt{FLUX\_DISK\_Y} & Disk F105W (Y) flux density [nJy]\\
\texttt{FLUX\_DISK\_Y\_{ERR}} & Uncertainty in FLUX\_DISK\_Y [nJy]\\
\texttt{FLUX\_BULGE\_J} & Bulge F125W (J) flux density [nJy]\\
\texttt{FLUX\_BULGE\_J\_{ERR}} & Uncertainty in FLUX\_BULGE\_J [nJy]\\
\texttt{FLUX\_DISK\_J} & Disk F125W (J) flux density [nJy]\\
\texttt{FLUX\_DISK\_J\_ERR} & Uncertainty in FLUX\_DISK\_J [nJy]\\
\texttt{FLUX\_BULGE\_H} & Bulge F160W (H) flux density [nJy]\\
\texttt{FLUX\_BULGE\_H\_{ERR}} & Uncertainty in $FLUX\_BULGE\_H$\\
\texttt{FLUX\_DISK\_H} & Disk F160W (H) flux density [nJy]\\
\texttt{FLUX\_DISK\_H\_{ERR}} & Uncertainty in $FLUX\_DISK\_H$ [nJy]\\
\texttt{CHI2\_BULGE\_DISK\_MIN} & $\chi^2$ bulge$+$disk fit BVIZYJH\\
\texttt{THETA\_DEG} & Inclination angle [deg]\\
\texttt{POS\_ANGLE} & Position angle [deg]\\
\texttt{E\_BULGE} & Ellipticity of bulge\\
\texttt{E\_DISK} & Ellipticity of disk\\
\texttt{E\_GALAXY} & Ellipticity of summed bulge+disk model\\
\texttt{NFREEDOM\_BULGE\_DISK} & Degrees of freedom (2 component fits)\\
\texttt{RE\_GALAXY} & Best fit $R_{e}^{g}$ (single S\'ersic fit) [arcseconds]\\
\texttt{N\_GALAXY} & Best fit S\'ersic index for single S\'ersic fit\\   
\texttt{CHI2\_SERSIC\_MIN} & $\chi^2$ for best fit model from single S\'ersic fit\\  
\texttt{NFREEDOM\_SERSIC} & Degrees of freedom for single S\'ersic fit\\
\hline
\end{tabular}
\caption{Catalog of fitted parameters from 3D model-fitting to HST
  galaxies in GOODS-S}
\label{table:catalog}
\end{table*}

\label{lastpage}

\end{document}